\documentclass[12pt,a4paper,twocolumn,aps,fleqn,nofootinbib,superscriptaddress,longbibliography]{revtex4}
\usepackage[utf8]{inputenc}
\usepackage[english]{babel}
\usepackage{epsf}
\usepackage{latexsym,amssymb,euscript}
\usepackage[dvips]{graphicx}
\usepackage[fleqn]{amsmath}
\usepackage{nicefrac}
\usepackage{slashed}
\usepackage{booktabs}
\usepackage{braket}
\usepackage{chngcntr}
\usepackage{bbold}
\usepackage{graphics}
\usepackage{graphicx}
\usepackage{mciteplus}
\usepackage{pdfpages}
\usepackage{setspace}
\usepackage[titletoc]{appendix}
\graphicspath{{./figures/}}
\usepackage[left=1.5cm, right=1.5cm, top=3.0cm, bottom=2.5cm]{geometry}
\usepackage{pstricks}
\usepackage{color}
\usepackage{xcolor}
\definecolor{urlblue}{rgb}{0.2,0.4,0.7}
\definecolor{citegreen}{rgb}{0,0.4,0.2}
\definecolor{linkred}{rgb}{0.9,0.2,0.1}
\usepackage{float}
\usepackage{academicons}
\definecolor{orcidlogocol}{HTML}{A6CE39}
\usepackage{fancyhdr}
\usepackage{hyperref}
\hypersetup{
 linktocpage = true,
 urlcolor = urlblue,
 colorlinks = true,
 linkcolor = urlblue,
 anchorcolor = urlblue,
 citecolor = urlblue,
 pdfstartview = {XYZ null null 1.25} 
           }
\newcommand{\drv}{{\rm d}}

\newcommand{\as}{\alpha_s}
\newcommand{\LQCD}{\Lambda_{\rm QCD}}
\newcommand{\MSb}{\overline{\rm MS}}

\newcommand{\CnLL}{{\cal C}_n^{\rm LL}}

\newcommand{\CnNLLstar}{{\cal C}_n^{{\rm NLL}^*}}

\newcommand{\CnHENLOstar}{{\cal C}_n^{{\rm HE}\text{-}{\rm NLO}^*}}

\newcommand{\DY}{\Delta Y}

\newcommand{\tcite}[1]{~\cite{#1}}
\newcommand{\tref}[1]{~\ref{#1}}
\newcommand{\eref}[1]{~\eqref{#1}}

\newcommand{\tarr}{
\begin{array}}
\newcommand{\earr}{\end{array}}

\newcommand{\orcidFGC}{\href{https://orcid.org/0000-0003-3299-2203}{\includegraphics[scale=0.1]{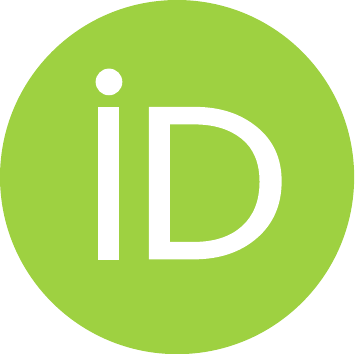}}}

\begin{document}

\newcounter{appcnt}

\setcounter{secnumdepth}{4}
\setcounter{tocdepth}{4}


\title{
  {\Large \bf High-energy emissions of light mesons plus heavy flavor \\ at the LHC and the Forward Physics Facility}
}

\author{Francesco Giovanni Celiberto \orcidFGC}
\email{fceliberto@ectstar.eu}
\affiliation{\setstretch{1.0}European Centre for Theoretical Studies in Nuclear Physics and Related Areas (ECT*), I-38123 Villazzano, Trento, Italy}
\affiliation{\setstretch{1.0}Fondazione Bruno Kessler (FBK), I-38123 Povo, Trento, Italy}
\affiliation{\setstretch{1.0}INFN-TIFPA Trento Institute of Fundamental Physics and Applications, I-38123 Povo, Trento, Italy}

\begin{abstract}
\vspace{0.35cm}
\begin{center}
 {\bf Abstract}
\end{center}
\vspace{0.35cm}
\hrule \vspace{0.50cm}
\footnotesize \setstretch{1.0}

The study of the dynamics of strong interactions in the high-energy regime is a core line of frontier researches at the LHC as well as at new-generation colliding facilities.
Here, the enhancement of energy logarithms due to diffractive semi-hard final states spoils the convergence of the perturbative series in the QCD running coupling, thus calling for an improvement of the pure collinear factorization that accounts for an all-order resummation of these large logarithmic contributions.
Motivated by the recent discovery that inclusive emissions of heavy-flavored particles allow for clear signals of a stabilization of high-energy resummed differential distributions under higher-order corrections and scale variations, we provide novel predictions of rapidity and azimuthal-angle observables for the inclusive hadroproduction of a light meson ($\eta$ or $\pi^\pm$) in association with a heavy-flavored hadron ($\Lambda_c$ or $b$-flavored hadron).
We calculate our observables in a hybrid high-energy and collinear factorization framework, where next-to-leading BFKL-resummed partonic cross sections are convoluted with collinear parton distributions and fragmentation functions. We consider kinematic ranges typically covered by acceptances of LHC detectors, and new ones coming from the combined tag of an ultra-forward particle at the future Forward Physics Facility (FPF) and of a central one at ATLAS via a tight timing-coincidence setup.
By performing a detailed study on uncertainties associated to collinear inputs via a replica-driven analysis, as well as on ones intrinsically coming from the high-energy resummation, we highlight the challenges and the steps required to gauge the feasibility of precision studies of our processes. 

\vspace{0.50cm} \hrule
\vspace{0.75cm}
{
 \setlength{\parindent}{0pt}
}
\vspace{2.00cm}
\end{abstract}


\maketitle

\newpage

\begingroup
 \hypersetup{linktoc = page, 
             }
 \phantom{.}\\\phantom{.}\\\phantom{.}
 { 
 \tableofcontents
 }
\endgroup

\vspace{0.95cm}
\hrule \vspace{0.50cm}


\section{Introductory remarks}
\label{sec:intro}

Novel opportunities in the exploration of the dynamics of fundamental interactions at new-generation colliders\tcite{Chapon:2020heu,Anchordoqui:2021ghd,Feng:2022inv,Hentschinski:2022xnd,Accardi:2012qut,AbdulKhalek:2021gbh,Khalek:2022bzd,Acosta:2022ejc,AlexanderAryshev:2022pkx,Brunner:2022usy,Arbuzov:2020cqg,Abazov:2021hku,Bernardi:2022hny,Amoroso:2022eow,Celiberto:2018hdy,2064676} herald the dawn of a new era for particle physics.
Accessing kinematic sectors so far uncharted will open us a window of possibilities to make stringent analyses of the Standard Model (SM) as well as (in)direct searches for deviations from SM predictions.

Important challenges inside the SM come from the strong-interaction sector. Here, the duality between perturbative and nonperturbative aspects of Quantum Chromodynamics (QCD) leads to yet unresolved puzzles in the answer of fundamental questions, such as the origin of hadrons' mass and spin, as well as the behavior of QCD observables in relevant kinematic corners of the phase space.

Precise tests of the dynamics underlying strong interactions essentially rely upon two core ingredients. 
On the one hand, a key role is played by our ability of performing more and more accurate calculations of high-energy parton scatterings via higher-order perturbative QCD techniques.
On the other hand, high-energy physics \emph{assumes} the knowledge of proton structure, whose inner dynamics is determined by motion and spin interactions among constituent partons.

A long series of successes in the description of data for hadron, lepton and lepton-hadron initiated reactions has been collected by the \emph{collinear factorization} (for a review, see Ref.\tcite{Collins:1989gx} and references therein), where partonic cross sections calculated within the pure perturbative-QCD framework are convoluted with collinear parton distribution functions (PDFs).

These densities carry information about the probability of finding a parton inside the struck hadron with a certain longitudinal momentum fraction, $x$. They evolve according to the Dokshitzer--Gribov--Lipatov--Altarelli--Parisi (DGLAP) equation\tcite{Gribov:1972ri,Gribov:1972rt,Lipatov:1974qm,Altarelli:1977zs,Dokshitzer:1977sg}. Collinear PDFs are suited to describe inclusive or semi-inclusive observables that are weekly sensitive to low-transverse momentum ($\kappa_T$) regimes.

Analogously to PDFs that depict hadronic initial states, the production mechanism of identified hadrons is portrayed by collinear fragmentation functions (FFs), which tell us about the probability of generating a given final-state hadron with momentum fraction $z$ from an outgoing collinear parton with longitudinal fraction $\zeta = x/z$.

By making use of collinear PDFs one neglects the information about the transverse-space distribution and motion of partons. Therefore, the description afforded by collinear factorization can be interpreted as a one-dimensional hadron mapping.

\emph{Vice versa}, getting a correct description of low-$\kappa_T$ observables requires a stretch towards a three-dimensional vision, which allows us to catch intrinsic effects of transverse motion and spin of partons and their interplay with the polarization state of the parent hadron.
Such a tomographic hadron imaging is naturally provided by the \emph{transverse-momentum-dependent} (TMD) factorization (see Refs.\tcite{Collins:1981uk,Collins:2011zzd} and references therein).
More in general, since parton densities and fragmentation functions have a nonperturbative nature, they have to be extracted from data via \emph{global fits} on combinations of hadronic processes.

Despite the remarkable achievements obtained at the hands of the pure collinear approach, there exist kinematic sectors where the genuine fixed-order, DGLAP-driven description must be complemented by the inclusion of enhanced logarithmic contributions that enter the perturbative expansion in the strong running coupling, $\alpha_s$, with a power that increases with the order.

Thus, to restore the convergence of the perturbative series, these logarithms need to be accounted for to all orders via \emph{ad hoc} procedures, known as \emph{resummations}.
Depending on the kinematic regimes covered, different kinds of logarithms appear and this brings us to the adoption of one or more specific all-order resummations.

As an example, the correct description of differential distributions for the inclusive production of hadrons, bosons or Drell--Yan leptons at low $\kappa_T$ relies on the use of the transverse-momentum (TM) resummation (see, \emph{e.g.}, Refs.\tcite{Catani:2000vq,Bozzi:2005wk,Bozzi:2008bb,Catani:2010pd,Catani:2011kr,Catani:2013tia,Catani:2015vma} and references therein).
TM-resummed predictions have been recently proposed for the hadroproduction of photon\tcite{Cieri:2015rqa,Alioli:2020qrd,Becher:2020ugp,Neumann:2021zkb}, Higgs\tcite{Ferrera:2016prr} and $W$-boson\tcite{Ju:2021lah} pairs, and for boson-plus-jet\tcite{Monni:2019yyr,Buonocore:2021akg} and $Z$-plus-photon\tcite{Wiesemann:2020gbm} systems.
Within the same framework, third-order fiducial predictions for Drell--Yan and Higgs spectra were provided in Refs.\tcite{Ebert:2020dfc,Re:2021con,Chen:2022cgv} and\tcite{Billis:2021ecs,Re:2021con}, respectively.

Conversely, when a physical observable is defined and/or measured close to the edges of its phase space, the so-called Sudakov effects coming from the emissions of soft and collinear gluons near \emph{threshold} become relevant and they need to be included via an appropriate resummation.
Several approaches exist in the literature to achieve the threshold resummation for the inclusive rates\tcite{Sterman:1986aj,Catani:1989ne,Catani:1996yz,Bonciani:2003nt,deFlorian:2005fzc,Ahrens:2009cxz,deFlorian:2012yg,Forte:2021wxe}.
In recent past, there have been several studies to perform resummation for the rapidity distributions as well (see, \emph{e.g.}, Refs.\tcite{Mukherjee:2006uu,Bolzoni:2006ky,Becher:2006nr,Becher:2007ty,Bonvini:2010tp,Ahmed:2014era,Banerjee:2018vvb}).

The standard fixed-order description of cross sections has been improved via the inclusion of the threshold resummation for a notable number of processes.
An incomplete list includes the following: Drell--Yan\tcite{Moch:2005ky,Idilbi:2006dg,Catani:2014uta,Ajjath:2020rci,Ajjath:2020lwb,Ajjath:2021pre,Ahmed:2020nci,Ajjath:2021lvg}, scalar Higgs in gluon fusion\tcite{Kramer:1996iq,Catani:2003zt,Moch:2005ky,Bonvini:2012an,Bonvini:2014joa,Catani:2014uta,Bonvini:2014tea,Bonvini:2016frm,Beneke:2019mua,Ajjath:2020sjk,Ajjath:2020lwb,Ahmed:2020nci,Ajjath:2021bbm} (see Refs.\tcite{deFlorian:2007sr,Schmidt:2015cea,Ahmed:2016otz,Bhattacharya:2019oun,Bhattacharya:2021hae} for the pseudo-scalar case) and in bottom annihilation\tcite{Bonvini:2016fgf,Ajjath:2019neu,Ahmed:2020nci}, deep inelastic scattering (DIS)\tcite{Moch:2005ba,Das:2019btv,Ajjath:2020sjk,Abele:2022wuy}, electron-positron single inclusive annihilation (SIA)\tcite{Ajjath:2020sjk}, and $\mbox{spin-}2$ boson\tcite{Das:2019bxi,Das:2020gie} productions.
A combined TM and threshold resummation for $\kappa_T$-distributions of colorless final states was developed in Ref.\tcite{Muselli:2017bad}.

From a partonic point of view, the threshold regime corresponds to the $\mbox{large-}x$ limit. A first determination of $\mbox{large-}x$ improved collinear PDFs was obtained in Ref.\tcite{Bonvini:2015ira}.
Furthermore, when the struck-parton $x$ approaches one, the effect of \emph{target-mass} power corrections (see, \emph{e.g.}, Refs.\tcite{Nachtmann:1973mr,Georgi:1976ve,Barbieri:1976rd,Ellis:1982wd,Ellis:1982cd,Schienbein:2007gr,Accardi:2008ne,Accardi:2008pc,Accardi:2013pra}) also becomes relevant.
The interplay between threshold resummation and target-mass corrections for $\mbox{large-}x$ DIS events was investigated in Ref.\tcite{Accardi:2014qda}.

Another relevant kinematic regime, sensitive to logarithmic enhancements, is the so-called \emph{semi-hard} (or Regge--Gribov) sector\tcite{Gribov:1983ivg,Celiberto:2017ius}, where the scale hierarchy
\begin{equation}
\label{semi-hard_regime}
 \LQCD \ll \{Q\} \ll \sqrt{s}
\end{equation}
is stringently preserved. In Eq.\eref{semi-hard_regime} $\LQCD$ is the QCD characteristic scale, $\{Q \}$ represents one or a set of perturbative scale typical of the process, and $s$ stands for the squared center-of-mass energy.
Here, large energy logarithms of the form $\ln (s/Q^2)$ compensate the smallness of $\alpha_s$, up to spoil the convergence of the pure perturbative series.

As it happens for the previously mentioned cases, also these logarithms must be resummed to all orders via a suited procedure. The most adequate and powerful mechanism to perform such a resummation is the Balitsky--Fadin--Kuraev--Lipatov (BFKL) formalism\tcite{Fadin:1975cb,Kuraev:1976ge,Kuraev:1977fs,Balitsky:1978ic} (see Ref.\tcite{Fadin:1998sh} for a review of theoretical advancements), whose validity is established up to the leading approximation (LL), which accounts for all contributions proportional to $\alpha_s^n \ln (s/Q^2)^n$, and the next-to-leading approximation (NLL), which indicates the inclusion of all terms proportional to $\alpha_s^{n+1} \ln (s/Q^2)^n$.

Within BFKL, the imaginary part of amplitudes (thence cross sections, in the case of inclusive reactions, thanks to the \emph{optical theorem}\tcite{optical_theorem_Newton}) takes the form of a high-energy factorization, where the building blocks are $\kappa_T$-dependent functions\tcite{Catani:1990xk,Catani:1990eg,Catani:1993ww}.

More in particular, the BFKL amplitude comes out as a convolution between two \emph{impact factors}, describing the transition from each parent particle to the outgoing object(s) produced in its \emph{fragmentation region}, and a Green's function. The latter is process universal and energy dependent. It evolves according to the BFKL integral equation, whose kernel was calculated with next-to-leading order (NLO) accuracy for forward scatterings\tcite{Fadin:1998py,Ciafaloni:1998gs}, namely for $t = 0$ and $t$-channel color-singlet exchanges, as well as for any fixed $t$ (not growing with $s$) and any possible two-gluon color exchange\tcite{Fadin:1998jv,Fadin:2000kx,Fadin:2004zq,Fadin:2005zj}.

Impact factors are instead final-state sensitive. They also depend on $\{Q\}$, but not on $s$.
Therefore, they represent the most challenging part of a BFKL study, and only few of them have been calculated within NLO so far. They are as follows: (\emph{i}) quark and gluon impact factors\tcite{Fadin:1999de,Fadin:1999df,Ciafaloni:1998kx,Ciafaloni:1998hu,Ciafaloni:2000sq}, which are key ingredients for the computation of (\emph{ii}) the forward light-jet impact
factor\tcite{Bartels:2001ge,Bartels:2002yj,Caporale:2011cc,Ivanov:2012ms,Colferai:2015zfa} and (\emph{iii}) the forward light-hadron one\tcite{Ivanov:2012iv}, (\emph{iiii}) the impact factor for the leading-twist ($\gamma^* \to {\cal V}$)
sub-process\tcite{Ivanov:2004pp}, where ${\cal V}$ is a light vector meson, (\emph{v}) the ($\gamma^* \to \gamma^*$) impact factor\tcite{Bartels:2000gt,Bartels:2001mv,Bartels:2002uz,Bartels:2003zi,Bartels:2004bi,Fadin:2001ap,Balitsky:2012bs}, and (\emph{vi}) the forward Higgs impact factor\tcite{Hentschinski:2020tbi,Celiberto:2022fgx} obtained in the large top-mass limit.

A first class of processes that serve as probe channels for the high-energy resummation consists in single forward emissions. Here, the cross section for an inclusive process takes the usual BFKL-factorized form. In particular, when at least one hadron is involved in the initial state, the impact factor depicting the production of the forward identified particle is convoluted with the BFKL Green's function and a nonperturbative quantity, called the hadron impact factor.

\begin{figure*}[t]
\centering
\includegraphics[width=0.30\textwidth]{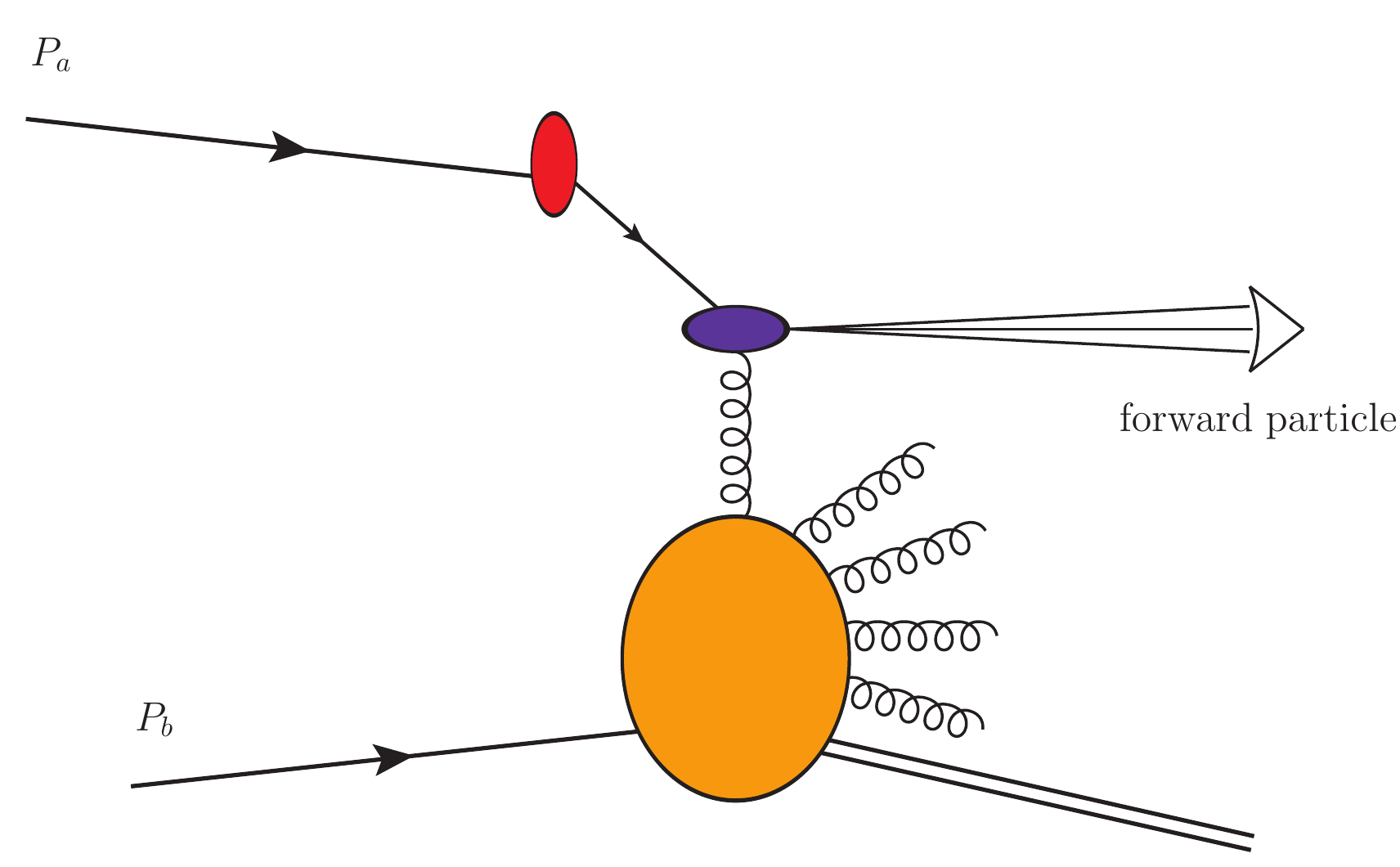}
\hspace{0.50cm}
\includegraphics[width=0.30\textwidth]{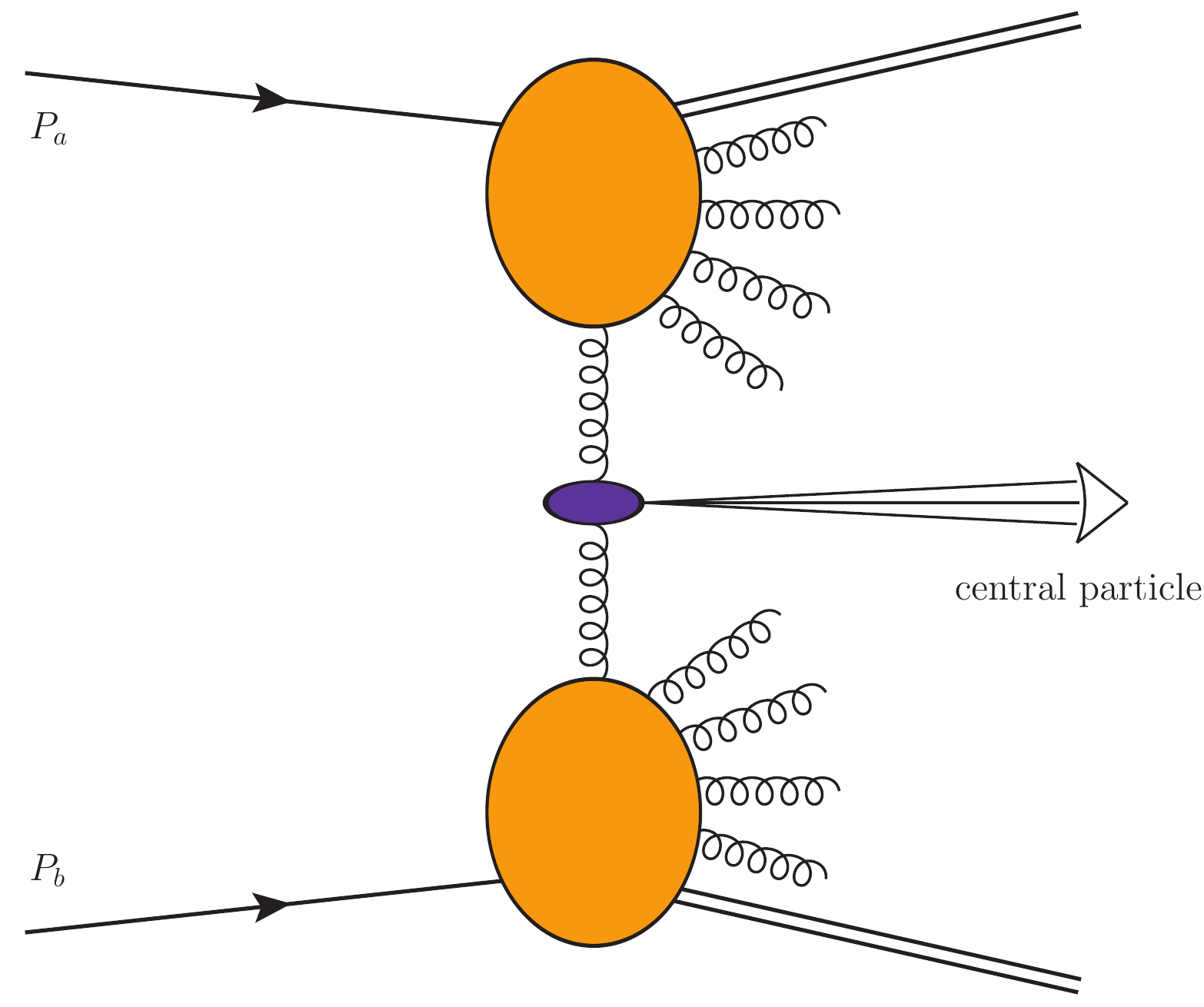}
\hspace{0.50cm}
\includegraphics[width=0.30\textwidth]{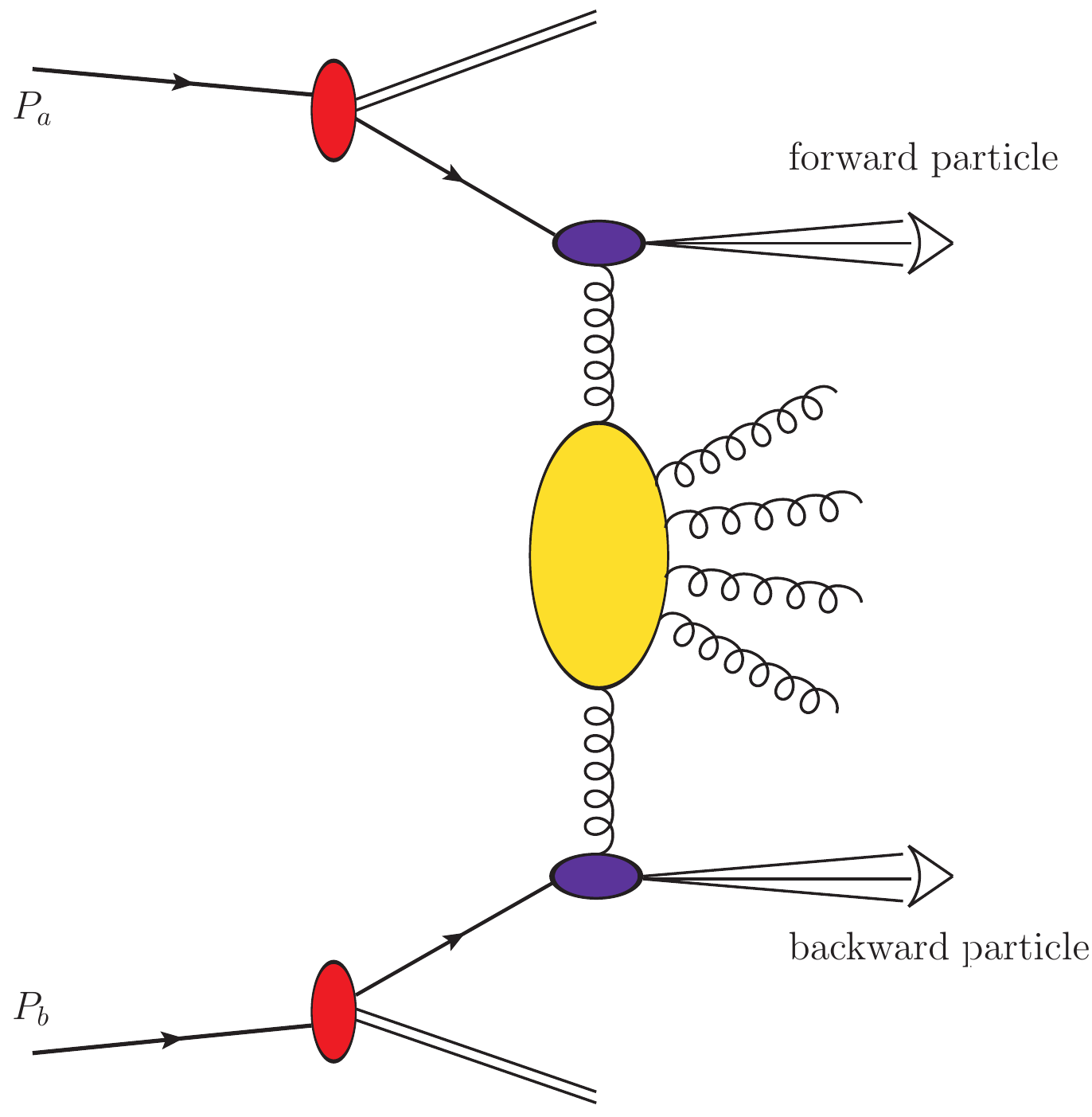}
\\ \vspace{0.25cm}
\hspace{0.10cm}
($a$) Single forward \hspace{2.75cm}
($b$) Single central \hspace{2.50cm}
($c$) Forward-backward
\caption{Diagrams for ($a$) an inclusive single forward process in hybrid high-energy and collinear factorization, ($b$) an inclusive single central emission in pure high-energy factorization, ($c$) an inclusive forward-backward reaction in hybrid factorization. The red blobs in diagrams ($a$) and ($c$) is for collinear PDFs, while violet ones in all diagrams depict the hard part of the off-shell impact factor describing the emission of a given particle in forward, central and/or backward regions of rapidity. Orange blobs in diagrams ($a$) and ($b$) portray the UGD, which embodies the nonperturbative gluon content of the proton at high energies/small~$x$. The BFKL Green’s function is given in terms of the yellow oval. This figure is complemented by corresponding lepton-hadron initiated channels and exclusive counterparts, not shown here. Gluon-induced emissions from the collinear regions in diagrams ($a$) and ($c$) (not shown) are embodied in sea-green blobs. Diagrams were obtained via {\tt JaxoDraw 2.0}\tcite{Binosi:2008ig}.}
\label{fig:HE-factorization}
\end{figure*}

The sub-convolution of the last two ingredients gives us a operational definition of the BFKL \emph{unintegrated gluon distribution} (UGD). The hadron impact factor represents the initial-scale UGD, while the Green's function regulates its small-$x$ evolution. Indeed, due to forward kinematics, the struck parton is mostly a gluon extracted with a small longitudinal-momentum fraction.
Therefore, in this context the high-energy resummation is \emph{de facto} a small-$x$ resummation.

An analogous high-energy factorization formula holds for the imaginary part of the amplitude of exclusive single forward processes. This is possible because, in the forward limit, skewness effects are suppressed and the same UGD can be employed. In more general, off-forward configurations one should consider rather small-$x$ improved \emph{generalized parton distributions} (GPDs)\tcite{Diehl:2003ny,Diehl:2015uka,Muller:1994ses,Belitsky:2005qn}.

An interesting sub-class of inclusive forward reactions is given by the proton-initiated ones. Here, a \emph{hybrid} high-energy and collinear factorization is established, where the forward object stems from a fast parton with a moderate $x$, portrayed by a collinear PDF, and the other proton is described by the UGD (see diagram ($a$) of Fig.\tref{fig:HE-factorization}).

Studies on the BFKL UGD historically began with the rise of interest on forward physics at HERA. DIS structure functions at small $x$ were investigated in Refs.\tcite{Hentschinski:2012kr,Hentschinski:2013id}. Then, results obtained with different UGD models were compared with HERA data for the exclusive light vector-meson electroproduction\tcite{Anikin:2011sa,Besse:2013muy,Bolognino:2018rhb,Bolognino:2018mlw,Bolognino:2019bko,Bolognino:2019pba,Celiberto:2019slj}.
Clear evidences of the onset of small-$x$ dynamics are expected to come out from $\rho$-meson studies at the Electron-Ion Collider (EIC)\tcite{Bolognino:2021niq,Bolognino:2021gjm,Bolognino:2022uty,Celiberto:2022fam} and from photoemissions of quarkonium states\tcite{Bautista:2016xnp,Garcia:2019tne,Hentschinski:2020yfm,GayDucati:2013sss,GayDucati:2016ryh,Goncalves:2017wgg,Goncalves:2018blz,Cepila:2017nef,Guzey:2020ntc,Jenkovszky:2021sis,Flore:2020jau,ColpaniSerri:2021bla}. Hadronic probes for the UGD are the forward Drell--Yan reaction at LHCb~\cite{Motyka:2014lya,Brzeminski:2016lwh,Motyka:2016lta,Celiberto:2018muu} and the single inclusive $b$-quark tag at the LHC~\cite{Chachamis:2015ona}.

In addition to single forward emissions, the small-$x$ regime can be accessed also via gluon-induced single central productions (see diagram ($b$) of Fig.\tref{fig:HE-factorization}). Cross sections for those inclusive small-$x$ channels are written in a pure high-energy factorized form, namely as a convolution between two BFKL UGDs and a central-production impact factor, also known as \emph{coefficient function}. 
Being a doubly off-shell quantity (($g^*g^*$), the gluon virtualities being driven by their $\kappa_T$), its calculation turns out to be much more complicated than forward-case ones. According to our knowledge, the central light-jet vertex is the only coefficient function known with NLO accuracy\tcite{Bartels:2006hg}.

A powerful method to improve standard fixed-order calculations for central processes via the resummation of small-$x$ logarithms is the Altarelli--Ball--Forte (ABF) formalism\tcite{Ball:1995vc,Ball:1997vf,Altarelli:2001ji,Altarelli:2003hk,Altarelli:2005ni,Altarelli:2008aj,White:2006yh}, where the $\kappa_T$-factorization theorems\tcite{Catani:1990xk,Catani:1990eg,Collins:1991ty,Catani:1993ww,Ball:2007ra} are used to consistently incorporate both the DGLAP and BFKL inputs. Then, the high-energy series is stabilized by enforcing consistency conditions based on duality aspects, symmetrizing the BFKL kernel in collinear and anti-collinear regions of the phase space, and embodying these contributions to running coupling which affect small-$x$ singularities.

Striking progresses have been made in the context of small-$x$ studies within the ABF formalism on inclusive central emissions of Higgs-bosons in gluon fusion\tcite{Marzani:2008az,Caola:2010kv,Caola:2011wq,Forte:2015gve} and in higher-order corrections to top-quark pair productions\tcite{Muselli:2015kba}.
Notably, the same framework was employed to extract for the first time $\mbox{small-}x$ improved collinear PDFs\tcite{Ball:2017otu,Abdolmaleki:2018jln,Bonvini:2019wxf}, whose information was subsequently used to fix parameters of initial-scale unpolarized and helicity gluon TMDs\tcite{Bacchetta:2020vty}.
We mention, for completeness, a study on the inclusion of $\mbox{small-}x$ dynamics in the \emph{parton branching} method\tcite{Hautmann:2017xtx,Hautmann:2017fcj} to TMD distributions\tcite{Monfared:2019uaj}.

\begin{figure*}[tb]
\centering

\includegraphics[width=0.45\textwidth]{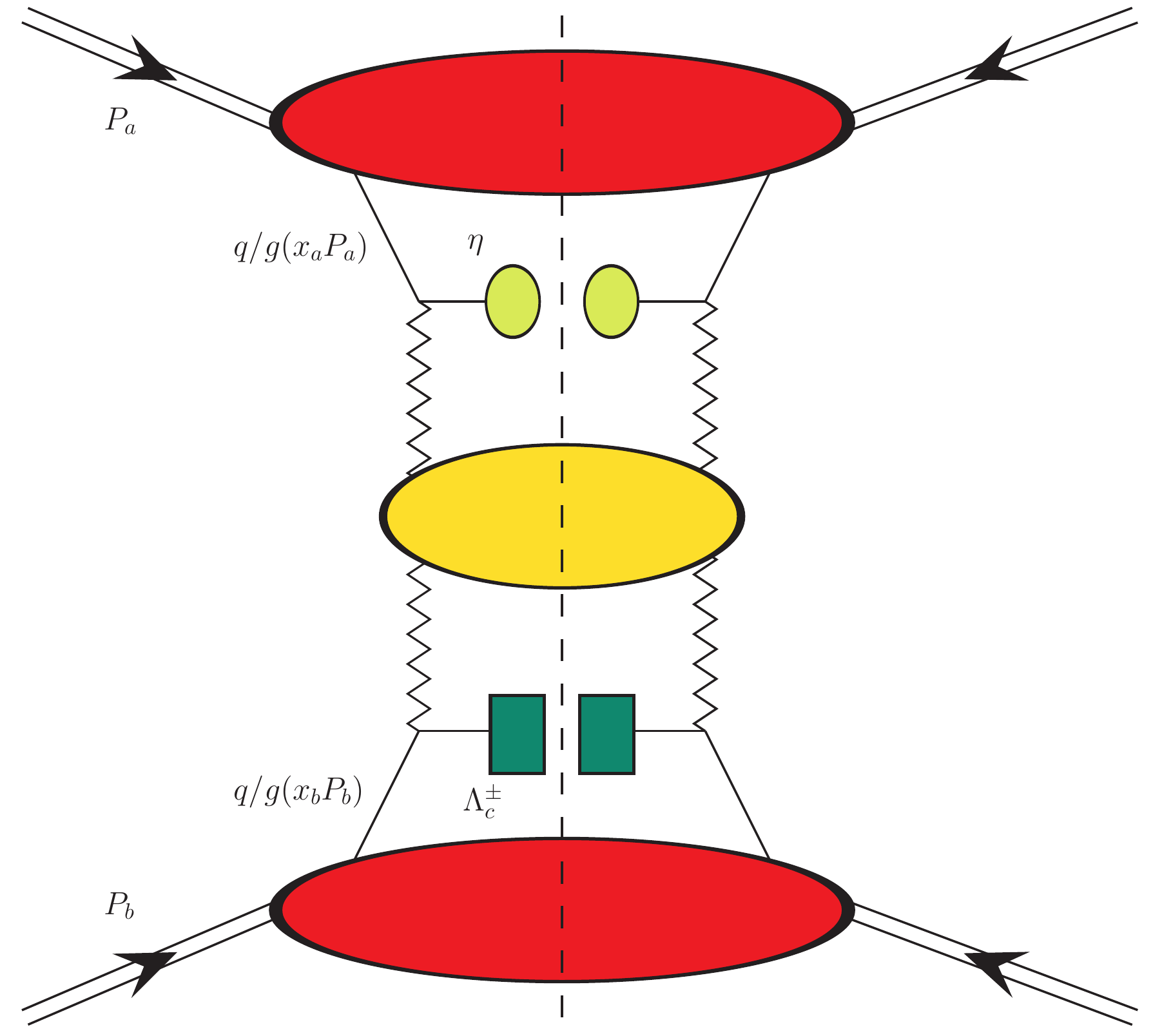}
\hspace{0.25cm}
\includegraphics[width=0.45\textwidth]{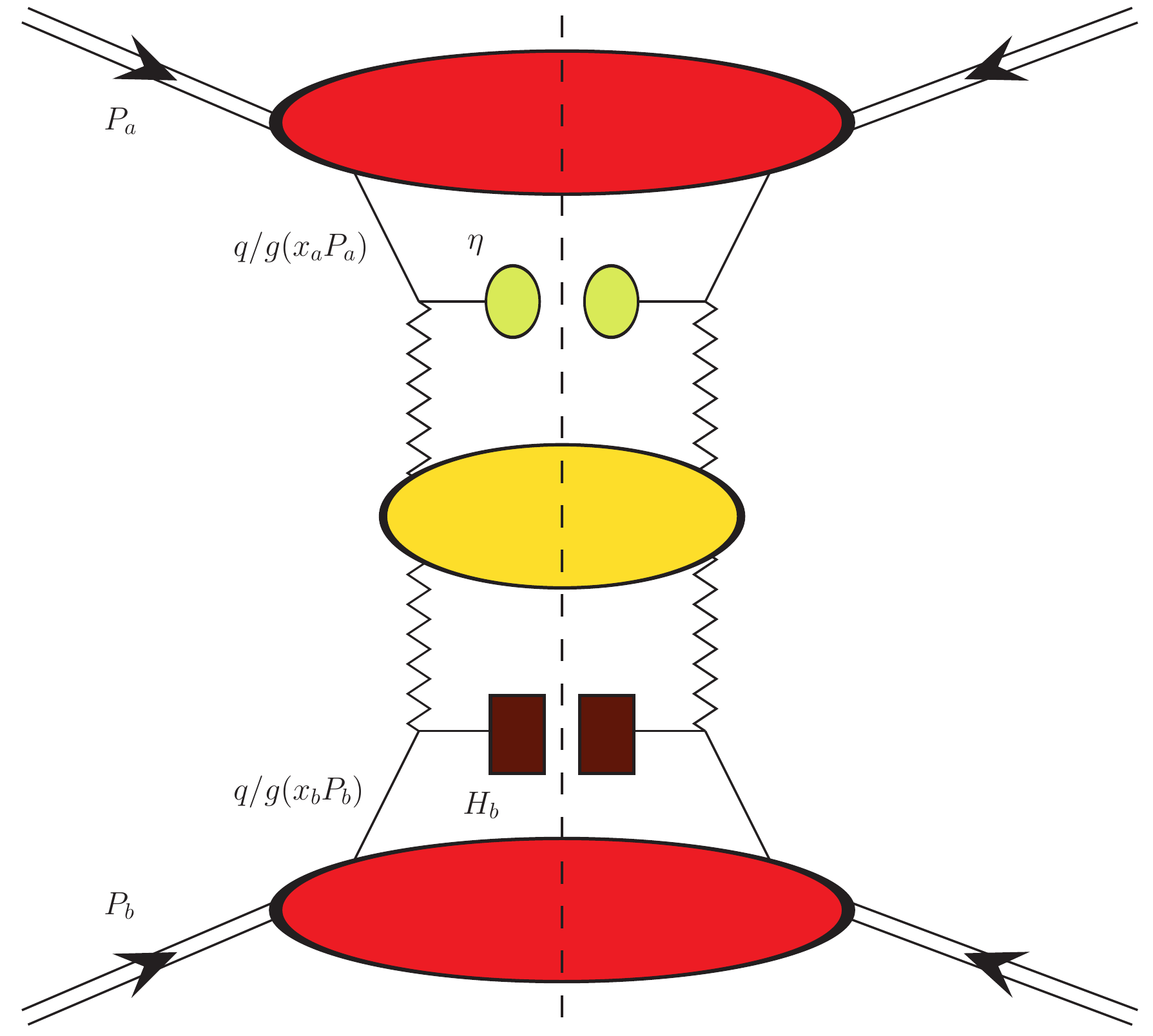}
\\ \vspace{0.25cm}
\hspace{-0.50cm}
($a$) $\eta$ $+$ $\Lambda_c^\pm$ \hspace{6.50cm}
($b$) $\eta$ $+$ $H_b$

\vspace{0.90cm}

\includegraphics[width=0.45\textwidth]{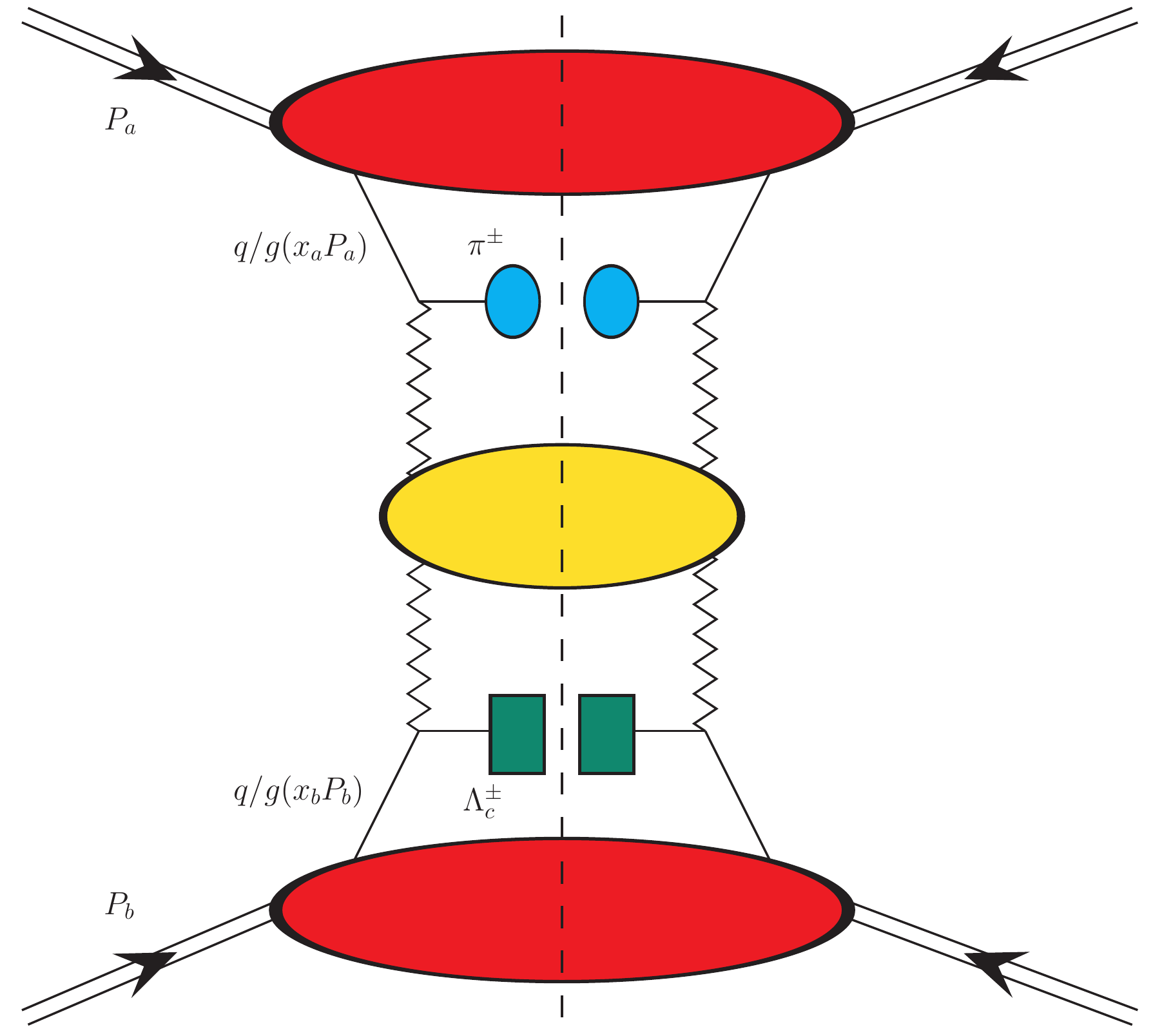}
\hspace{0.25cm}
\includegraphics[width=0.45\textwidth]{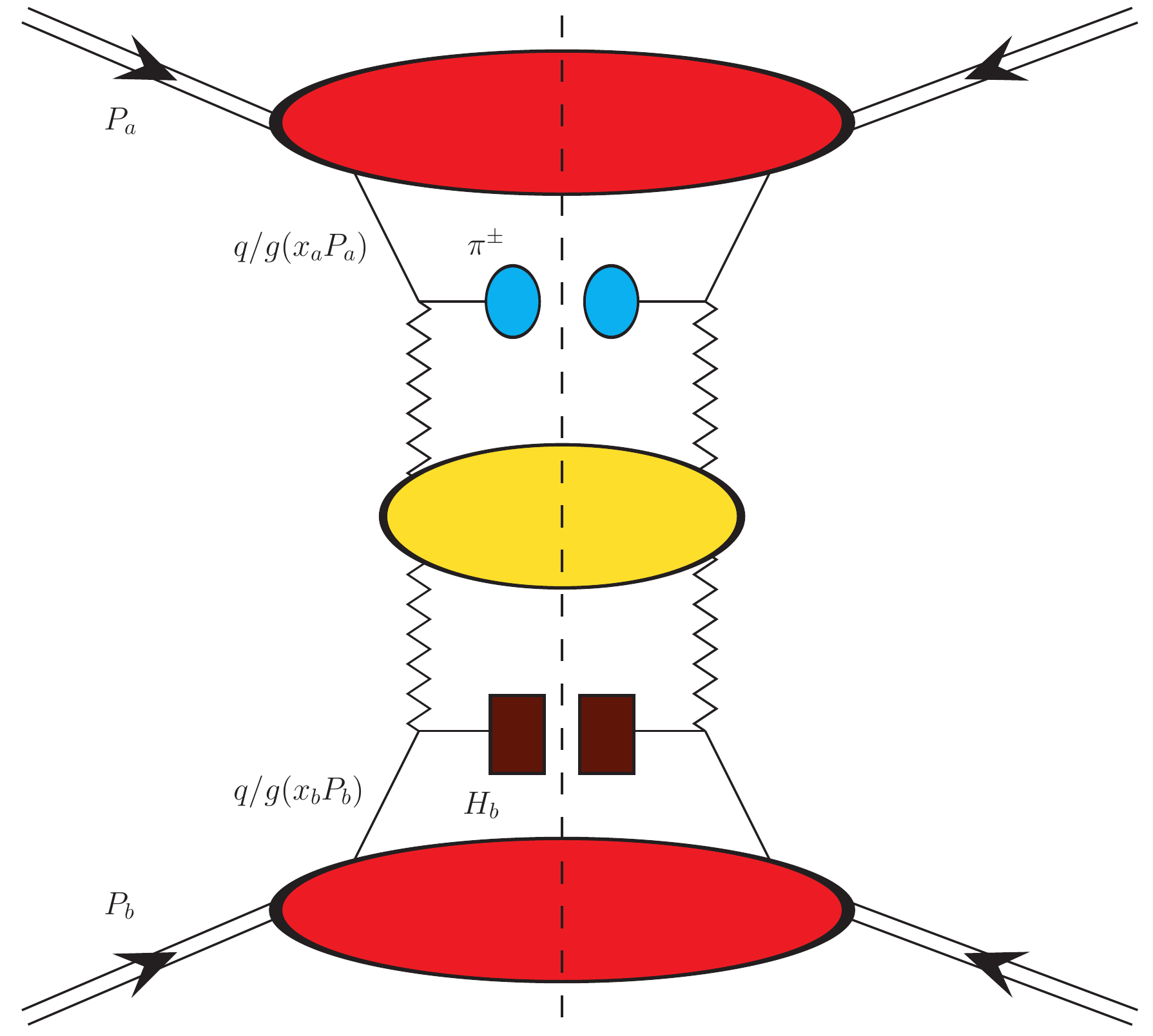}
\\ \vspace{0.25cm}
\hspace{-0.50cm}
($c$) $\pi^\pm$ $+$ $\Lambda_c^\pm$ \hspace{6.50cm}
($d$) $\pi^\pm$ $+$ $H_b$

\caption{Hybrid high-energy and collinear factorization in play. Schematic representation of the four channels under analysis. Red blobs stand for proton collinear PDFs. Lime and turquoise ovals represent $\eta$-meson and charged-pion collinear FFs. Bordeaux rectangles and green rectangles, respectively, portray $b$~hadron and $\Lambda_c$ baryon collinear FFs. The BFKL Green's function, depicted by the big yellow blob, is connected to impact factors via Reggeon waggle lines. Diagrams were obtained via {\tt JaxoDraw 2.0}\tcite{Binosi:2008ig}.}
\label{fig:process}
\end{figure*}

Another relevant testing ground for the manifestation of distinctive signals of high-energy QCD dynamics is represented by inclusive hadroproductions of two objects featuring transverse masses well above the QCD scale and widely separated in rapidity,\footnote{Striking evidences of BFKL dynamics were observed also in photon-initiated reactions, such as the ($\gamma^*\gamma^*$) process\tcite{Brodsky:2002ka,Chirilli:2014dcb,Ivanov:2014hpa}, the double exclusive vector-meson electroproduction\tcite{Segond:2007fj,Ivanov:2005gn,Ivanov:2006gt}, and the inclusive heavy-quark pair photoproduction\tcite{Celiberto:2017nyx}. High-energy effects in observables accessible at new-generation lepton colliders\tcite{AlexanderAryshev:2022pkx,Brunner:2022usy} are expected be sizable.} see diagram ($c$) of Fig.\tref{fig:HE-factorization}.
More precisely, these forward final states are inclusive and \emph{diffractive} at the same time, since the undetected-gluon emissions are condensed in the central-rapidity region between the two detected particles, and summed over. This permits the use of the optical theorem to relate the differential cross section to the imaginary part of a purely diffractive amplitude, characterized by the absence of any central activity.

At variance with single forward and central processes (first two diagrams of Fig.\tref{fig:HE-factorization}), the formal description of forward-backward two-particle hadroproductions is sensitive to enhanced energy logarithms even outside the $\mbox{small-}x$ domain.
Indeed, on the one side kinematic ranges in transverse momentum and rapidity currently covered by acceptances of LHC detectors lead to moderate-$x$ values. Therefore, a collinear PDF-based description remains valid.

At the same time, however, high rapidity intervals ($\DY$) translate in large $\kappa_T$~exchanges in the $t$~channel, which in turn bring to the rise of energy logarithms. This calls for a $\kappa_T$-factorization treatment, genuinely afforded by the BFKL formalism.
Therefore, another kind of \emph{hybrid} high-energy and collinear factorization is established, where high-energy resummed partonic cross sections are natively obtained from BFKL, and then they are convoluted with collinear PDFs.

More in general, the use of collinear inputs together with building blocks of the high-energy factorization, as the Green's function and the off-shell impact factors, was proposed by different Collaborations and in the context of distinct final states, which include single forward, multiple forward and forward-plus-backward emissions
Another formalism, which is close in spirit with our hybrid one, was proposed in Ref.\tcite{Deak:2009xt} to study forward jets at the LHC. Then, it was employed to investigate $Z_0$-plus-jet final states\tcite{vanHameren:2015uia,Deak:2018obv} as well as topologies of three-jet events\tcite{VanHaevermaet:2020rro}.

From a phenomenological viewpoint, the ``mother" reaction of inclusive forward-backward hadroproductions is the Mueller--Navelet\tcite{Mueller:1986ey} emission of two light jets at large $\kappa_T$ and $\DY$, for which many phenomenological analyses have appeared so far\tcite{Marquet:2007xx,Colferai:2010wu,Caporale:2012ih,Ducloue:2013hia,Ducloue:2013bva,Caporale:2013uva,Caporale:2014gpa,Ducloue:2015jba,Celiberto:2015yba,Celiberto:2015mpa,Caporale:2015uva,Mueller:2015ael,Celiberto:2016ygs,Celiberto:2016vva,Caporale:2018qnm} and they have been compared with CMS data at $\sqrt{s} = 7\mbox{ TeV}$\tcite{Khachatryan:2016udy}. 

Further observables, sensitive to more exclusive final states, were proposed as suitable channels where to hunt for clues of the onset of the BFKL dynamics in a deeper and complementary way with respect to what provided by Mueller--Navelet channels. A noninclusive list is made of: light di-hadron\tcite{Celiberto:2016hae,Celiberto:2016zgb,Celiberto:2017ptm,Celiberto:2017uae,Celiberto:2017ydk}, hadron-jet\tcite{Bolognino:2018oth,Bolognino:2019cac,Bolognino:2019yqj,Celiberto:2020wpk,Celiberto:2020rxb}, multi-jet\tcite{Caporale:2015vya,Caporale:2015int,Caporale:2016soq,Caporale:2016vxt,Caporale:2016xku,Celiberto:2016vhn,Caporale:2016djm,Caporale:2016pqe,Chachamis:2016qct,Chachamis:2016lyi,Caporale:2016lnh,Caporale:2016zkc,Chachamis:2017vfa,Caporale:2017jqj} and Drell--Yan\tcite{Golec-Biernat:2018kem} azimuthal distributions, Higgs-jet rapidity and transverse-momentum distributions\tcite{Celiberto:2020tmb,Celiberto:2021fjf,Celiberto:2021tky,Celiberto:2021txb,Celiberto:2021xpm}, heavy-flavored jet\tcite{Bolognino:2021mrc,Bolognino:2021hxx} and hadron\tcite{Boussarie:2017oae,Celiberto:2017nyx,Bolognino:2019ouc,Bolognino:2019yls,Bolognino:2019ccd,Celiberto:2021dzy,Celiberto:2021fdp,Bolognino:2021zco,Bolognino:2022wgl,Celiberto:2022dyf,Celiberto:2022zdg} cross sections.

Studies on azimuthal-angle correlations for light-jet and/or light-hadron detections were particularly relevant to decisively discriminate between high-energy resummed and fixed-order calculations thanks to the use of asymmetric $\kappa_T$~ranges\tcite{Celiberto:2015yba,Celiberto:2015mpa,Celiberto:2020wpk}.
At the same time, however, they highlighted that large instabilities associated to higher-order BFKL corrections rise in the theoretical description of those observables.
More in particular, NLL contributions are of the same size but with opposite sign of pure LL terms. This makes the high-energy series unstable and very sensitive to the choice of renormalization ($\mu_R$) and factorization ($\mu_F$) scales.

The adoption of some scale-optimization methods, such as the Brodsky--Lepage--Mackenzie (BLM) prescription\tcite{Brodsky:1996sg,Brodsky:1997sd,Brodsky:1998kn,Brodsky:2002ka} in its semi-hard designed version\tcite{Caporale:2015uva} allowed us to partially dampen these instabilities on azimuthal correlations. Unfortunately, it turned out to be ineffective on cross sections, since the found optimal scales were much larger than the natural ones afforded by kinematics\tcite{Celiberto:2020wpk}, with a consequent substantial and unphysical lowering of statistics. Thus any attempt at reaching precision in the study of inclusive forward-backward light-flavored objects was unsuccessful.

A first evidence of the existence of inclusive semi-hard reactions whose intrinsic features lead to a fair stabilization of the NLL BFKL series came out quite recently in the context of forward tags of heavy-light hadrons, such as $\Lambda_c^\pm$ baryons\tcite{Celiberto:2021dzy} or bottom-flavored ($H_b$) hadrons\tcite{Celiberto:2021fdp}, and of vector quarkonia, $J/\psi$ or $\Upsilon$, emitted at large $\kappa_T$\tcite{Celiberto:2022dyf}.
This stabilization pattern is the net effect of the convolution of a nondecreasing with $\mu_F$ gluon FF with proton PDFs. The peculiar behavior of the heavy-hadron FFs act as a \emph{natural stabilizer} of the high-energy resummation (see Section~3.4 of Ref.\tcite{Celiberto:2021dzy} and the Appendix of Ref.\tcite{Celiberto:2021fdp} for technical details on the connection between heavy-flavor FFs and the stability of energy-resummed cross sections under scale variations).

Stabilization effects of cross sections and azimuthal correlations emerged also in semi-hard final states, studied with partial NLL accuracy so far, featuring the emission of an object with a large transverse mass, such as a Higgs boson\tcite{Celiberto:2020tmb} or a heavy-flavored jet\tcite{Bolognino:2021mrc}.
Here, large transverse masses regulate the all-order growth-with-energy of logarithms, thus being \emph{natural stabilizers} for these reactions. Full NLL analyses are however needed to corroborate that statement.

In this article we will focus on rapidity and azimuthal-angle distributions for a novel selection of forward-backward two-particle semi-hard reactions, whose final states are characterized by identified hadrons only (see diagrams of Fig.\tref{fig:process}).
The first hadron is light flavored. It can be a pion, whose detection in CMS typical ranges has been already considered in previous studies\tcite{Celiberto:2020wpk,Celiberto:2020rxb}, or a $\eta$ meson, whose tag is included for the first time in the context of semi-hard phenomenology.
The second hadron is a heavy-light meson, namely a $\Lambda_c$ baryon or a $b$-flavored particle (an inclusive state consisting in the sum of fragmentation channels to noncharmed $B$~mesons and $\Lambda_b^0$ baryons, see Ref.\tcite{Celiberto:2021fdp}).

The use of light mesons instead of light jets permits to partially reduce the aforementioned instabilities that rise in both the NLO impact factors portraying the emission of corresponding objects, but they are stronger in the NLO-jet case\tcite{Celiberto:2017ptm,Celiberto:2017ius,Celiberto:2020wpk}.
Furthermore, it offers us a complementary channel to constrain collinear FFs describing the hadronization mechanism of these light particles. This is particularly true in the case of pions, for which detailed analyses on uncertainties coming from global-fit procedures have already undertaken the first steps of a path towards precision\tcite{Bertone:2017tyb,Khalek:2021gxf,Borsa:2022vvp,Barry:2022itu}.
Conversely, single-charmed and single-bottomed hadrons will serve as stabilizers of the high-energy series entering the description of our distributions.

The adoption of disjoint windows for the transverse momenta of the light meson and the heavy hadron is helpful not only to better disengage high-energy imprints from DGLAP background\tcite{Celiberto:2015yba,Celiberto:2015mpa,Celiberto:2020wpk}, but also to avoid Sudakov logarithmic contaminations rising from almost back-to-back final states that would call for another appropriate resummation\tcite{Mueller:2012uf,Mueller:2013wwa,Marzani:2015oyb,Mueller:2015ael,Xiao:2018esv}.

A detailed study of all the potential uncertainty sources is a \emph{conditio sine qua non} for an accurate description of our observables.
With the aim of comparing our predictions obtained within the hybrid factorization with fixed-order results, and pursuing the goal of tracing a path towards precision, we will assess the impact of a comprehensive set of uncertainties. 
Some of them are related to collinear-factorization phenomenology, such as replica-based studies on PDFs and FFs.
Other ones are originated from intrinsic effects belonging to the high-energy resummation.

A systematic high-energy versus DGLAP analysis would rely on the comparison between distributions calculated by the hands of our hybrid factorization and pure fixed-order calculations.
However, according to our knowledge, a numerical tool for the calculation of NLO cross sections for the inclusive production in proton collisions of two identified hadrons widely separated in rapidity has not yet been developed.
Although the leading-order (LO) limit for this class of reactions can be extracted from higher-order analyses (see, \emph{e.g.}, Ref.\tcite{Chiappetta:1996wp,Owens:2001rr,Binoth:2002ym,Binoth:2002wa,Almeida:2009jt,Hinderer:2014qta}), it cannot be compared with our calculations due to kinematics. Indeed, any LO two-particle computation not supplemented by a resummation prescribes a back-to-back final state, which is not compatible with our asymmetric windows for the observed transverse momenta.

Therefore, to gauge the impact of the high-energy resummation on top of the DGLAP approach, we will compare BFKL-driven predictions with the corresponding ones obtained via a high-energy fixed-order treatment, originally developed in the context of light di-jet\tcite{Celiberto:2015yba,Celiberto:2015mpa} and hadron-jet\tcite{Celiberto:2020wpk} azimuthal correlations.
It is based on the truncation of the high-energy series up to the NLO accuracy, thus allowing us to mimic the high-energy signal of a pure NLO calculation.

Concerning rapidity ranges, we will consider two distinct kinematic configurations.
In the first case both the light meson and the heavy hadron are reconstructed inside current acceptances of CMS or ATLAS barrel and detectors.
This choice offers us symmetric rapidity window for both the particles, thus representing a suitable channel where to further test the high-energy QCD dynamics in a similar way to what has been done through the two-particle semi-hard processes investigated so far.
In the second case we allow for a simultaneous tag of the light meson in the \emph{ultra-forward} rapidity ranges accessible at the planned Forward Physics Facility (FPF)\tcite{Anchordoqui:2021ghd,Feng:2022inv}, and of the heavy-flavored hadron in the standard ATLAS-barrel ranges.
Our interest in unveiling the feasibility of high-energy studies via the FPF~$+$~ATLAS coincidence setup relies on multiple reasons.

First, the concurrent detection of a far-forward particle together with a central one\footnote{We remark that, although being detected in the central-rapidity region of ATLAS, the heavy-flavored hadron still remains backward with respect to the light meson, due to the large $\DY$ separating the two objects. Thus, final states reconstructed at FPF~$+$~ATLAS fall in the class of forward-backward semi-hard reactions.} leads to an asymmetric configuration between the longitudinal-momentum fractions of the two corresponding incoming partons, one of them being large and the other one assuming more moderate values.
Therefore, the FPF~$+$~ATLAS coincidence represents a unique venue where to explore not only high-energy effects thanks to the very large rapidity intervals accessible, but also the interplay between the threshold and the BFKL resummation.

Studies on combined large- and small-$x$ effects for the central hadroproduction of Higgs scalars\tcite{Bonvini:2018ixe,Ball:2013bra} have shown that the weight of such a double-logarithmic resummation is small at ongoing LHC energies, whereas it becomes more sizeable at the nominal ones of the Future Circular Collider (FCC)\tcite{Mangano:2016jyj}.
Conversely, the high-energy description of transverse-momentum distributions for the Higgs-plus-jet hadroproduction already deviates from the pure NLO pattern at current LHC configurations\tcite{Celiberto:2020tmb}.
Therefore, our two-particle final states are expected to exhibit a strong sensitivity to the co-action of the two resummation mechanisms.

Then, future analyses at the FPF will be relevant to deepen our knowledge of perturbative QCD and of proton and nuclear structure in regimes so far unexplored. The FPF will be sensitive to the very forward production of light hadrons and charmed mesons, granting us access to BFKL effects and gluon-recombination dynamics. 
TeV-scale neutrino-induced DIS experiments doable at the FPF will be valuable probes of the proton structure as well as of production mechanisms of heavy or light decaying hadrons.
Our work on light mesons at the FPF via the hybrid factorization can provide a common basis for the description of production and decays of these particles.

Finally, QCD studies represent one of the founding pillars of the multi-frontier activity constituting the FPF research program.
Searches for long-lived particles, dark-matter indirect detections and sterile neutrinos, as well as explorations of the muon puzzle, the lepton universality and the connection between high-energy particle physics and modern astroparticle physics certainly rely on a profound knowledge of the SM.
Progresses on the path toward precision QCD in the FPF kinematic sector are core elements to fuel the interest of the scientific Community toward novel and engaging directions.

This article reads as follows: the general structure of the cross section for our processes is presented in Section\tref{sec:theory}, together with our choice of perturbative and nonperturbative ingredients; results for our energy-resummed observables are shown and discussed in Section\tref{sec:pheno}; Section\tref{sec:conclusions} contains our conclusions.

\section{Hybrid factorization at work}
\label{sec:theory}

In this Section we give theoretical details on our hybrid high-energy and collinear factorization for the inclusive light-meson plus heavy-flavor production. After a brief overview on process and kinematics~(Section\tref{ssec:process}), we present our high-energy resummed cross section~(Section\tref{ssec:sigma}). Choices for the running coupling and the renormalization scheme are given in Section\tref{ssec:pert}, while and our selection for collinear PDFs and FFs is discussed in\tref{ssec:PDFs_FFs}.

\subsection{Process and kinematics}
\label{ssec:process}

We investigate the inclusive set of reactions (Fig.\tref{fig:process})
\begin{equation}
\label{process}
 \hspace{-0.15cm}
 p(P_a) + p(P_b) \rightarrow {\cal M}(\kappa_{\cal M}, y_{\cal M}) + {\cal X} + {\cal H}(\kappa_{\cal H}, y_{\cal H}) \; ,
\end{equation}
where a light meson ${\cal M}$ (a $\eta$ meson with mass $m_\eta = 547.85$~MeV, or a $\pi^\pm$ with mass $m_\pi = 139.57$~MeV), having four-momentum $\kappa_{\cal M}$ and rapidity $y_{\cal M}$ is emitted in association with a heavy-flavored hadron with four-momentum $\kappa_{\cal H}$ and rapidity $y_{\cal H}$. 
The two incoming partons have four-momentum $x_a P_a$ and $x_b P_b$, with $P_{a,b}$ the parent protons' momenta.
In our analysis we consider two possibilities for the heavy hadron: (\emph{i}) the detection of a charmed-flavored species, namely a $\Lambda_c^{\pm}$ baryon (${\cal H} \equiv \Lambda_c^{\pm}$, with $m_{\Lambda_c} = 2.286$~GeV); (\emph{ii}) the inclusive tag a combination of bottom-flavored hadrons $H_b$ comprehending noncharmed $B$ mesons and $\Lambda_b^0$ baryons (${\cal H} \equiv H_b$). The $\cal X$ term in Eq.\eref{process} depicts all the undetected produced objects.
The two incoming-proton four-momenta are selected as Sudakov-basis vectors satisfying $P_a^2 = P_b^2 \equiv 0$ and $ (P_a \cdot P_b) = s/2$, and the observed four-momenta are decomposed in the following way
\begin{equation}\label{sudakov}
\kappa_{{\cal M},{\cal H}} = x_{{\cal M},{\cal H}} P_{a,b} + \frac{\vec \kappa_{{\cal M},{\cal H}}^{\,2}}{x_{{\cal M},{\cal H}}}\frac{P_{b,a}}{s} + \kappa_{\perp_{{\cal M},\cal H}} \ ,
\end{equation}
where
\begin{equation}\label{kT}
\kappa_{\perp_{{\cal M},\cal H}}^2 \equiv -\vec \kappa_{T_{{\cal M},{\cal H}}}^{\,2}\;.
\end{equation}
The outgoing-particle longitudinal momentum fractions, $x_{{\cal M},{\cal H}}$, are linked to their rapidities through the relation
\begin{equation}\label{yMH}
y_{{\cal M},{\cal H}} = 
\pm \ln \left( \frac{x_{{\cal M},{\cal H}}}{|\vec \kappa_{T_{\cal M},{\cal H}}|} \sqrt{s} \right)
\,,
\end{equation}
with
\begin{equation}\label{dyMH}
 \drv y_{{\cal M},{\cal H}} = \pm \frac{\drv x_{{\cal M},{\cal H}}}{x_{{\cal M},{\cal H}}}
\;.
\end{equation}
The diffractive semi-hard nature of the final state
is ensured by requiring transverse momenta respecting the hierarchy $\Lambda_{\rm QCD} \ll |\vec \kappa_{T_{\cal M,\cal H}}| \ll \sqrt{s}$, a large rapidity separation between the ${\cal M}$ meson and the heavy hadron, $\DY = y_{\cal M} - y_{\cal H}$. 
Moreover, to warrant the validity of a variable-flavor number-scheme (VFNS) description for the heavy-hadron production\tcite{Mele:1990cw,Cacciari:1993mq}, the $\kappa_{T_{\cal H}}$ range needs to stay sufficiently over the DGLAP-evolution thresholds given by charm and bottom masses.

\subsection{Resummed cross section}
\label{ssec:sigma}

As already mentioned, a pure QCD collinear approach for the LO cross section of our reaction (Eq.\eref{process}) would rely on the convolution between the partonic hard factor, the incoming-nucleon PDFs and the outgoing-hadron FFs
\begin{widetext}
\begin{equation}
\label{sigma_collinear_LO}
\begin{split}
\hspace{-0.05cm}
\frac{\drv \sigma^{[p + p \,\rightarrow\, {\cal M} + {\cal H} + {\cal X}]}_{\rm LO,\,coll.}}{\drv x_{\cal M}\drv x_{\cal H}\drv ^2\vec \kappa_{T_{\cal M}}\drv ^2\vec \kappa_{T_{\cal H}}}
&\,=\, \sum_{i,j=q,{\bar q},g}\int_0^1 \drv x_a \int_0^1 \drv x_b\ f_i\left(x_a,\mu_F\right) f_j\left(x_b,\mu_F\right)
\\
&\,\times\, \int_{x_{\cal M}}^1\frac{\drv \zeta_1}{\zeta_1}\int_{x_{\cal H}}^1\frac{\drv \zeta_2}{\zeta_2}D^{\cal M}_{i}\left(\frac{x_{\cal M}}{\zeta_1},\mu_F\right)D^{\cal H}_{j}\left(\frac{x_{\cal H}}{\zeta_2},\mu_F\right)
\frac{\drv {\hat\sigma}_{i,j}\left(\hat s,\mu_F,\mu_R\right)}
{\drv x_{\cal M}\drv x_{\cal H}\drv ^2\vec \kappa_{T_{\cal M}}\drv ^2\vec \kappa_{T_{\cal H}}}\;.
\end{split}
\end{equation}
\end{widetext}
In Eq.\eref{sigma_collinear_LO} the $i$ and $j$ indices run over all the parton species, (anti)-quarks and gluon, whereas $f_{i,j}\left(x_{1,2}, \mu_F \right)$ are the colliding-proton PDFs and $D^{{\cal M},{\cal H}}_{i,j}\left(x_{{\cal M},{\cal H}}/\zeta_{1,2}, \mu_F \right)$ represent the ${\cal M}$ meson and ${\cal H}$ particle FFs; $x_{a,b}$ stand for the longitudinal momentum fractions of the partons entering the hard sub-process and $\zeta_{1,2}$ the longitudinal fractions of partons fragmenting to observed hadrons. The partonic cross section $\hat \sigma_{i,j} \left( \hat s,\mu_F,\mu_R \right)$ depends on the squared center-of-mass energy of the partonic collision, $\hat s \equiv x_a x_b s$, and on factorization ($\mu_F$) and renormalization ($\mu_R$) scales.

Conversely, to build differential cross sections in  hybrid high-energy and collinear factorization we first account for BFKL resummation of energy logarithms rising due to transverse-momentum exchanges in the $\mbox{\emph{t}-channel}$. Then we encode in the formalism collinear inputs, \emph{i.e.} proton PDFs and emitted hadrons' FFs.
We suitably represent the differential cross section as a Fourier series of azimuthal-angle coefficients
\begin{widetext}
\begin{equation}
 \label{dsigma_Fourier}    
 \hspace{-0.19cm}
 \frac{\drv \sigma^{[p + p \,\rightarrow\, {\cal M} + {\cal X} + {\cal H}]}}{\drv y_{\cal M} \drv y_{\cal H} \drv\vec\kappa_{T_{\cal M}} \drv\vec\kappa_{T_{\cal H}} \drv\phi_{{\cal M}} \drv\phi_{{\cal H}}} = 
 \frac{1}{(2\pi)^2} \left[{\cal C}_0 + 2 \sum_{n=1}^\infty \cos (n \varphi)\,
 {\cal C}_n \right]\, ,
\end{equation}
\end{widetext}
where $\phi = \phi_{\cal M} - \phi_{\cal H} - \pi$ is the distance between the azimuthal angles of the light and the heavy hadron.
The azimuthal coefficients ${\cal C}_n$ are calculated within the BFKL framework and they embody the resummation of energy logarithms. A NLL-consistent formula obtained in the $\MSb$ renormalization scheme\tcite{PhysRevD.18.3998} is cast as follows (for details on the derivation see, \emph{e.g.}, Ref.\tcite{Caporale:2012ih})
\begin{widetext}
\begin{equation}
\label{Cn_NLLstar_MSb}
\begin{split}
 \CnNLLstar &\,=\, \int_0^{2\pi} \drv \phi_{\cal M} \int_0^{2\pi} \drv \phi_{\cal H}\,
 \cos (n \varphi) \,
 \frac{\drv \sigma^{[p + p \,\rightarrow\, {\cal M} + {\cal X} + {\cal H}]}_{{\rm NLL}^*}}{\drv y_{\cal M} \drv y_{\cal H}\, \drv |\vec \kappa_{T_{\cal M}}| \, \drv |\vec \kappa_{T_{\cal H}}| \drv \phi_{\cal M} \drv \phi_{\cal H}}\;
\\
 &\,=\, \frac{e^{\DY}}{s} 
 \int_{-\infty}^{+\infty} \drv \nu \, e^{{\DY} \bar \alpha_s(\mu_R) \chi^{\rm NLL}(n,\nu)}
 \alpha_s^2(\mu_R)
\\
 &\,\times\, 
 \left[
 c_{\cal M}^{\rm NLO}(n,\nu,|\vec \kappa_{T_{\cal M}}|, x_{\cal M})[c_{\cal H}^{\rm NLO}(n,\nu,|\vec \kappa_{T_{\cal H}}|,x_{\cal H})]^*\,
 + \bar \alpha_s^2(\mu_R) 
 \, \DY
 \frac{\beta_0}{4 N_c}\chi(n,\nu)\,\Xi(\nu)
 \right] \;,
\end{split}
\end{equation}
\end{widetext}
with $\bar \alpha_s(\mu_R) \equiv \alpha_s(\mu_R) N_c/\pi$, $N_c$ the number of colors and $\beta_0$ standing for the first coefficient of the QCD $\beta$-function.
The $\chi^{\rm NLL}\left(n,\nu\right)$ is the NLL BFKL kernel and its expression reads
\begin{widetext}
\begin{equation}
\chi^{\rm NLL}(n,\nu) = \chi(n,\nu) +\bar\alpha_s(\mu_R) \left[\bar\chi(n,\nu)+\frac{\beta_0}{8 N_c}\chi(n,\nu)\left[-\chi(n,\nu)+\frac{10}{3}+2\ln \frac{\mu_R^2}{|\vec \kappa_{T_{\cal M}}| |\vec \kappa_{T_{\cal H}}|} \right]\right] \;,
\label{chi_NLO}
\end{equation}
with
\begin{equation}
\chi\left(n,\nu\right)=2\left\{\psi\left(1\right)-{\rm Re} \left[\psi\left( i\nu+\frac{n+1}{2} \right)\right] \right\}
\label{chi}
\end{equation}
\end{widetext}
the LL BFKL eigenvalue and $\psi(z) = \Gamma^\prime(z)/\Gamma(z)$ the logarithmic derivative of the Gamma function. 
The $\hat\chi(n,\nu)$ characteristic function in Eq.\eref{chi_NLO} was calculated in\tcite{Kotikov:2000pm} (see also Ref.\tcite{Kotikov:2002ab}).
Its expression is reported in Appendix~\hyperlink{app:NLL_kernel}{A}.
The function $c_h^{\rm NLO}(n,\nu,|\vec \kappa_T|, x)$ is the NLO impact factors for the production of a generic hadron $h$. It was calculated in Ref.\tcite{Ivanov:2012iv} in light-quark limit and contains the collinear inputs. The use of this impact factor also for heavy-hadron species is valid in the spirit of our VFNS treatment, namely provided that the values of the observed transverse momentum $|\vec \kappa_{T_{\cal H}}|$ are definitely higher than the charm (for a $\Lambda_c$ baryon) or bottom (for a $b$~hadron) masses.
Its expression reads
\begin{widetext}
\begin{equation}
\label{HIF}
c_h^{\rm NLO}(n,\nu,|\vec \kappa_T|, x) =
c_h(n,\nu,|\vec \kappa_T|, x) +
\alpha_s(\mu_R) \, \hat c_h(n,\nu,|\vec \kappa_T|, x) \; ,
\end{equation}
where
\begin{equation}
\label{LOHIF}
\hspace{-0.25cm}
c_h(n,\nu,|\vec \kappa_T|, x) 
= 2 \sqrt{\frac{C_F}{C_A}}
|\vec \kappa_T|^{2i\nu-1} \int_{x}^1\frac{\drv \xi}{\xi}
\left( \frac{\xi}{x} \right)
^{2 i\nu-1} 
 \left[\frac{C_A}{C_F}f_g(\xi)D_g^{h}\left(\frac{x}{\xi}\right)
 +\sum_{i=q,\bar q}f_i(\xi)D_i^{h}\left(\frac{x}{\xi}\right)\right] 
\end{equation}
is the LO part and $\hat c_h(n,\nu,|\vec \kappa_T|, x)$ is its NLO correction (see Appendix~\hyperlink{app:NLO_IF}{B} for the analytic formula).
The $\Xi(\nu)$ function in Eq.\eref{Cn_NLLstar_MSb} embodies the logarithmic derivative of the two LO impact factors
\begin{equation}
 \Xi(\nu) = \frac{i}{2} \, \frac{\drv}{\drv \nu} \ln\left(\frac{c_{\cal M}(n,\nu,|\vec \kappa_{T_{\cal H}}|, x_{\cal M})}{[c_{\cal H}(n,\nu,|\vec \kappa_{T_{\cal H}}|, x_{\cal H})]^*}\right) + \ln\left(|\vec \kappa_{T_{\cal M}}| |\vec \kappa_{T_{\cal H}}|\right) \;.
\label{fnu}
\end{equation}
\end{widetext}
From Eqs.~(\ref{Cn_NLLstar_MSb})-(\ref{Cn_LL_MSb}) we gather the way how our hybrid factorization is realized. The cross section comes a factorized formula \emph{à la} BFKL, where the Green's function is high-energy convoluted between the light- and the heavy-hadron impact factors. The latter ones are written in turn as a collinear convolution between collinear PDFs and FFs, and the hard-scattering term.
The NLL$^*$ label refers to the fact that our representation for azimuthal coefficients in Eq.\eref{Cn_NLLstar_MSb} contains terms beyond the NLL accuracy generated by the cross product of the NLO corrections to impact factors, $\hat c_{\cal M}(n,\nu,|\vec \kappa_{T_{\cal M}}|, x_{\cal M}) \, [\hat c_{\cal H}(n,\nu,|\vec \kappa_{T_{\cal H}}|,x_{\cal H})]^*$.
We will calculate our observables by employing another representation, labeled as NLL, where that next-to-NLL factor is not considered.
We will gauge the impact of passing from the NLL to the NLL$^*$ representation for a limited selection of rapidity-differential cross sections (see point~(\emph{v}) of Section\tref{ssec:jethad_settings} and discussion of results presented in Section\tref{ssec:C0}).

By expanding NLL azimuthal coefficients in Eq.~(\ref{Cn_NLLstar_MSb}) up to the ${\cal O}(\alpha_s^3)$ order, we come out with a formula that acts as an effective high-energy fixed-order (HE-NLO) counterpart of our BFKL-resummed expression.
In previous works this procedure was called high-energy DGLAP (see Refs.\tcite{Celiberto:2015yba,Celiberto:2015mpa,Celiberto:2020wpk,Celiberto:2020rxb,Celiberto:2021dzy}).
It permits us to pick the leading-power asymptotic signal of a pure NLO DGLAP calculation, concurrently removing those factors which are dampened by inverse powers of $\hat s$.
Our HE-NLO expression for azimuthal coefficients in the $\MSb$ scheme reads
\begin{widetext}
\begin{equation}
\label{Cn_HENLOstar_MSbar}
 \CnHENLOstar = \frac{e^{\Delta Y}}{s}
 \int_{-\infty}^{+\infty} \drv \nu \, \alpha_s^2(\mu_R)
 \left[ c_{\cal M}^{\rm NLO}(n,\nu,|\vec \kappa_{T_{\cal M}}|,x_{\cal M})[c_{\cal H}^{\rm NLO}(n,\nu,|\vec \kappa_{T_{\cal H}}|,x_{\cal H})]^* + \bar \alpha_s(\mu_R) \DY \chi \right] \,
\end{equation}
\end{widetext}
where an expansion up to terms proportional to $\alpha_s(\mu_R)$ replaces the BFKL exponentiated kernel.
In analogy to Eq.\eref{Cn_NLLstar_MSb}, our high-energy fixed order formula is labeled as HE-NLO$^*$ because it encodes contributions beyond the NLL accuracy due to the cross product of NLO corrections to impact factors.
Also in this case we will mainly employ a HE-NLO representation, where those higher-order terms are not included, and we will assess the effect of switching them on for a limited selection of rapidity-differential cross sections.

Finally, by neglecting all NLO terms in the BFKL kernel and impact factors in Eq.\eref{Cn_NLLstar_MSb}, we can write a pure LL expression of the azimuthal coefficients
\begin{widetext}
\begin{equation}
\label{Cn_LL_MSb}
  \CnLL = \frac{e^{\DY}}{s} 
 \int_{-\infty}^{+\infty} \drv \nu \, e^{{\DY} \bar \alpha_s(\mu_R)\chi(n,\nu)} \, \alpha_s^2(\mu_R) \, c(n,\nu,|\vec \kappa_{T_{\cal M}}|, x_{\cal M})[c(n,\nu,|\vec \kappa_{T_{\cal H}}|,x_{\cal H})]^* \,,
\end{equation}
\end{widetext}
which we will use for comparisons with corresponding NLL and HE-NLO calculations.

In our phenomenological analysis (see Section\tref{sec:pheno}) we consider observables built in terms of NLL$^{(*)}$, HE-NLO$^{(*)}$, and LL azimuthal coefficients. We fix renormalization and factorization scales at the \emph{natural} energies provided by the given final state. Thus we have $\mu_R = \mu_F = \mu_N \equiv m_{\perp_{\cal M}} + m_{\perp_{\cal H}}$, with $m_{h \perp} = \sqrt{ m_h^2 + |\vec \kappa_{T_{h}}|^2}$ being the $h$-hadron transverse mass. To guess the size of higher-order corrections, scales will be varied as specified in point~(\emph{i}) of Section\tref{ssec:jethad_settings}.

\subsection{Perturbative ingredients: running coupling and renormalization scheme}
\label{ssec:pert}

We adopt in our analysis a two-loop running-coupling choice with $\alpha_s\left(M_Z\right)=0.11707$ and five quark $n_f = 5$ flavors active. Working in the $\MSb$ renormalization scheme, one has
\begin{equation}
\label{as_MSb}
 \hspace{-0.41cm}
 \as(\mu_R) \equiv \as^{\MSb}(\mu_R) = \frac{\pi}{\beta_0 \, {\cal L}_R} \left( 4 - \frac{\beta_1}{\beta_0^2} d \frac{\ln {\cal L}_R}{{\cal L}_R} \right) 
\end{equation}
with
\begin{equation}
\label{as_parameters}
 {\cal L}_R(\mu_R) = \ln \frac{\mu_R^2}{\LQCD^2} \;,
\end{equation}
\[
 \beta_0 = 11 - \frac{2}{3} n_f \;, \qquad
 \beta_1 = 102 - \frac{38}{3} n_f \;.
\]
The corresponding expression for the strong coupling in the momentum (MOM) renormalization scheme\tcite{Barbieri:1979be,PhysRevD.20.1420,PhysRevLett.42.1435}, whose definition is related to the QCD three-gluon vertex, is obtained by inverting the finite renormalization given below
\begin{equation}
\label{as_MOM}
 \as^{\MSb} = \as^{\rm MOM} \left( 1 +  \frac{T_{[\alpha_s]}^{\beta} + T_{[\alpha_s]}^{\rm conf}}{\pi} \as^{\rm MOM} \right) \;,
\end{equation}
where
\begin{equation}
\label{T_bc}
T_{[\alpha_s]}^\beta = - \left(\frac{I}{3} +  \frac{1}{2} \right) \beta_0
\end{equation}
and
\begin{equation}
\label{T_conf}
\begin{aligned}
T_{[\alpha_s]}^{\rm conf} & = \frac{C_A}{8}\left[ \frac{17}{2}I +\frac{3}{2}\left(I-1\right)\zeta_{{\rm conf}} \right. \\
 & \left. - \, \left(\frac{I}{3} - 1\right)\zeta_{{\rm conf}}^2-\frac{1}{6}\zeta_{{\rm conf}}^3 \right] \; ,
\end{aligned}
\end{equation}
with $C_A \equiv N_c$ is the Casimir factor associated to a the emission of a gluon from a gluon and
\begin{equation}
 \label{I_scheme}
 I = -2 \int_0^1 \drv \tau \, \frac{\ln \tau}{(1 - \tau)^2 + \tau} \simeq 2.3439 \; ,
\end{equation}
the gauge parameter $\zeta_{{\rm conf}}$ being fixed to zero in the following.

We remark that in our treatment energy scales are strictly related to transverse masses of observed particles. Therefore they always fall in the perturbative region, so that no infrared improvement of the running coupling (see, \emph{e.g.}, Ref.\tcite{Webber:1998um}) is needed. Furthermore, large scale values protect us from a region where the \emph{diffusion pattern}\tcite{Bartels:1993du} (see also Refs.\tcite{Caporale:2013bva,Ross:2016zwl}) becomes important.

Main calculations of our observables are done in the $\MSb$ scheme. We assess the weight of the systematic uncertainty arising when passing from the $\MSb$ to the MOM scheme (see point~(\emph{iiii}) of Section\tref{ssec:jethad_settings}).
This change of scheme can be done by operating the following replacement of the running coupling in Eqs.~(\ref{Cn_NLLstar_MSb}), (\ref{HIF}), (\ref{Cn_HENLOstar_MSbar}), and~(\ref{Cn_LL_MSb})
\begin{equation}
 \label{MSb_2_MOM}
 \alpha_s^{\MSb}(\mu^{\rm BLM}_R) \,\to\, \alpha_s^{\rm MOM}(\mu^{\rm BLM}_R) \;.
\end{equation}
In particular, we pass from the analytic formula of the strong coupling in the $\MSb$ scheme~(Eq.\eref{as_MSb}) to the MOM one obtained via Eq.\eref{as_MOM}, without altering the value of the renormalization scale, $\mu_R$.
We stress, however, that a complete MOM study of our distributions would rely on collinear PDFs and FFs whose evolution has been determined in the MOM scheme. Therefore, our approximated way to gauge the size of a renormalization-scheme variation can be thought as an upper limit of the full one, whose overall effect still needs to be quantified.

\subsection{Nonperturbative ingredients: collinear PDFs and FFs}
\label{ssec:PDFs_FFs}

As already mentioned, due to the moderate parton $x$~values, in our analysis we rely on collinear PDF and FF inputs which evolve via DGLAP, while the high-energy resummation is accounted for by the BFKL Green's function.
We mainly employ central values of NLO {\tt MMHT14} proton PDF sets\tcite{Harland-Lang:2014zoa}, while the novel {\tt NNPDF4.0} NLO determination\tcite{NNPDF:2021uiq,NNPDF:2021njg} is used for uncertainty studies (see point~(\emph{ii}) of Section\tref{ssec:jethad_settings}).

Concerning light mesons, we describe $\eta$ emissions via the only FF set thus far available, namely the NLO {\tt AESSS11} one\tcite{Aidala:2010bn} obtained from a global fit on data for SIA events at various center-of-mass energies and for proton-proton collisions at BNL-RHIC in a wide range of transverse momenta.
As for charged pions, we can benefit from a wider choice at NLO.
{\tt NNFF1.0} parametrizations\tcite{Bertone:2017tyb} were extracted from SIA data via a neural-network approach, the gluon FF being generated at NLO.
{\tt DEHSS14} functions\tcite{deFlorian:2014xna} were obtained from data on SIA, lepton-nucleon semi-inclusive deep-inelastic scattering (SIDIS) and proton-proton collisions.
They assume a partial $SU(2)$ isospin symmetry that leads to $D_u^{\pi^+} + D_{\bar u}^{\pi^+} \propto D_d^{\pi^+} + D_{\bar d}^{\pi^+}$.
{\tt JAM20}\tcite{Moffat:2021dji} FFs include SIA and SIDIS dataset and were simultaneously determined together with collinear PDFs extracted from DIS and fixed-target Drell--Yan measurements. They rely on a full $SU(2)$ isospin symmetry which turns into $D_u^{\pi^+} = D_{\bar d}^{\pi^+}$.
{\tt MAPFF1.0}\tcite{Khalek:2021gxf} determinations were recently obtained from SIA and SIDIS data by using neural-network techniques.
They rely on two independent parametrizations for $D_u^{\pi^+}$ and $D_{\bar d}^{\pi^+}$, thus allowing for a violation of the isospin symmetry that depends on the hadron momentum fraction, $z$. Gluon is generated at NLO, but data are taken at lower energies, where the gluon content has a larger size. Quite recently, the technology built for the extraction of {\tt MAPFF1.0} FFs was used to study the fragmentation of the $\Xi^-/{\bar \Xi}^+$ octet baryon from SIA data\tcite{Soleymaninia:2022qjf} and to determine a novel FF set to describe an unidentified charged light hadron from SIA and SIDIS data\tcite{Soleymaninia:2022alt}.

As regards heavy hadrons, we employ the novel {\tt KKSS19} set\tcite{Kniehl:2020szu} to describe parton fragmentation into $\Lambda_c$ baryons.
These FFs were extracted from OPAL and Belle data for SIA and mainly rely on a description \emph{\`a la} Bowler\tcite{Bowler:1981sb} for charm and bottom flavors.
We depict emissions of $b$~flavored hadrons in terms of the {\tt KKSS07} parametrization\tcite{Kniehl:2008zza} based on data of the inclusive $B$-meson production in SIA events at CERN LEP1 and SLAC SLC and portrayed by a simple, three-parameter power-like \emph{Ansatz}\tcite{Kartvelishvili:1985ac} for heavy-quark species.
The {\tt KKSS19} and {\tt KKSS07} determinations use the VFNS. 
We remark that the employment of given VFNS PDFs or FFs is admitted in our approach, provided that typical energy scales are much larger than thresholds for the DGLAP evolution of charm and bottom quarks. As highlighted in Section\tref{ssec:kinematics}, this requirement is always fulfilled.

Our way to estimate the uncertainty coming from FFs is explained in point~(\emph{iii}) of Section\tref{ssec:jethad_settings}).
We remark that in {\tt KKSS19} and {\tt KKSS07} datasets no quantitative information on the extraction uncertainty is provided. Future studies relying on possible novel $\Lambda_c$ and $H_b$ FF parametrizations including uncertainties are needed to complement our analysis on systematic errors of high-energy distributions.

\section{Numerical analysis}
\label{sec:pheno}

In this Section we present results of our phenomenological analysis done via the {\tt JETHAD} technology\tcite{Celiberto:2020wpk}. After explaining our strategy to gauge the weight of the main uncertainties entering the description of our observables~(Section\tref{ssec:jethad_settings}), we give details on the selected final-state ranges, which include also combined tags at ATLAS and FPF detectors~(Section\tref{ssec:kinematics}). 
Results for cross sections differential in the final-state rapidity distance, $\DY$, and for azimuthal-angle distributions are discussed in Sections\tref{ssec:C0} and\tref{ssec:phi}, respectively.

\subsection{Final-state kinematics}
\label{ssec:kinematics}

\begin{figure*}[tb]
\centering

\includegraphics[width=0.485\textwidth]{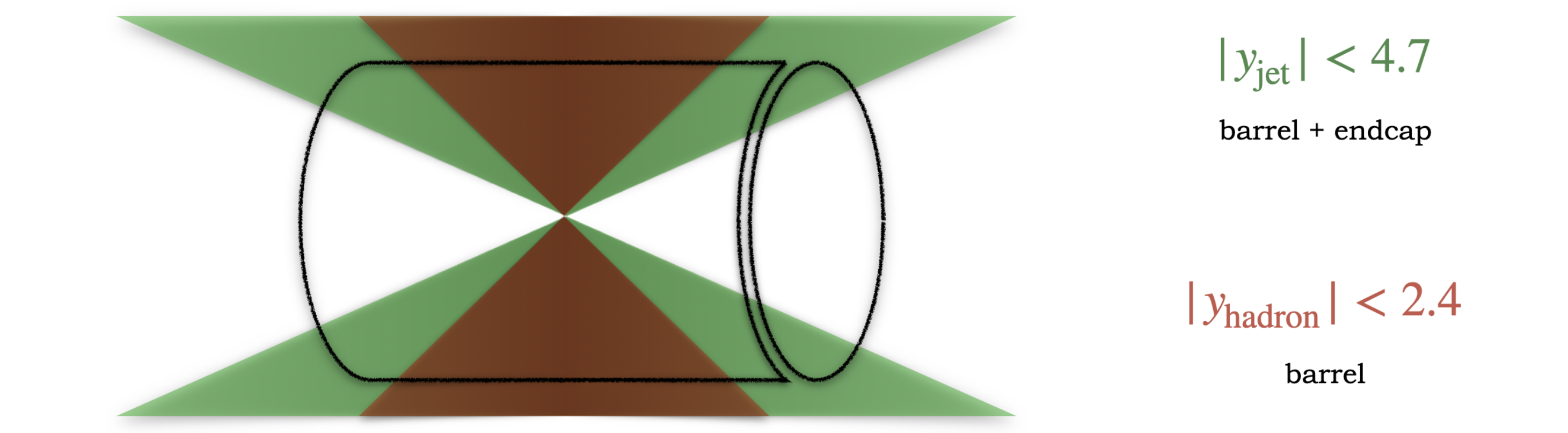}
\hspace{0.25cm}
\includegraphics[width=0.485\textwidth]{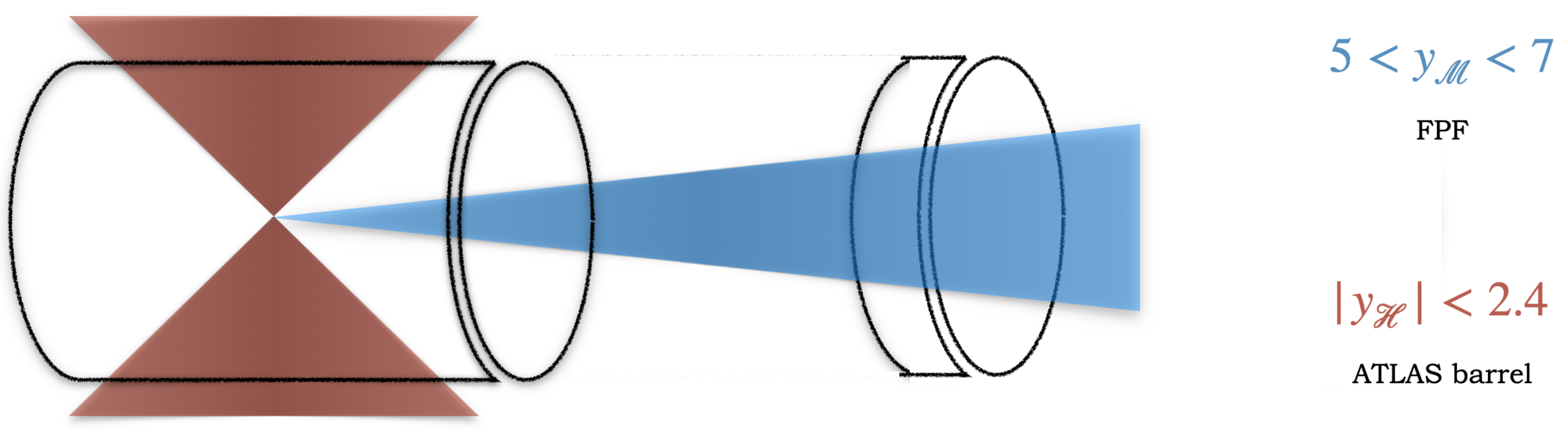}
\\ \vspace{0.25cm}
\hspace{-0.50cm}
($a$) CMS/ATLAS standard tag \hspace{3.50cm}
($b$) FPF + ATLAS coincidence

\caption{Left panel: example of a forward-backward two-particle tag in LHC standard configurations. Typical final-state objects\tcite{Bolognino:2018oth,Celiberto:2020wpk,Celiberto:2020rxb} are a light-flavored jet, detected by CMS or ATLAS barrel and end-cap detectors in the symmetric rapidity range $|y_{\rm jet}| < 4.7$, and a light-flavored hadron (pion, kaon, proton or a $\Lambda$ baryon), reconstructed only by CMS or ATLAS barrel in the symmetric rapidity range $|y_{\rm hadron}| < 2.4$.
Right panel: simultaneous tag, via a tight timing coincidence, of a light meson ($\eta$ meson or pion in our study) at the FPF\tcite{Anchordoqui:2021ghd,Feng:2022inv} in the \emph{ultra-forward} rapidity range $5 < y_{\cal M} < 7$, and of a heavy hadron ($\Lambda_c$ baryon or $b$~hadron in our study) inside the ATLAS barrel in the central rapidity range $|y_{\cal H}| < 2.4$.}
\label{fig:detectors}
\end{figure*}

Starting from Eqs.~(\ref{Cn_NLLstar_MSb}), (\ref{Cn_HENLOstar_MSbar}), and~(\ref{Cn_LL_MSb}), we build physical observables in terms of azimuthal coefficients \emph{integrated} over the final-state phase space objects, while their mutual rapidity separation, $\DY = y_{\cal M} - y_{\cal H}$, is kept fixed. We have
\begin{widetext}
\begin{equation}
 \label{DY_distribution}
 C_n (\DY; s) =
 \int_{\kappa_{T_{\cal M}}^{\rm min}}^{\kappa_{T_{\cal M}}^{\rm max}} \drv |\vec \kappa_{T_{\cal M}}|
 \int_{\kappa_{T_{\cal H}}^{\rm min}}^{\kappa_{T_{\cal H}}^{\rm max}} \drv |\vec \kappa_{T_{\cal H}}|
 \int_{y_{\cal M}^{\rm min}}^{y_{\cal M}^{\rm max}} \drv y_{\cal M}
 \int_{y_{\cal H}^{\rm min}}^{y_{\cal H}^{\rm max}} \drv y_{\cal H}
 \, \,
 \delta (\DY - (y_{\cal M} - y_{\cal H}))
 \, \,
 {\cal C}_n
 \, .
\end{equation}
\end{widetext}
Here, $C_n$ collectively represents all the LL, NLL$^{(*)}$ and HE-NLO$^{(*)}$ azimuthal coefficients. This permits us to impose and study different windows in transverse momenta and rapidities, based on realistic kinematic configurations used in current and forthcoming experimental analyses at the LHC.
We focus on the two following kinematic selections.

\subsubsection{LHC standard detection}
\label{sssec:LHC_standard}

We consider the emission of both the light- and the heavy-flavored hadron inside acceptances of CMS or ATLAS barrel calorimeters. At variance with the Mueller--Navelet channel, where typical CMS ranges\tcite{Khachatryan:2016udy} allow for jets tagged also in the end-caps (see panel ($a$) of Fig.\tref{fig:detectors}), thus having $|y_{\rm jet}| < 4.7$, hadrons can be easily detected only by barrels.
A realistic proxy for the rapidity window of hadron tags at the LHC can be taken from a recent analysis on $\Lambda_b$ particles at CMS\tcite{Chatrchyan:2012xg}, $|y_{\Lambda_b}| < 2$. In our study we admit a slightly wider range, namely the one covered by the CMS barrel detector, $|y_{{\cal M}, {\cal H}}| < 2.4$.
As done in previous studies\tcite{Celiberto:2017ptm,Celiberto:2020rxb,Celiberto:2021dzy}, we impose a $10 < |\vec \kappa_{T_{\cal M}}| / {\rm GeV} < 20$ window for the transverse momentum of the light hadron.
Conversely, the range of the heavy-particle transverse momentum is taken to be disjoint and larger than the light-meson one, namely $20 < |\vec \kappa_{T_{\cal M}}| / {\rm GeV} < 60$, as recently proposed in the context of high-energy emissions of $b$~hadrons\tcite{Celiberto:2021fdp}. This choice preserves the validity of our VFNS treatment, since energy scales are always much higher than thresholds for DGLAP evolution of the heavy-quark FFs (for further details see Section\tref{ssec:PDFs_FFs}).

On the one side, the fact that both the light and the heavy particle are tagged in a symmetric rapidity range allow us to apply our formalism in a ``known" sector, where stringent tests of the hybrid factorization can be conducted.
On the other side, as pointed out in Ref.\tcite{Celiberto:2020wpk}, the use of disjoint intervals for the two observed transverse momenta quenches the Born contribution, thus heightening effects of the additional undetected gluon radiation. This emphasizes imprints of the high-energy resummation with respect to the standard fixed-order treatment. Moreover, asymmetric $\kappa_T$-windows dampen possible instabilities rising in NLO calculations\tcite{Andersen:2001kta,Fontannaz:2001nq} as well as violations of the energy-momentum at NLL\tcite{Ducloue:2014koa}.
However, the $\DY$ values reachable with a LHC standard tag could be not so large to allow for a clear discrimination between BFKL and fixed-order signatures. This difficulty can be overcome by considering \emph{ultra-forward} emissions of one of the two particles, as suggested in Section\tref{sssec:FPF_ATLAS}.

\subsubsection{FPF + ATLAS coincidence}
\label{sssec:FPF_ATLAS}

In addition to the standard range described in Section\tref{sssec:LHC_standard}, we propose the simultaneous detection of an \emph{ultra-forward} light meson and of a more rapidity-central heavy-flavored bound state (see panel ($b$) of Fig.\tref{fig:detectors}).
Once the planned FPF\tcite{Anchordoqui:2021ghd,Feng:2022inv} will be operating, this unprecedented opportunity will become feasible by making FPF detectors work in coincidence with ATLAS. The chance of combining information from ATLAS and the FPF will rely on the ability of using ultra-forward events as a trigger for ATLAS. This will call for very precise timing procedures and will have an impact on the design of FPF detectors. Technical details on the FPF~$+$~ATLAS tight timing coincidence can be found in Section~VI~E of Ref.\tcite{Anchordoqui:2021ghd}.

As a motivational study, we consider the emission of a $\eta$ or a $\pi^{\pm}$ meson in the range $5 < y_{\cal M} < 7$, which we take as a proxy for forthcoming analyses at the FPF. Although larger rapidities could be reached, we opt for a more conservative choice, settling for a rapidity window disjoint and more forward than the range accessible by ATLAS end-caps. Studies on larger rapidity ranges are postponed to a next work.
The light meson is accompanied by a $\Lambda_c$ or a $H_b$ particle detected by the ATLAS barrel calorimeter in the standard rapidity spectrum, $|y_{\cal H}| < 2.4$.
The transverse momenta of the two final-state hadrons are the same as the ones given in Section\tref{sssec:LHC_standard}.

Combined FPF~$+$~ATLAS tags afford a hybrid and strongly asymmetric range selection that results in an excellent channel where to disentangle the onset of clear high-energy signals from the collinear background.
On the other hand, as pointed out in Refs.\tcite{Celiberto:2020rxb,Bolognino:2021mrc} and mentioned previously, the combined detection of a very forward particle together with a central one brings to an asymmetric configuration between the longitudinal fractions of the two corresponding incoming partons, which strongly restricts the weight of the undetected-gluon radiation at LO, and it has a sizable impact also at NLO.
This kinematic limitation translates in an incomplete cancellation between virtual and real contributions coming from gluon emissions, which leads to the emergence of large Sudakov-type logarithms (\emph{threshold} logarithms) in the perturbative series.
Since the BFKL formalism accounts for the resummation of energy-type single logarithms and systematically neglects the threshold ones, a partial worsening of the convergence of our resummed calculation is expected in this timing-coincidence setup with respect the LHC standard one (see Section\tref{ssec:C0}). Although being challenging, these features motivate us to next formal developments in embodying the threshold resummation inside our formalism.

Studies via the FPF~$+$~ATLAS coincidence method offer us a peerless opportunity for stringent and deeper tests of the dynamics of strong interactions the high-energy regime. In this sense, the advent of the FPF might complement the reach of the ATLAS detector, thus permitting us to (\emph{i}) gauge the feasibility of \emph{precision} analyses at the hands of the hybrid high-energy and collinear factorization, and (\emph{ii}) explore possible common ground among distinct resummations.

\subsection{{\tt JETHAD} settings and uncertainties}
\label{ssec:jethad_settings}

The main features of the {\tt JETHAD} multi-modular interface, aimed at the management, calculation and processing of observables defined in different approaches were presented for the first time in Ref.\tcite{Celiberto:2020wpk}.
In order to perform the studies presented in this article, {\tt JETHAD} has been sensibly upgraded.

Principal updates are as follows: the automation of the scale-variation analysis, the inclusion of a \textsc{Python}-based analyzer suited to the elaboration of results, and the possibility of using different sets of functions (collinear PDFs and FFs, TMDs, $\mbox{small-}x$ UGD, and so on) for each incoming and outgoing particle.
The implementation of this last feature was done by taking advantage of the \emph{structure}-based, smart-management system natively incorporated in {\tt JETHAD}, where physical particles are portrayed by \emph{object} prototypes, namely \textsc{Fortran} structures.
Particle structures contain all information about basic and kinematic properties of their physical \emph{Doppelg\"anger}, like mass, charge, transverse momentum and rapidity. They are first loaded from a master database via a dedicated \emph{particle-generation} routine, then \emph{cloned} to a final-state particle array, and finally \emph{injected} to the observable-related routine, differential in the final-state variables, to the corresponding impact-factor module by means of an \emph{impact-factor controller}.

Each particle structure has a \emph{particle-ascendancy} attribute, which allows {\tt JETHAD} to recognize from which particles the process was initiated (protons, leptons, or heavy nuclei) and automatically selects which modules must be initialized (PDFs, FFs, etc...).
With the aim of providing the scientific Community with a standard software for the analysis of different kinds of reactions via distinct approaches, we plan to release soon a first public version of {\tt JETHAD}.

An accurate description of our observables relies on identifying all the potential sources of uncertainty.
Pursuing the goal of comparing high-energy resummed predictions with next-to-leading calculations and highlighting the steps required to reach the precision level, we gauge distinct and combined effects of a selection of uncertainties. 
Our choice fall on those uncertainties which are typically considered in collinear-factorization phenomenology, but they turn out to be novel in the semi-hard one. We also include systematic effects intrinsically coming the high-energy resummation. More in particular, we have the following:
\begin{itemize}

 \item[(\emph{i})]
 \textbf{Scale-variation uncertainty}.
 As usually done in perturbative calculations, we assess the sensitivity of our distributions when the renormalization and factorization scales, $\mu_R$ and $\mu_F$, are varied around their \emph{natural} values up to a factor ranging from 1/2 to 2. Such an analysis permits to guess the size of higher-order corrections with respect to the considered accuracy. The $C_{\mu}$ parameter entering plots of Sections\tref{ssec:C0} and\tref{ssec:phi} stands for the ratio $C_\mu = \mu_{R,F}/\mu_N$, with $\mu_N$ being defined at the end of Section\tref{ssec:sigma};

 \item[(\emph{ii})]
 \textbf{PDF uncertainty}.
 Previous studies on semi-hard distributions have shown that the selection of different collinear PDF sets as well as of different members inside the same set does not have a sizable impact (see, \emph{e.g.}, Refs.\tcite{Bolognino:2018oth,Celiberto:2020wpk,Celiberto:2021fdp}).
 However, as mentioned in Section\tref{sssec:FPF_ATLAS}, the adoption of a FPF~$+$~ATLAS coincidence setup leads to an asymmetric configuration where one of the two parton longitudinal-momentum fractions is always large. Thus, we enter the so-called \emph{threshold} region, where PDFs could not be well constrained.
 For a limited selection of predictions, we perform a systematic study via the \emph{replica} method applied to the {\tt NNPDF4.0} set. Nowadays widely employed in QCD analyses, this bootstrap procedure was originally proposed in Ref.\tcite{Forte:2002fg} in the context of neural-network inspired techniques. Its key point is the possibility to generate a large number of replicas of the central value of a distribution by randomly altering its central value with a multi-variate Gaussian background featuring the original standard deviation.
 As a general digression on the study on PDF-based uncertainties, we remark that the analysis of the \emph{correlations} between different PDFs sets hides some ambiguities that still need to be clarified. As pointed out in Ref.\tcite{Ball:2021dab}, employing data-driven correlations to different PDFs into a unique, joint set can be dangerous, since the estimated uncertainty could miss the so-called \emph{functional-correlation} part. A more reliable method would rely in building joint sets via a statistical combination. The {\tt PDF4LHC15} PDF parametrization\tcite{Butterworth:2015oua} (see Ref.\tcite{Ball:2022hsh} for the new {\tt PDF4LHC21} set), based on this approach, was used for the first time in the context of semi-hard phenomenology to assess the weight of uncertainties in Mueller--Navelet azimuthal-angle correlations (see Section~3.3 of Ref.\tcite{Celiberto:2020wpk}).

 \item[(\emph{iii})]
 \textbf{FF uncertainty}.
 At variance with PDFs, our knowledge of collinear FFs is much more limited. This is true in particular for heavy-hadron FFs, for which the collected statistics is quite low. Moreover, data at low transverse momenta cannot be used to extract VFNS functions. 
 Therefore, studies on hadroproductions of different hadron species in semi-hard configurations at the LHC are relevant to better constrain FFs. The FPF~$+$~ATLAS setup offers us an unprecedented opportunity to extend these analyses to ranges complementary to the currently accessible ones.
 Unfortunately, the list of heavy-flavor VFNS functions is quite short, none of them containing sufficient information on statistical uncertainties. For this reason, we include in the calculation of our observables only the central values of {\tt KKSS19} $\Lambda_c$ and {\tt KKSS07} $H_b$ FFs. The same strategy is employed also for light $\eta$ mesons, for which, according to our understanding, {\tt AESSS11} is the only available set.
 Conversely, the current knowledge of $\pi^\pm$ fragmentation is much more robust, and we can rely on several determinations (see Section\tref{ssec:PDFs_FFs}) which incorporate a detailed analysis of systematic uncertainties.
 We compare the behavior of rapidity-interval and azimuthal-angle differential distributions calculated by employing the four $\pi^\pm$ FF determinations mentioned in Section\tref{ssec:PDFs_FFs}. Three of these sets, namely {\tt NNFF1.0}, {\tt JAM20}, and {\tt MAPFF1.0} were built via the replica method. We perform a replica-driven analysis of cross sections obtained with these three FFs.

 \item[(\emph{iiii})]
 \textbf{Change of renormalization scheme}.
 As anticipated in Section\tref{ssec:pert}, we gauge the systematic uncertainty on rapidity-differential distributions due to a change from the $\MSb$ to the MOM renormalization scheme.
 We remark again that a full MOM analysis counts on MOM-evolved collinear PDFs and FFs. Thus, our attempt at assessing the weight associated to such a scheme variation works as an upper limit for the full treatment, whose global size still needs to be evaluated.

 \item[(\emph{v})]
 \textbf{Inclusion of next-to-NLL terms}.
 We assess the weight of the systematic effect coming from the inclusion of next-to-NLL contributions in our distributions. This corresponds to gauging the impact of a change of representation of the high-energy resummed cross section. We pass from a pure NLL description to a NLL$^*$ one, where partial higher-order contributions are guessed via including the cross product of NLO corrections to BFKL impact factors (for further details see Section\tref{ssec:sigma}).

 \item[(\emph{vi})]
 \textbf{RG-improvement of BFKL kernel}.
 As a complementary analysis, we gauge the impact of modifying the NLL BFKL kernel through the so-called \emph{collinear improvement}\tcite{Salam:1998tj,Ciafaloni:2003kd,Ciafaloni:2003rd,Ciafaloni:2003ek,Ciafaloni:2002xk,Ciafaloni:2002xf,Ciafaloni:2000cb,Ciafaloni:1999au,Ciafaloni:1999yw,Ciafaloni:1998iv,SabioVera:2005tiv}. Based on the inclusion of terms generated by renormalization group (RG) to impose a compatibility with the DGLAP equation in the collinear limit, it operationally prescribes a modification of the kernel pole structure in the Mellin space.
 Therefore, although including it in this Section, we intend the collinear-improvement procedure as a tool to investigate a possible connection between the BFKL and the collinear resummation, rather than a systematic uncertainty internal to our approach.
 The precise expression of the RG-improved kernel is reported in Appendix~\hyperlink{app:NLL_kernel}{A}.

 \item[(\emph{vii})]
 \textbf{Error in the numerical integration}.
 The main source of numerical uncertainties comes from the multi-dimensional integration over the final-state phase space (see Eqs.~(\ref{DY_distribution}) and~(\ref{azimuthal_distribution})) and over the $\nu$ Mellin variable (see Eqs.~(\ref{Cn_NLLstar_MSb}), (\ref{Cn_HENLOstar_MSbar}), and~(\ref{Cn_LL_MSb})). It was directly performed through the integration routines natively implemented in {\tt JETHAD}, and the resulting error was constantly kept below 1\%.
 Secondary uncertainty sources are represented by the one-dimensional integration over the partons' longitudinal $x$ which defines the convolution between PDFs and FFs in the LO and NLO hadron impact factors (see Eq.\eref{LOHIF}) and the additional one-dimensional integration over the longitudinal momentum fraction, $\xi$, in the NLO impact-factor corrections (see Appendix~\hyperlink{app:NLO_IF}{B}). Preliminary tests have shown that these two uncertainties turn out to be negligible with respect to the main multi-dimensional integration.

\end{itemize}

\subsection{$\DY$-distributions}
\label{ssec:C0}

In this Section we present and discuss the behavior of $\DY$-distributions for our processes (see Fig.\tref{fig:process}). These observables correspond to $\phi$-summed cross sections, $C_0$, differential in the final-state rapidity distance, $\DY$. 
Assessing the feasibility of precision studies of these cross sections relies on a comprehensive analysis of systematic uncertainties. 
Therefore, we gauge the distinct effect of uncertainties presented in Section\tref{ssec:jethad_settings}. In some representative cases, we study the combined effect coming from a selection of two sources of uncertainty.

We remark that signals of a \emph{natural stabilization} of our distributions are strongly expected when LHC standard final-state kinematic cuts are imposed, and they are also awaited in a FPF~$+$~ATLAS coincidence setup.
When found, such a stability would represent a further reliability test for our hybrid high-energy and collinear factorization, and a core result in tracing the path towards future precision analyses.
Stabilizing effects can manifest at different levels.

The first level is the possibility of studying $\DY$-distributions for all the considered final states in Fig.\tref{fig:process} around the natural energy scales provided by kinematics.
This is a required condition to claim evidence of stability, which, as already mentioned, cannot be achieved when light hadrons and/or jets are considered\tcite{Ducloue:2013bva,Caporale:2014gpa,Celiberto:2017ptm,Celiberto:2020wpk}.
In those studies it was observed that large NLL contributions, which have the same size but the opposite sign of LL counterparts, lead to unphysical values for the $\DY$-distribution. More in particular, it can easily reach negative values for large values of $\DY$. Another evidence of instability emerges in the analysis of mean values of cosines of multiples of the azimuthal-angle distance, $\langle \cos (n \phi) \rangle$, which turn out to be larger than one.

The second level is a substantial reduction of the discrepancy between pure LL calculations and NLL-resummed ones with respect to what happens for semi-hard reactions involving the emission of lighter particles. 
Although the two stability levels are not independent, since the sensitivity to scale variation generally has an influence on the LL/NLL dichotomy\tcite{Celiberto:2020wpk,Celiberto:2021dzy,Celiberto:2022dyf}, we will see that the second level is fairly approached in LHC configurations, while factors external to the high-energy resummation prevent to completely achieve it in the FPF~$+$~ATLAS setup.

To gauge the impact of our resummed calculations on fixed-order predictions, we compare LL and NLL results for $C_0$ with the corresponding ones obtained by the hands of our HE-NLO formula~(Eq.\eref{Cn_HENLOstar_MSbar}), which consistently mimics the high-energy signal of a pure NLO computation.
We remark that, due to the absence of a numeric tool devoted to higher-order perturbative calculations of cross sections for the semi-hard hadroproduction of two identified hadrons, our HE-NLO approach remains the most valid and effective one.

\begin{figure*}[t]

   \includegraphics[scale=0.48,clip]{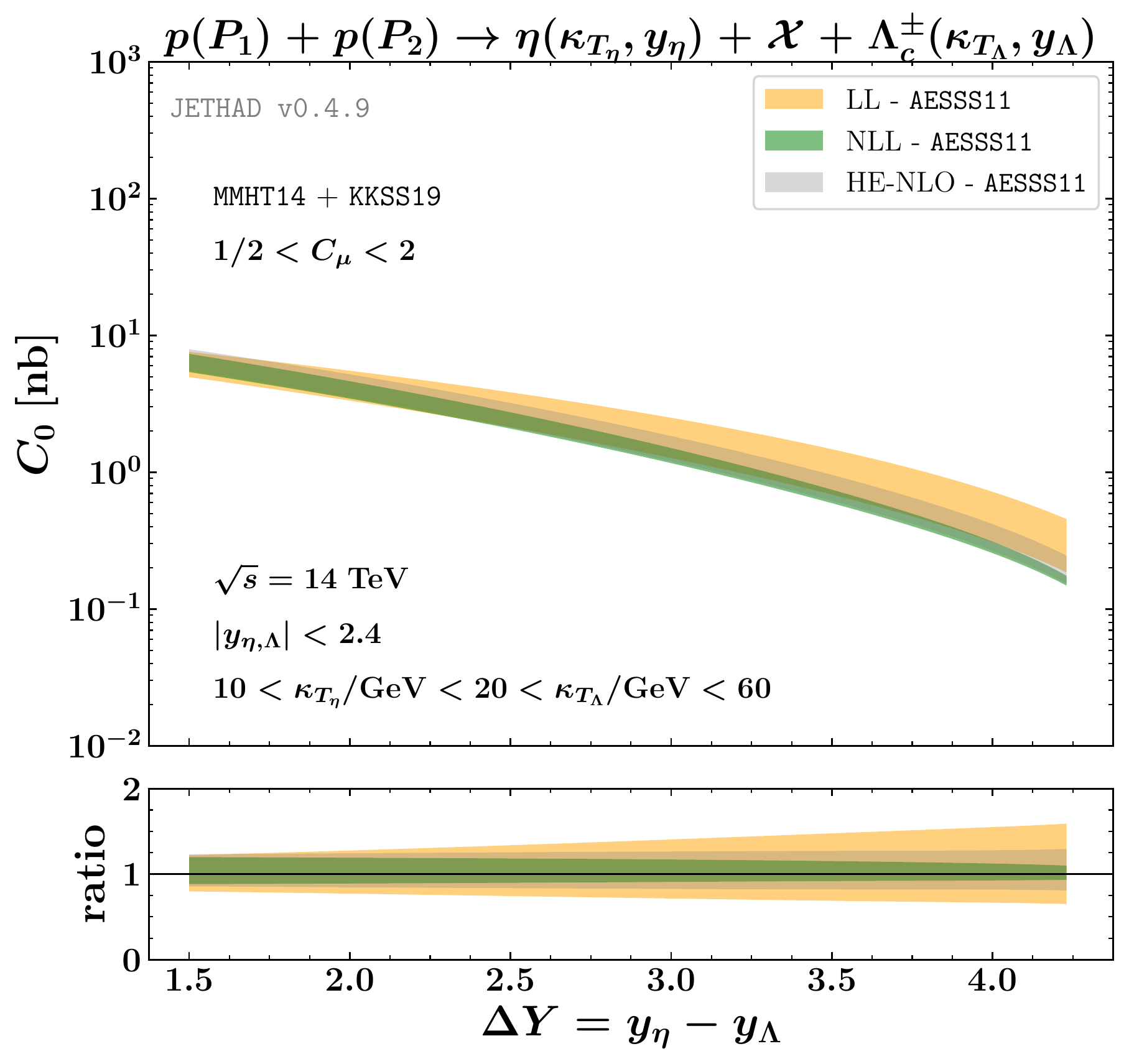}
   \includegraphics[scale=0.48,clip]{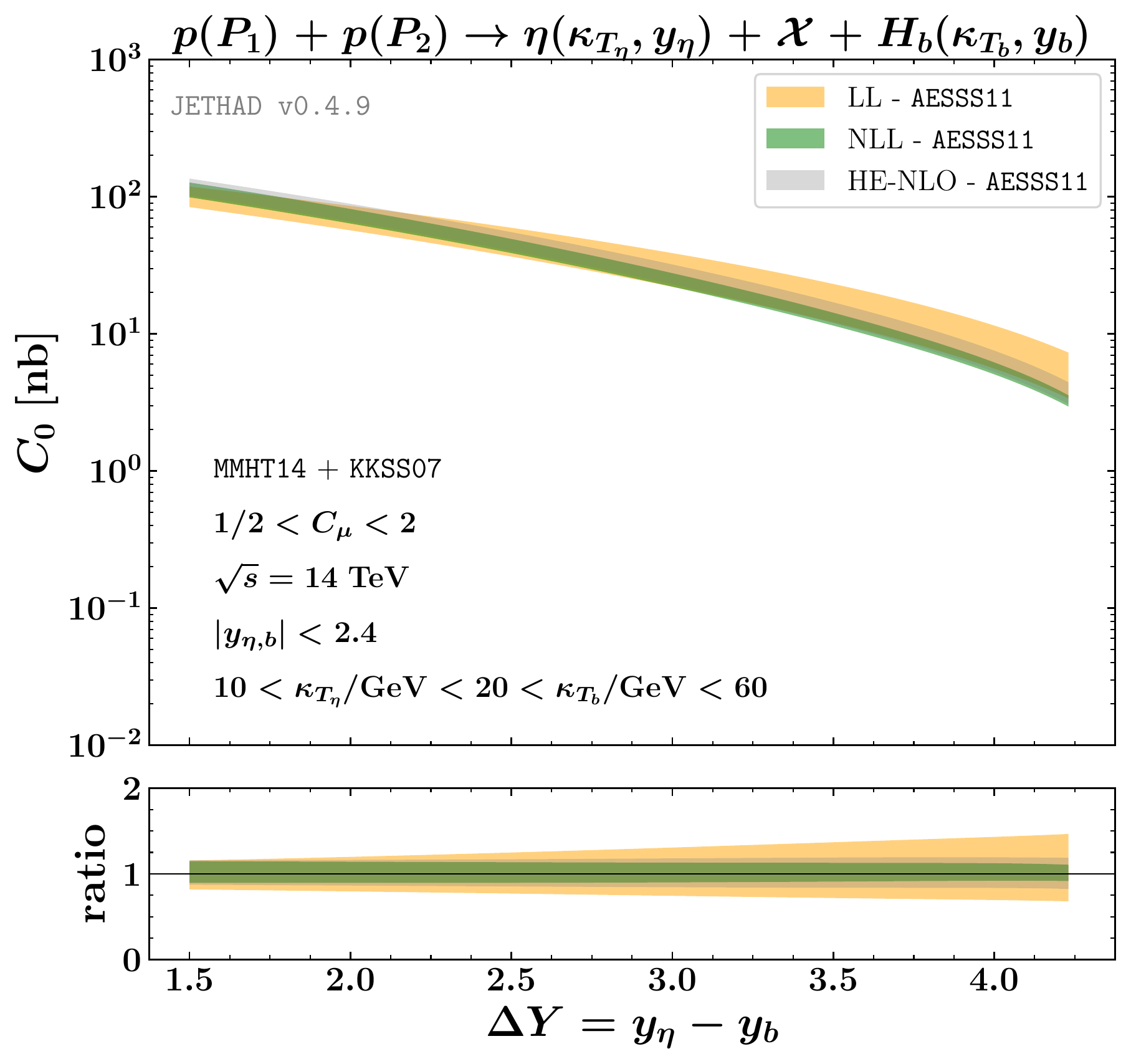}

   \includegraphics[scale=0.48,clip]{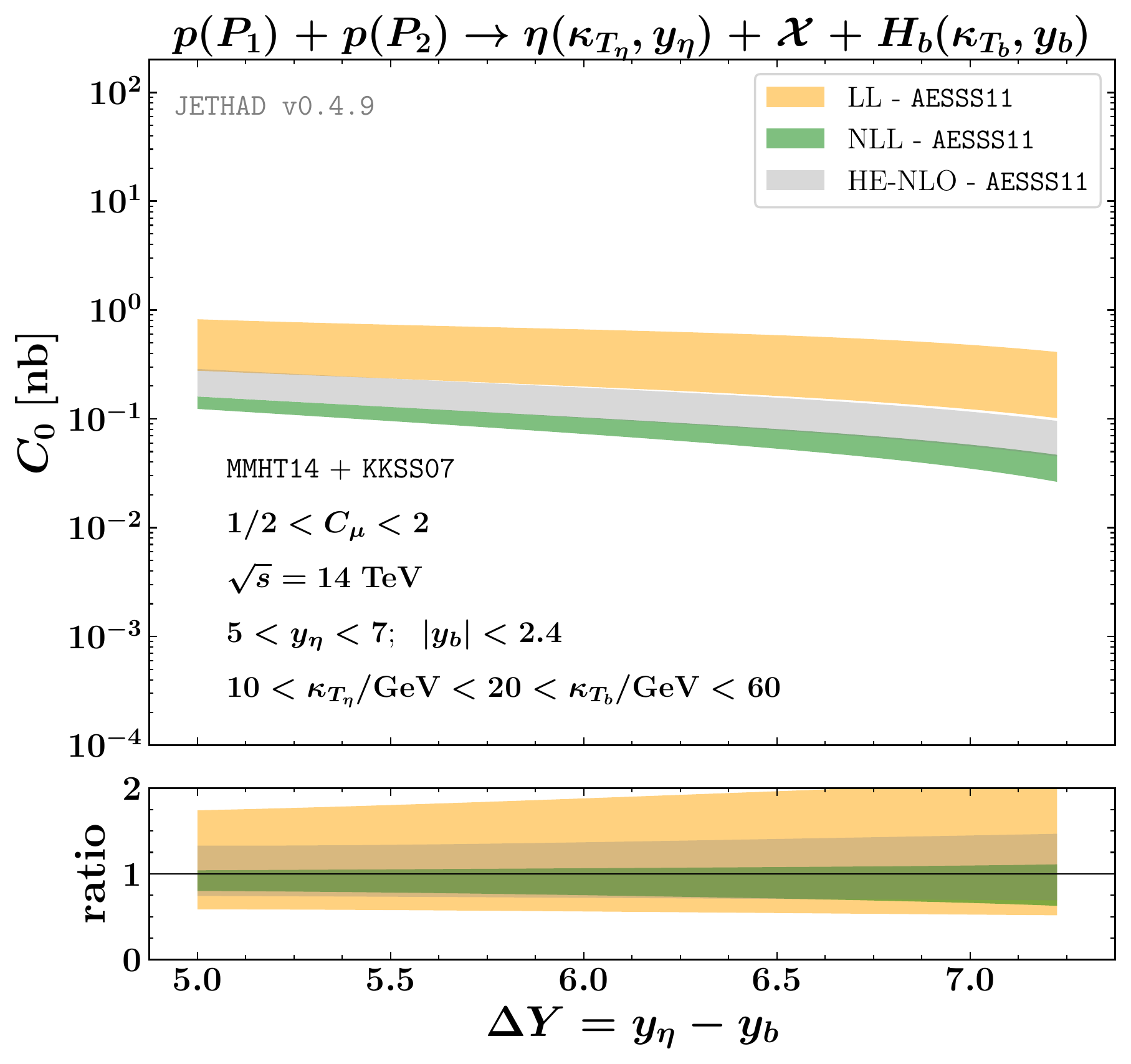}

\caption{$\DY$-distribution for the inclusive $\eta$-meson~$+$~heavy-flavor production at the LHC (upper panels) and at FPF~$+$~ATLAS (lower panel), for $\sqrt{s} = 14$ TeV. Text boxes inside plots show final-state kinematic ranges. Uncertainty bands encode the net effect of the scale variation and the multi-dimensional integration over the final-state phase space.
Ancillary panels below primary plots exhibit reduced cross sections, divided by themselves taken at natural scales.}
\label{fig:C0_eta}
\end{figure*}

\begin{figure*}[t]

   \includegraphics[scale=0.48,clip]{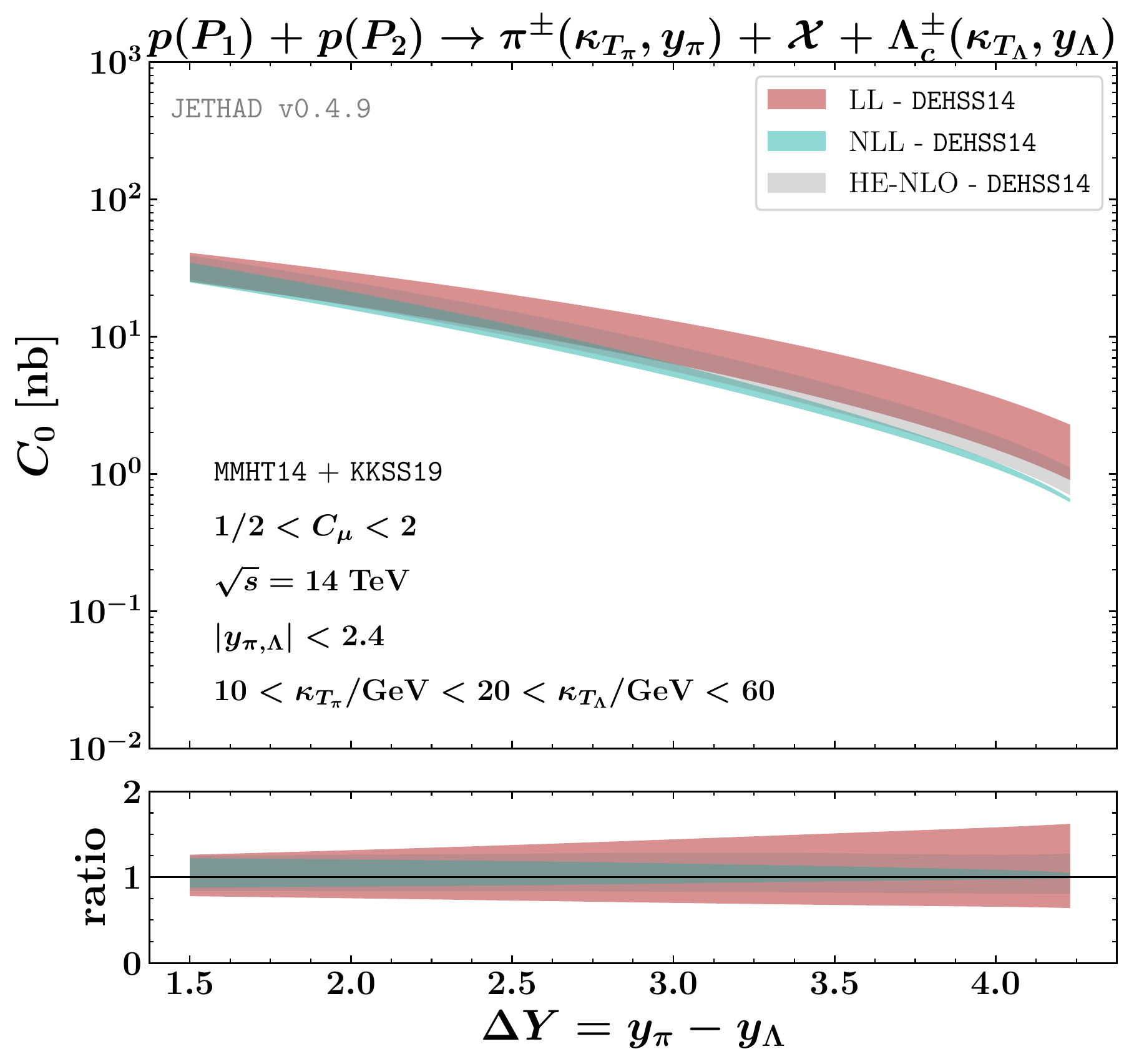}
   \includegraphics[scale=0.48,clip]{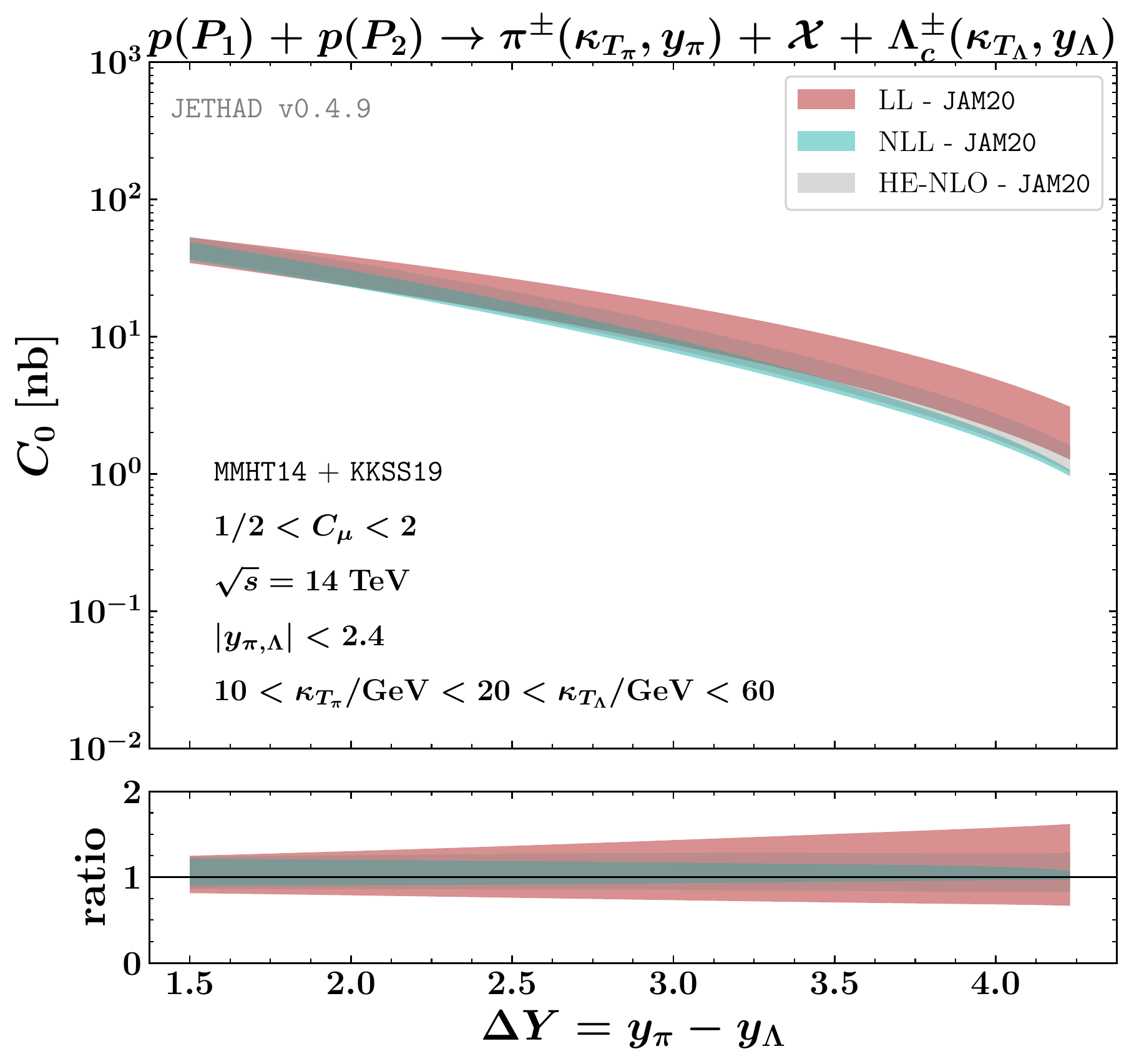}

   \includegraphics[scale=0.48,clip]{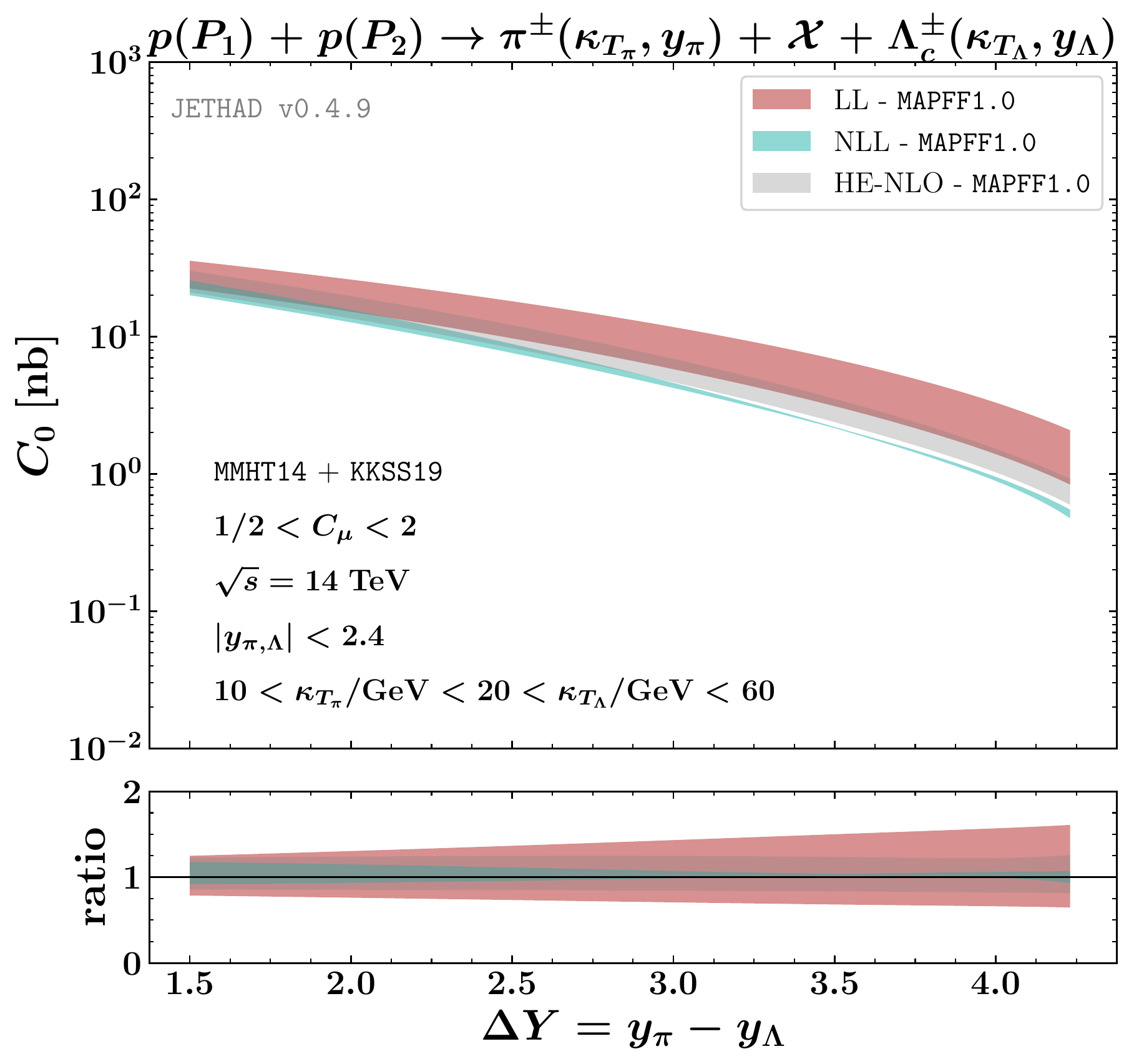}
   \includegraphics[scale=0.48,clip]{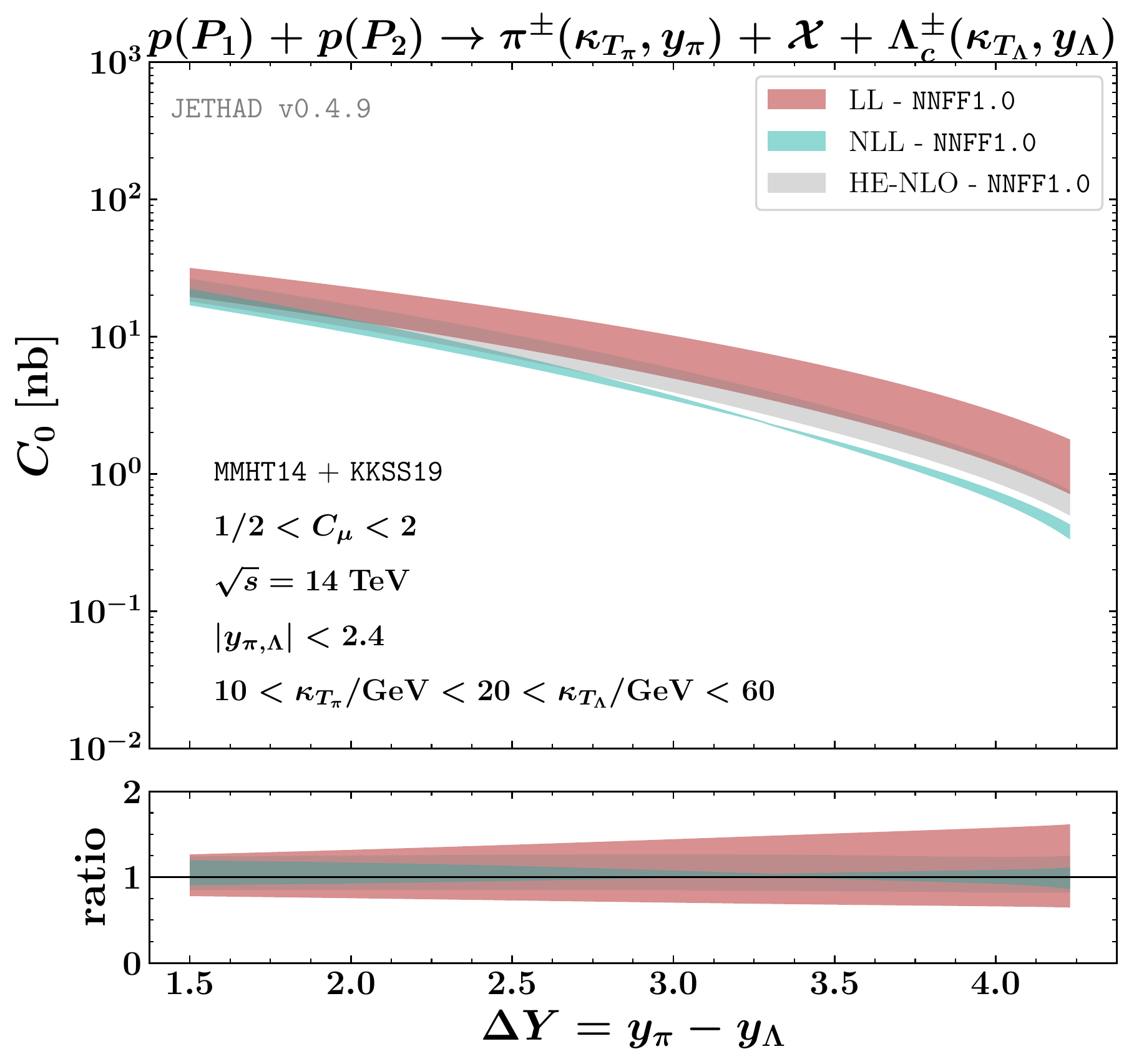}

\caption{$\DY$-distribution for the inclusive $\pi^\pm$~$+$~$\Lambda_c^\pm$ production at the LHC for four different choices of the pion collinear FF set (see Section\tref{ssec:PDFs_FFs}), and for $\sqrt{s} = 14$ TeV. Text boxes inside plots show final-state kinematic ranges. Uncertainty bands encode the net effect of the scale variation and the multi-dimensional integration over the final-state phase space.
Ancillary panels below primary plots exhibit the reduced $\DY$-distribution, namely the uncertainty bands divided by the corresponding values taken at natural scales.}
\label{fig:C0_PL_LHC}
\end{figure*}

\begin{figure*}[t]

   \includegraphics[scale=0.48,clip]{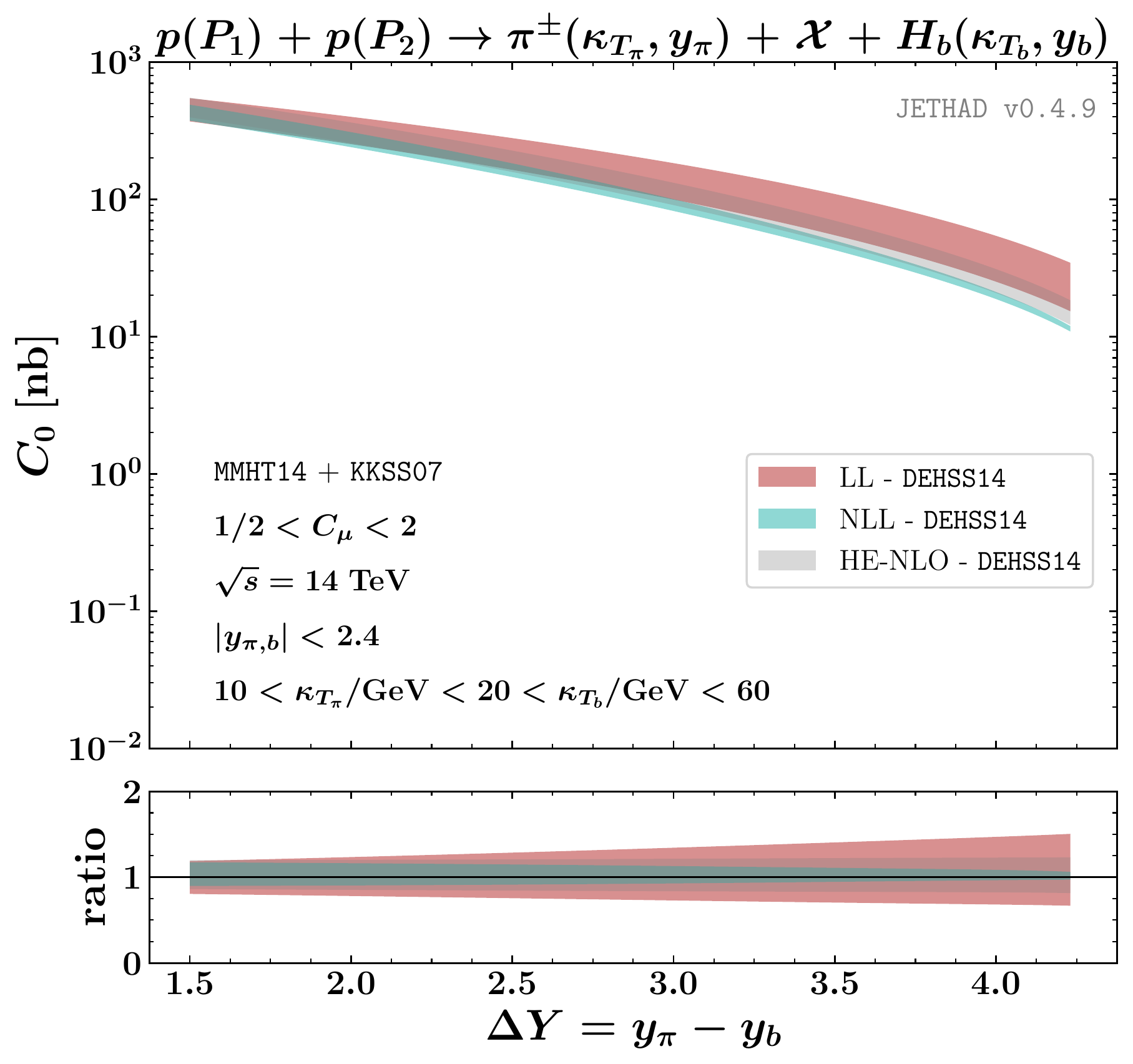}
   \includegraphics[scale=0.48,clip]{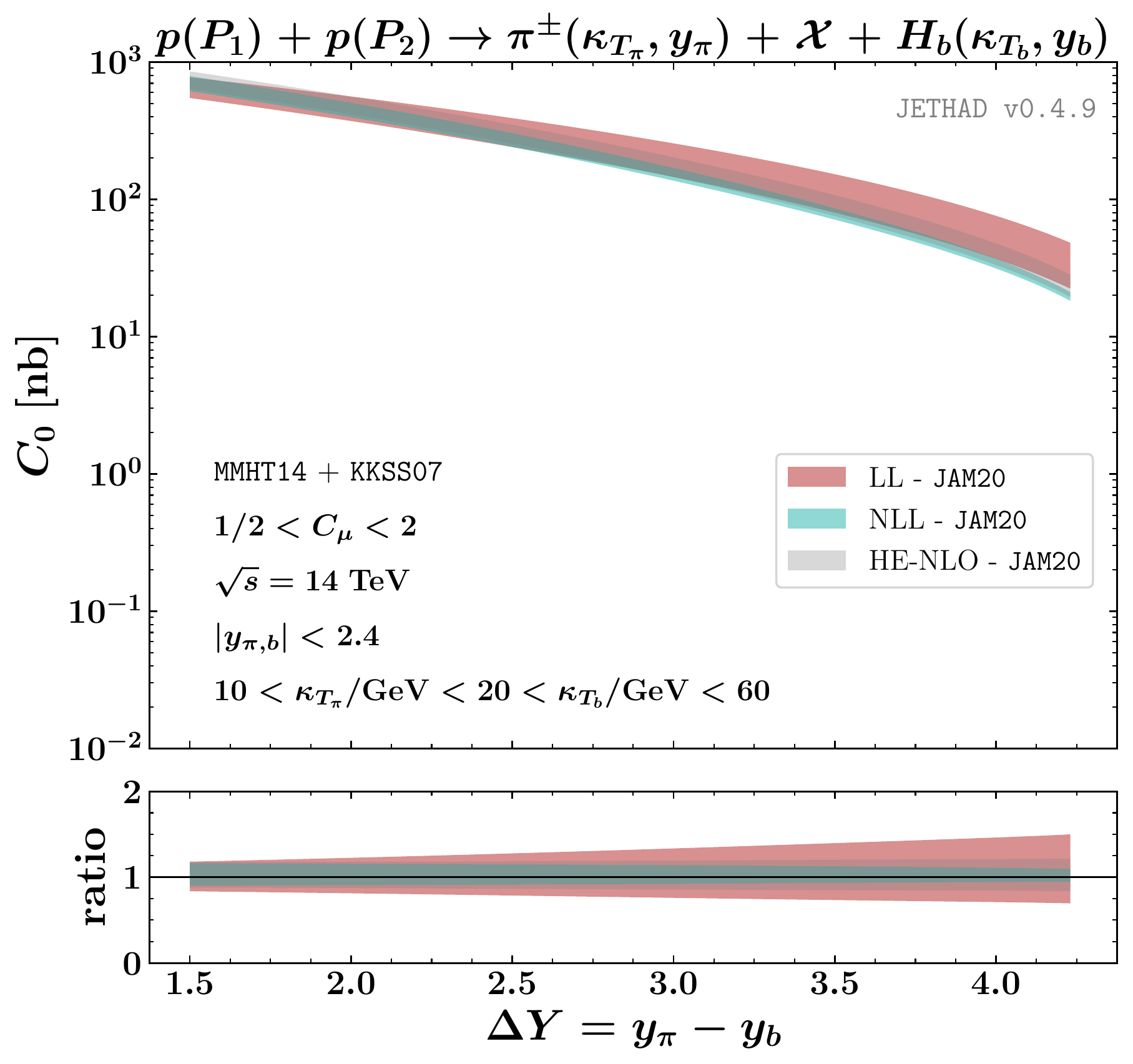}

   \includegraphics[scale=0.48,clip]{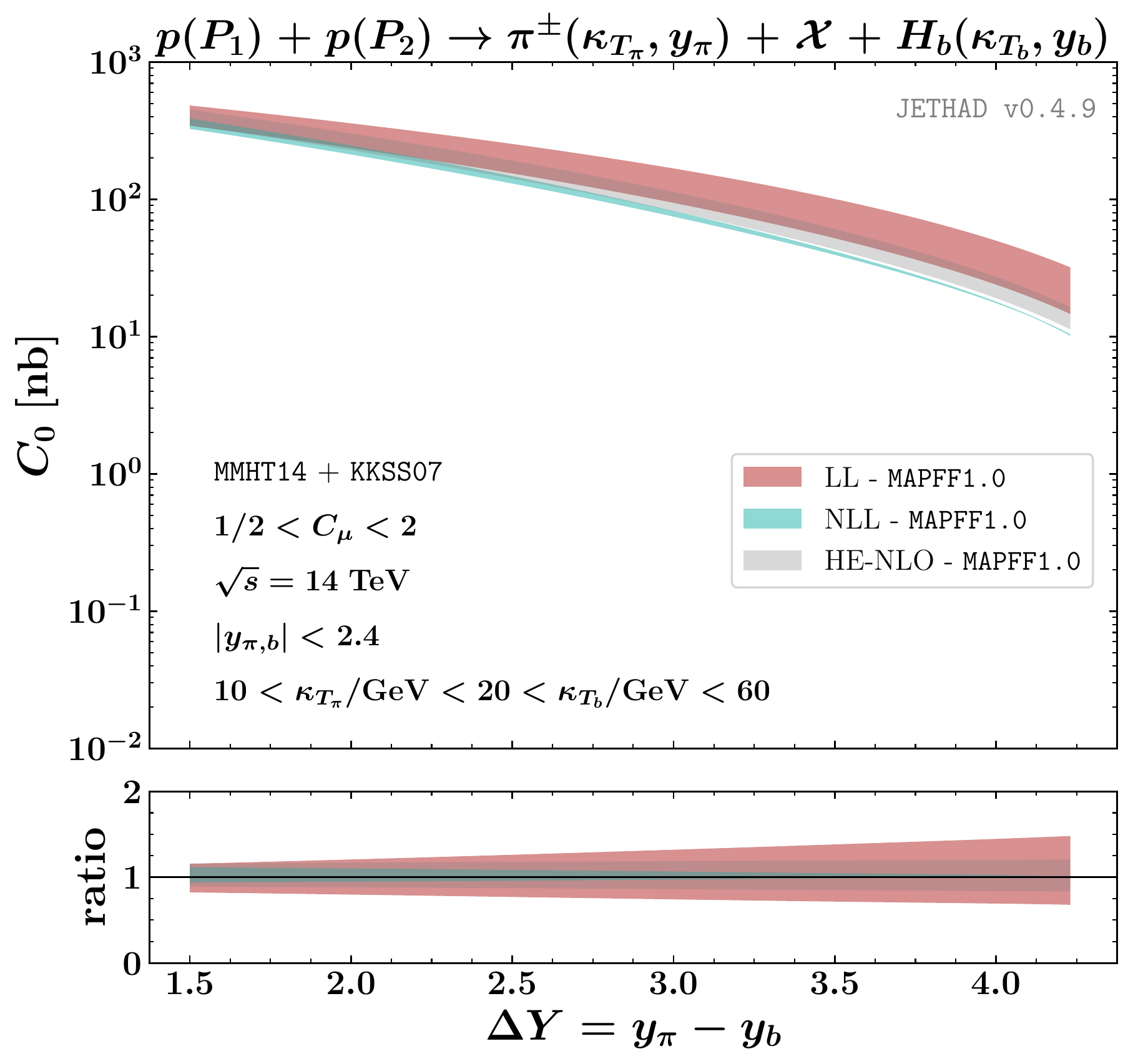}
   \includegraphics[scale=0.48,clip]{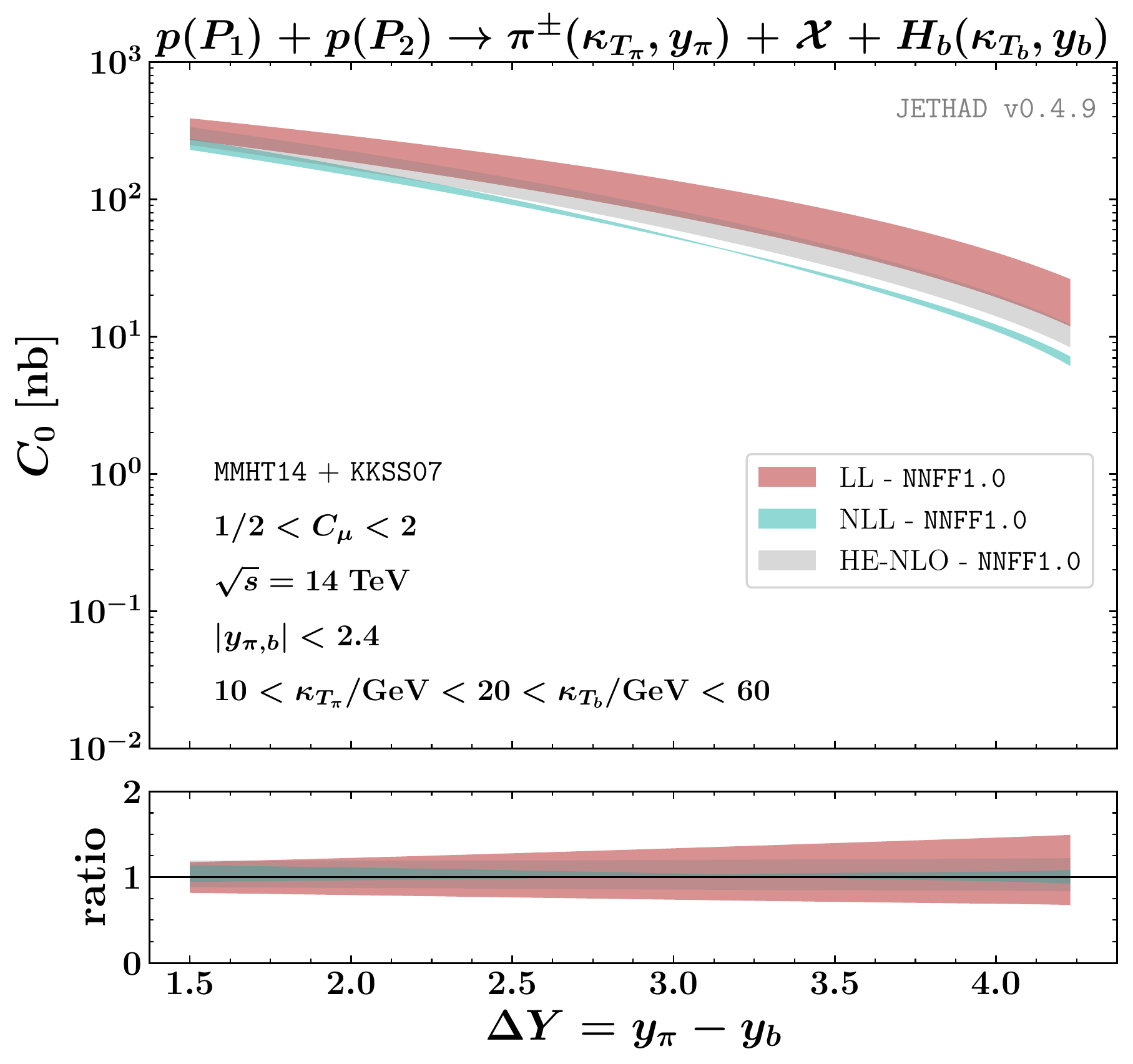}

\caption{$\DY$-distribution for the inclusive $\pi^\pm$~$+$~$b$~hadron production at the LHC for four different choices of the pion collinear FF set (see Section\tref{ssec:PDFs_FFs}), and for $\sqrt{s} = 14$ TeV. Text boxes inside plots show final-state kinematic ranges. Uncertainty bands encode the net effect of the scale variation and the multi-dimensional integration over the final-state phase space.
Ancillary panels below primary plots exhibit the reduced $\DY$-distribution, namely the uncertainty bands divided by the corresponding values taken at natural scales.}
\label{fig:C0_Pb_LHC}
\end{figure*}

\begin{figure*}[t]

   \includegraphics[scale=0.48,clip]{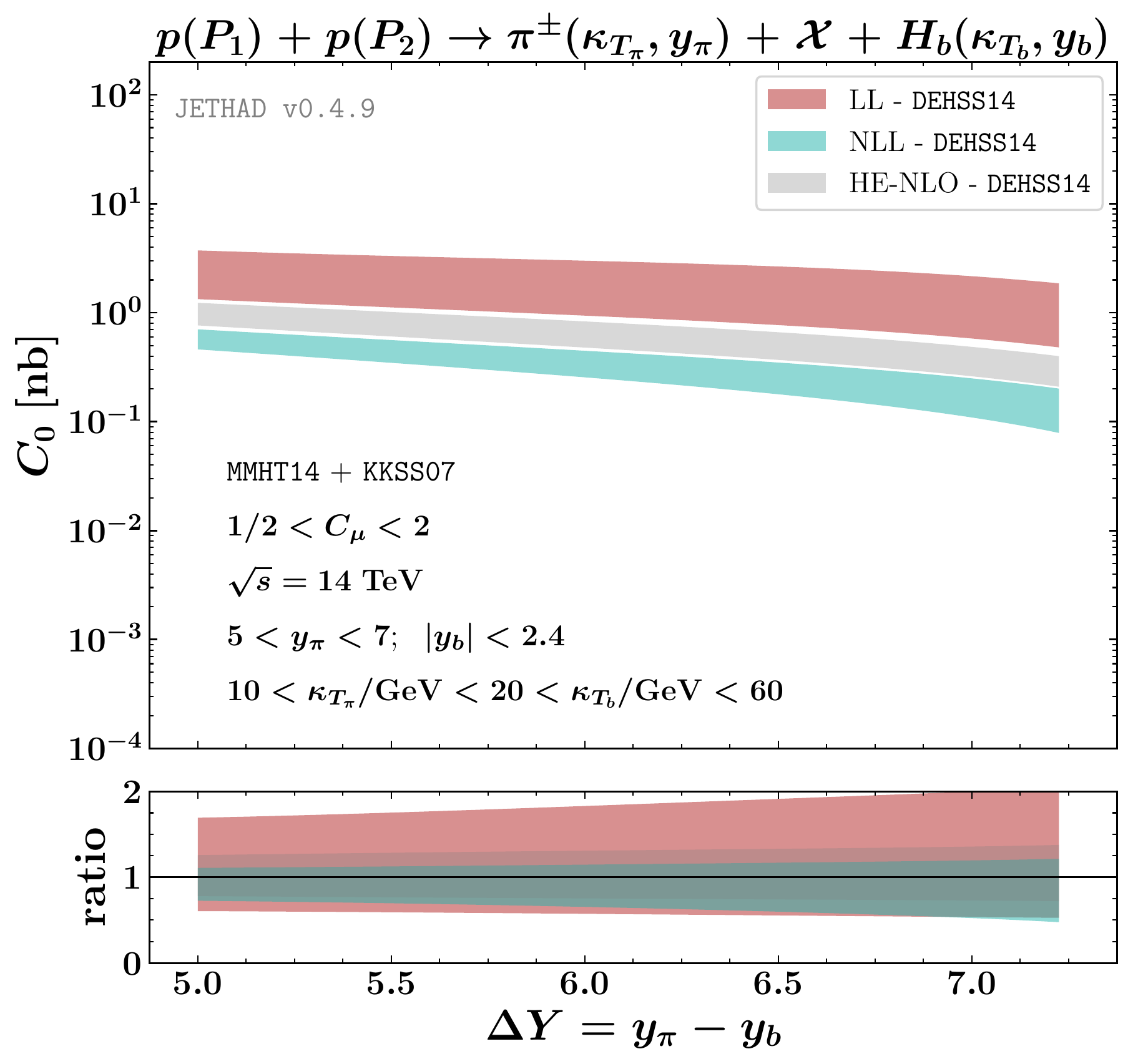}
   \includegraphics[scale=0.48,clip]{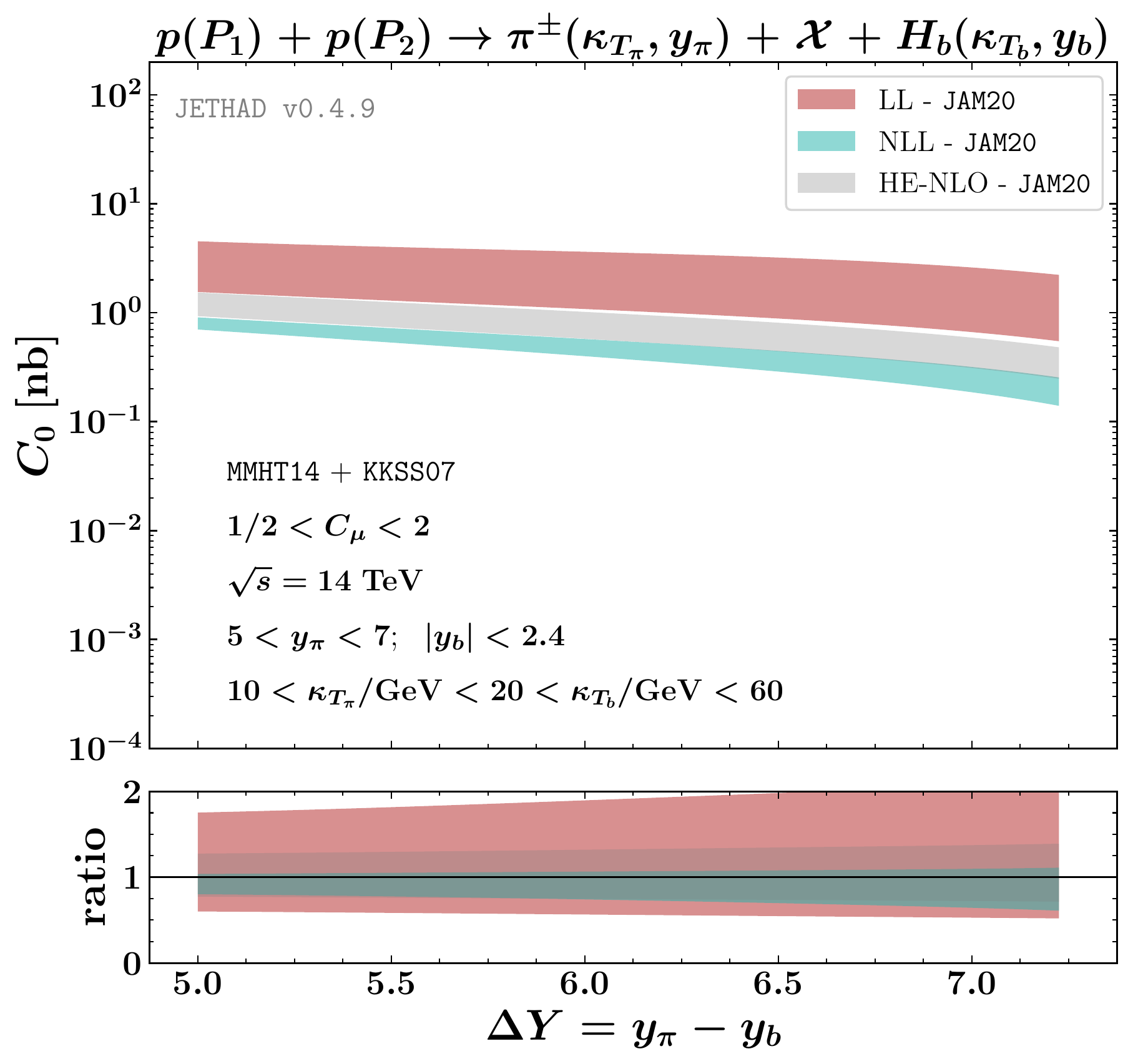}

   \includegraphics[scale=0.48,clip]{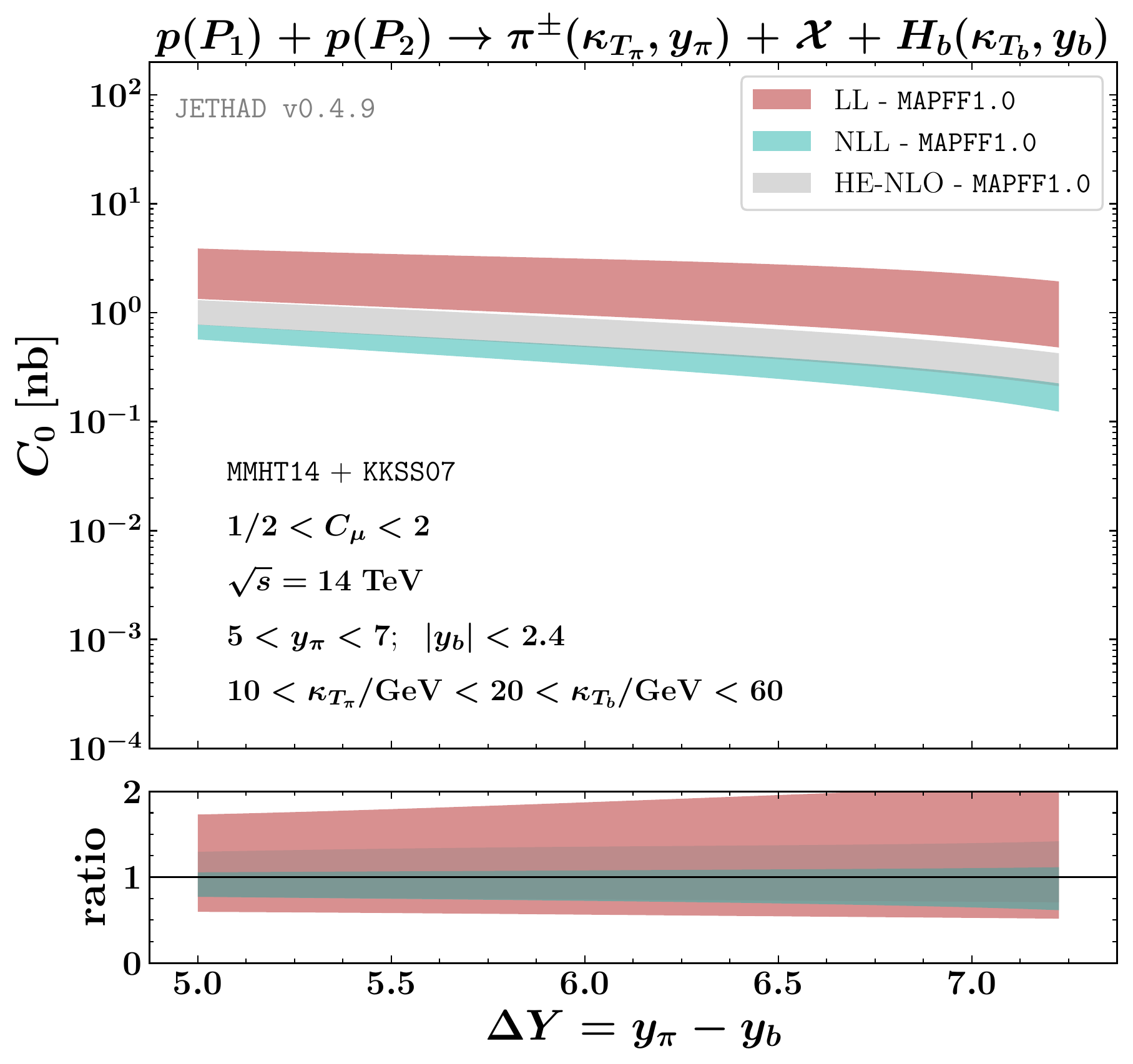}
   \includegraphics[scale=0.48,clip]{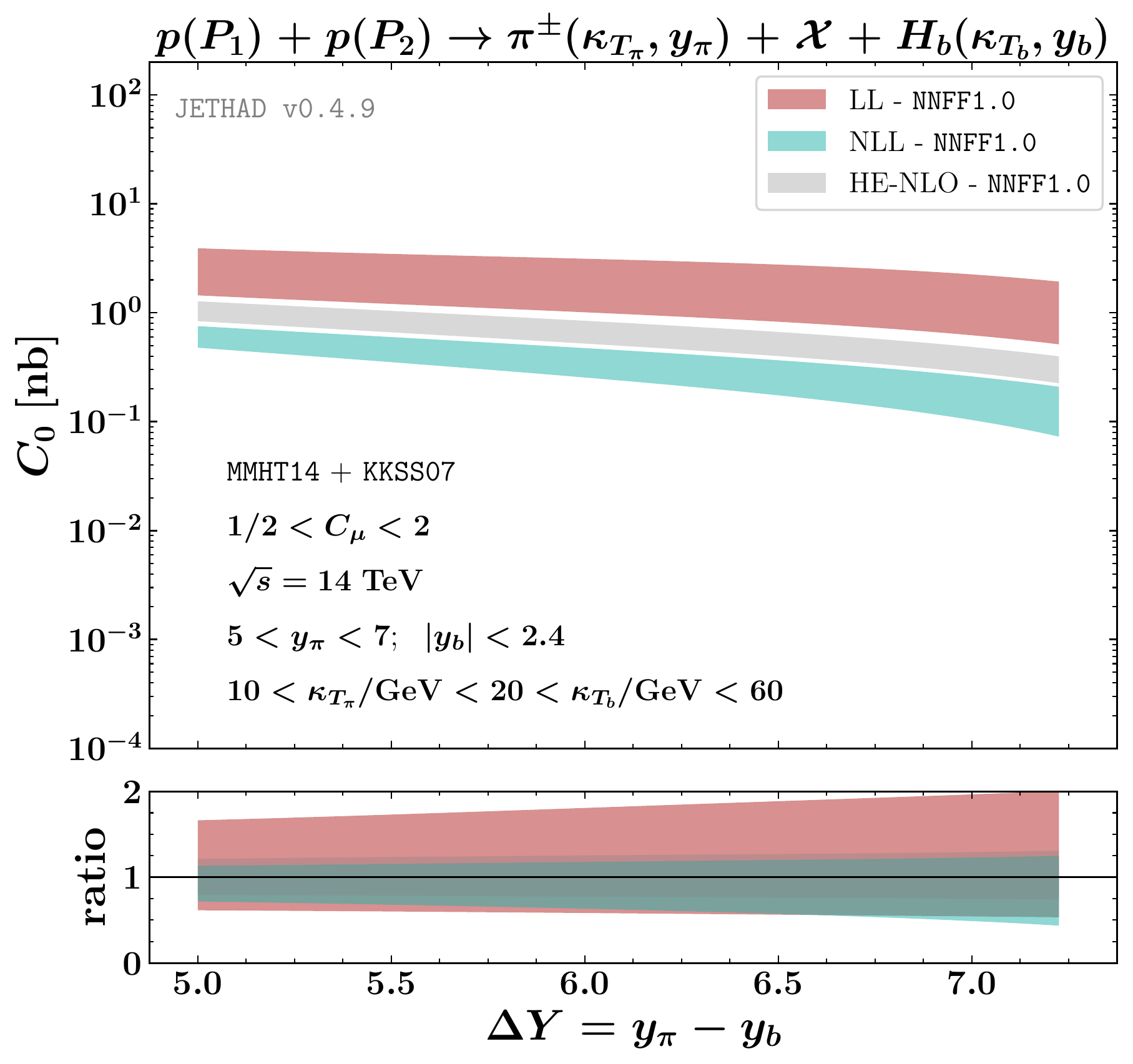}

\caption{$\DY$-distribution for the inclusive $\pi^\pm$~$+$~$b$~hadron production at FPF~$+$~ATLAS for four different choices of the pion collinear FF set (see Section\tref{ssec:PDFs_FFs}), and for $\sqrt{s} = 14$ TeV. Text boxes inside plots show final-state kinematic ranges. Uncertainty bands encode the net effect of the scale variation and the multi-dimensional integration over the final-state phase space.
Ancillary panels below primary plots exhibit the reduced $\DY$-distribution, namely the uncertainty bands divided by the corresponding values taken at natural scales.}
\label{fig:C0_Pb_FPF}
\end{figure*}

\begin{figure*}[t]

   \includegraphics[scale=0.48,clip]{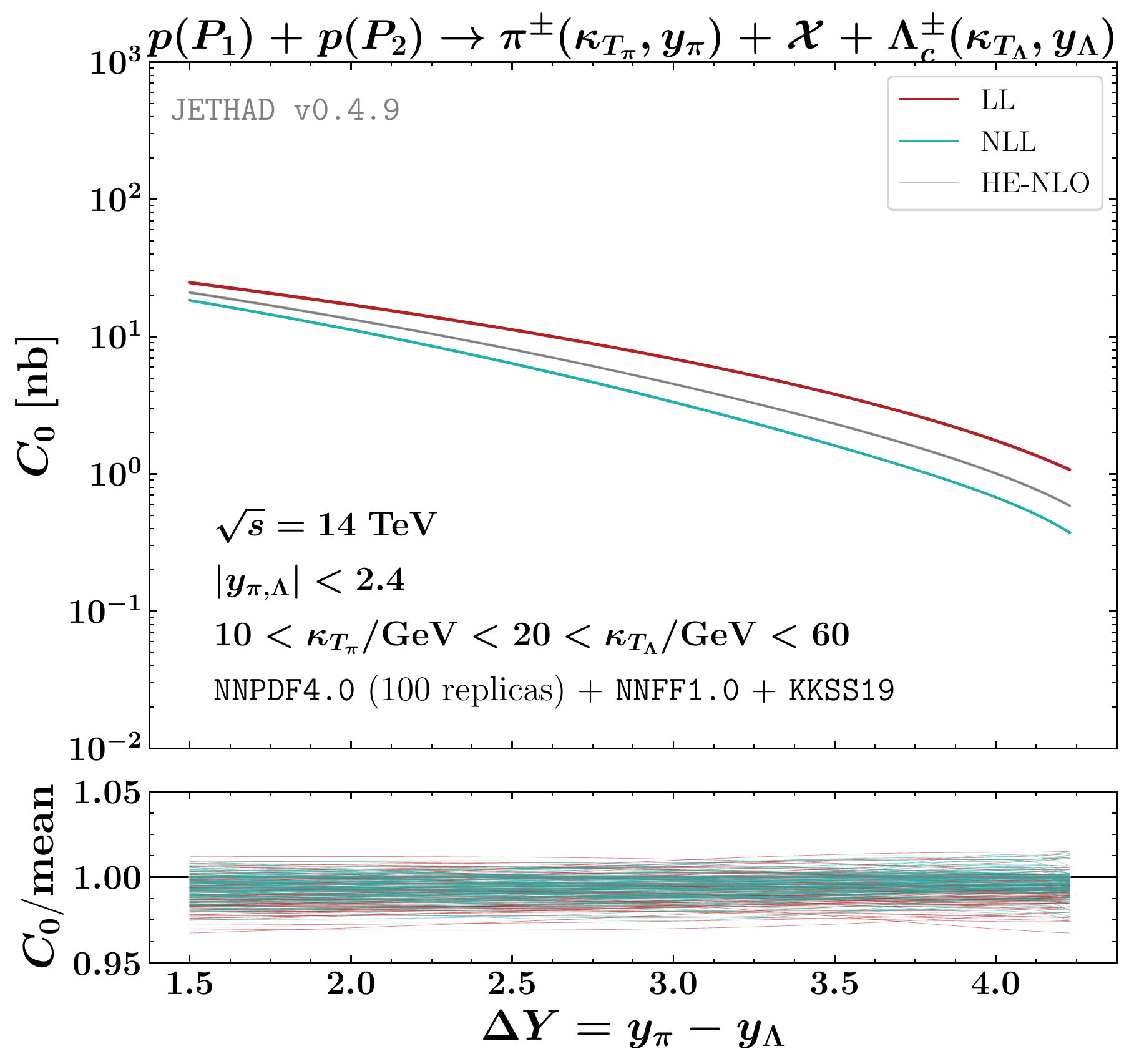}
   \includegraphics[scale=0.48,clip]{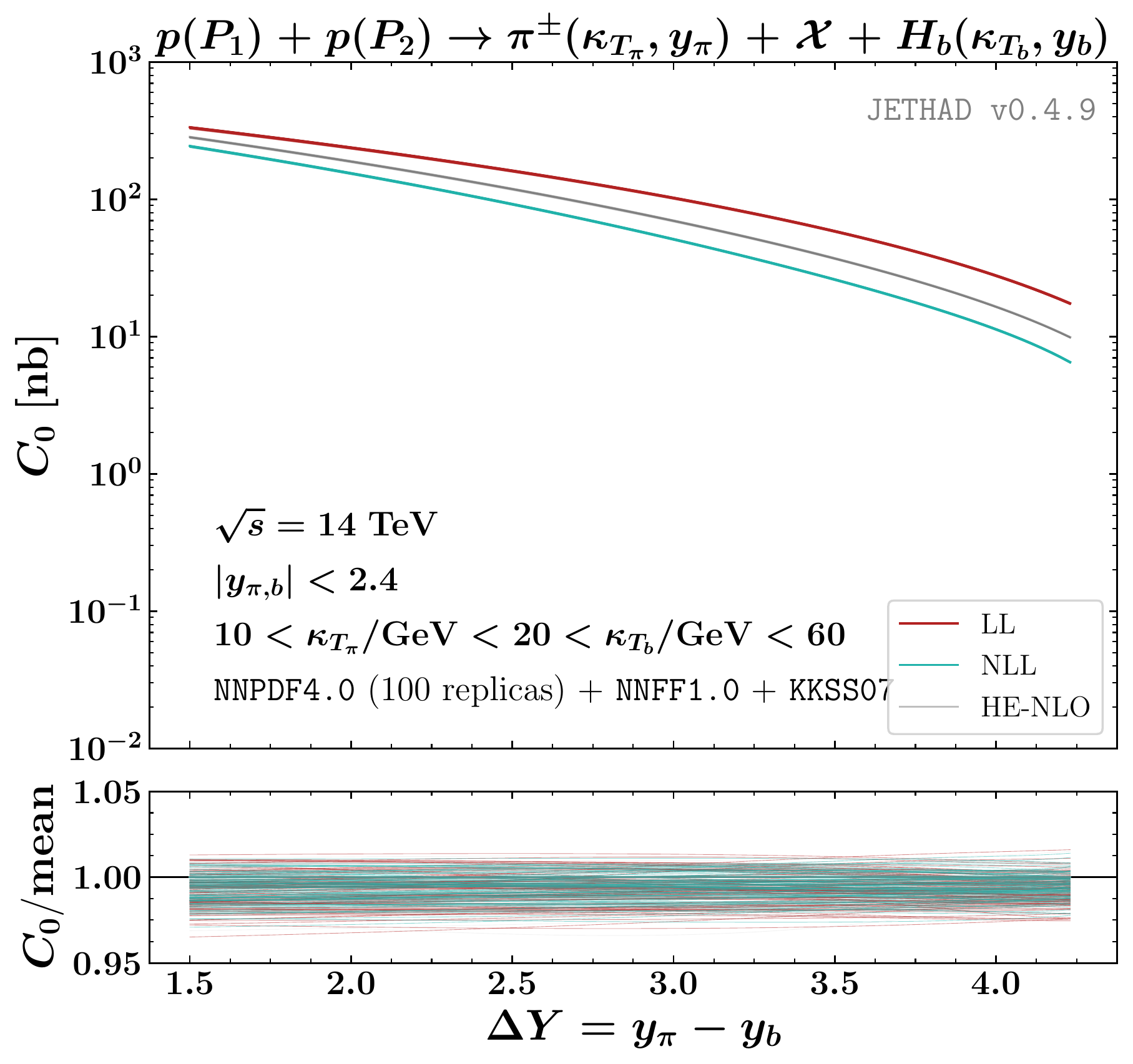}
   \includegraphics[scale=0.48,clip]{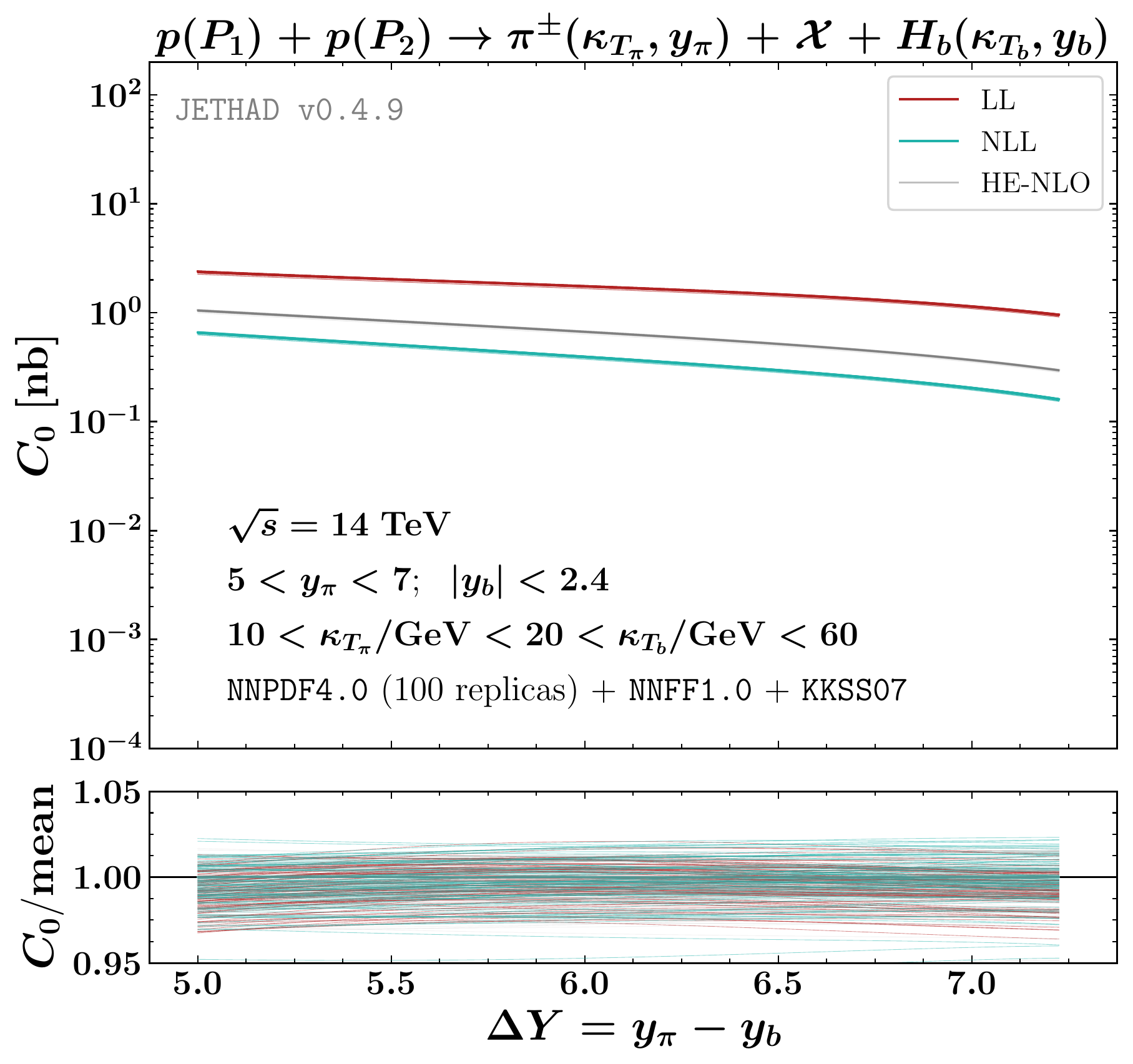}

\caption{$\DY$-distribution for the inclusive $\pi^\pm$~$+$~heavy-flavor production at the LHC (upper panels) and at FPF~$+$~ATLAS (lower panel), for $\sqrt{s} = 14$ TeV. Text boxes inside main plots show final-state kinematic ranges.
{\tt NNPDF4.0} proton collinear PDFs are used together with {\tt NNFF1.0} pion collinear FFs.
The envelope of main results is built in terms of a replica-driven study on {\tt NNPDF4.0} PDFs, while {\tt NNFF1.0} pion FFs are kept at their central value.
Ancillary panels below primary plots exhibit the reduced $\DY$-distribution, namely the envelope of replicas’ predictions divided by the mean value.}
\label{fig:C0_pion_PDF_rep}
\end{figure*}

\begin{figure*}[t]

   \includegraphics[scale=0.48,clip]{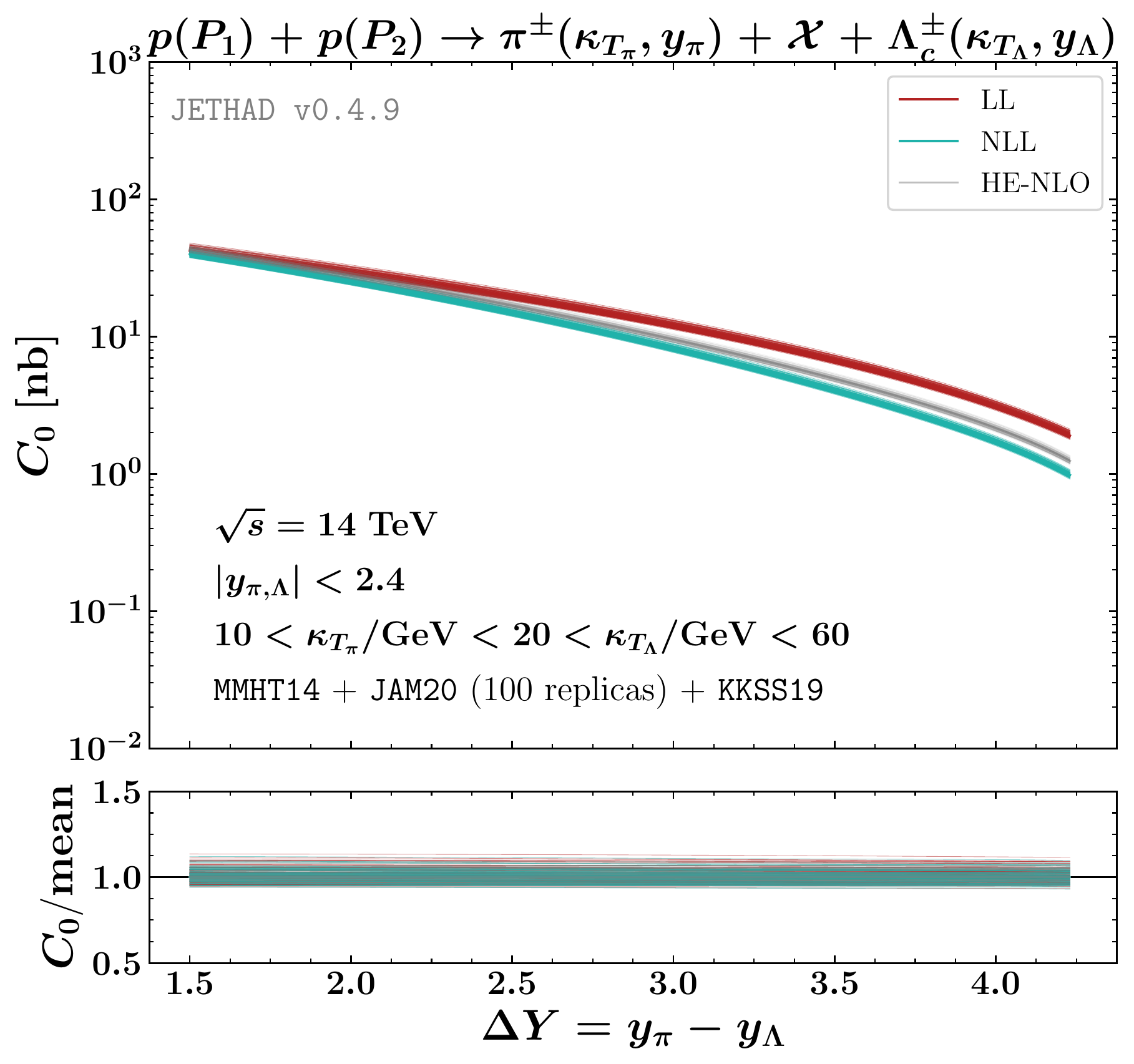}
   \includegraphics[scale=0.48,clip]{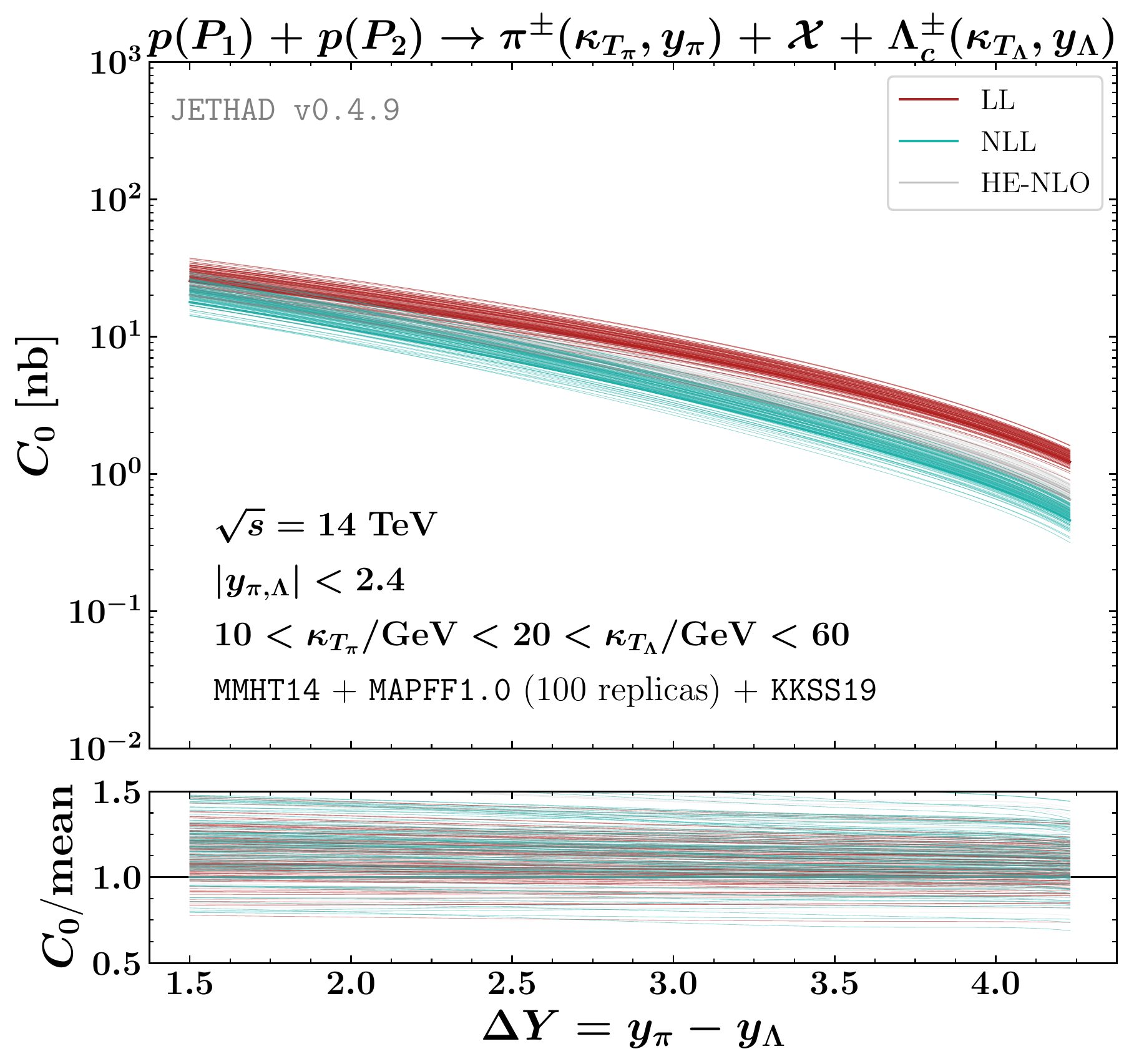}
   \includegraphics[scale=0.48,clip]{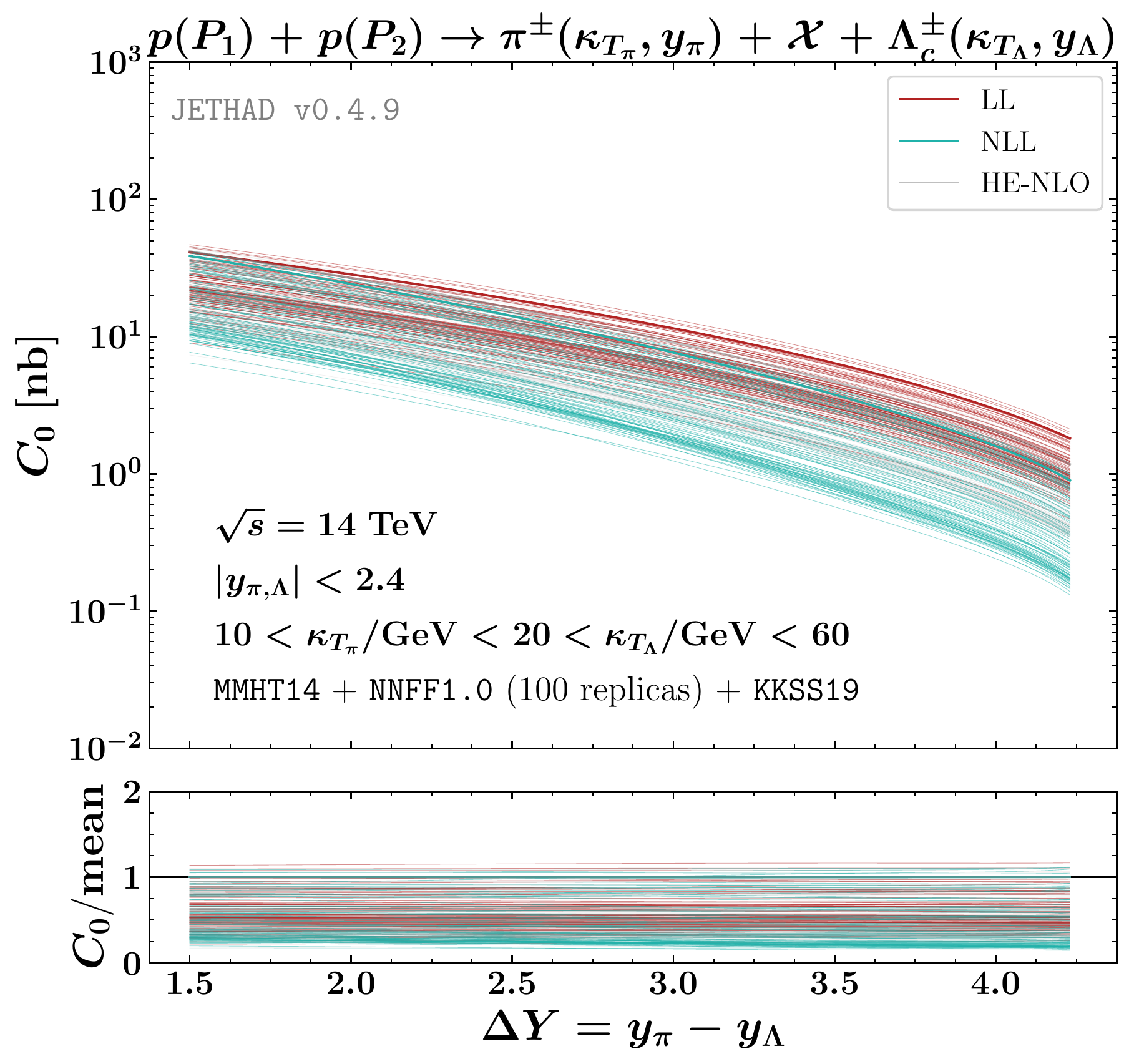}

\caption{$\DY$-distribution for the inclusive $\pi^\pm$~$+$~$\Lambda_c^\pm$ production at the LHC for three different choices of the pion collinear FF set, and for $\sqrt{s} = 14$ TeV. Text boxes inside main plots show final-state kinematic ranges. 
The envelope of main results is built in terms of a replica-driven study on pion collinear FFs.
Ancillary panels below primary plots exhibit the reduced $\DY$-distribution, namely the envelope of replicas’ predictions divided by the mean value.}
\label{fig:C0_PL_LHC_rep}
\end{figure*}

\begin{figure*}[t]

   \includegraphics[scale=0.48,clip]{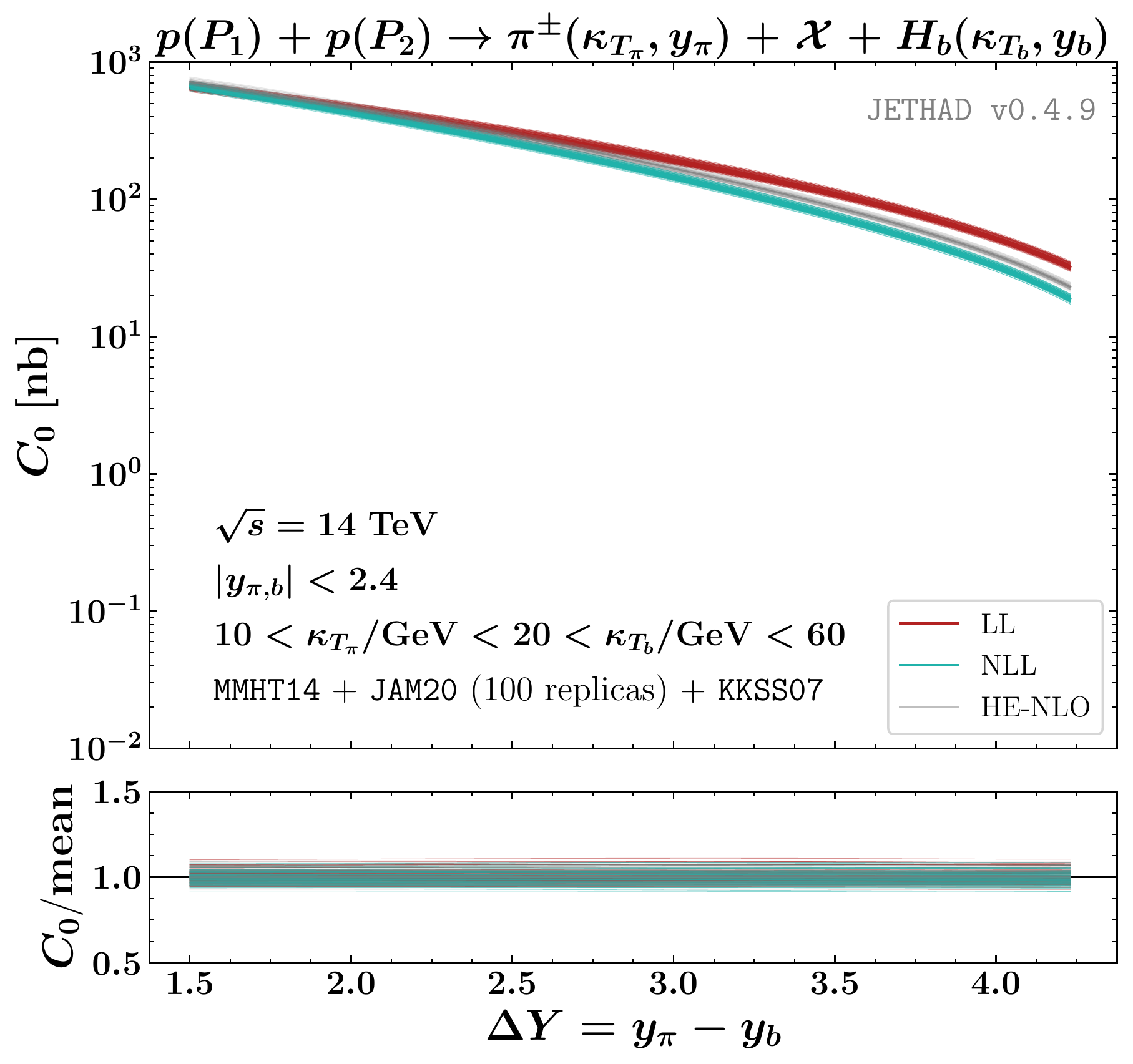}
   \includegraphics[scale=0.48,clip]{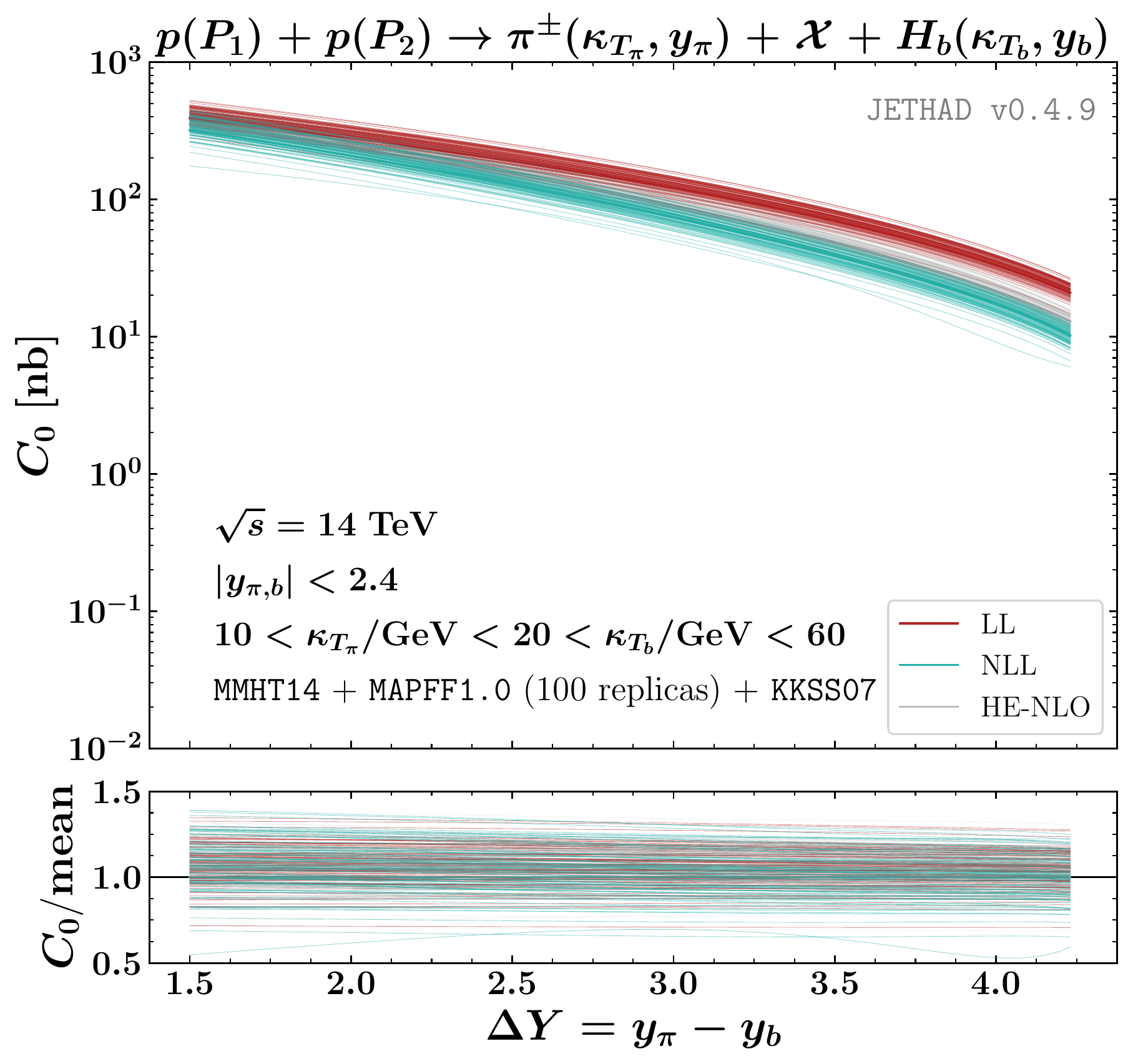}
   \includegraphics[scale=0.48,clip]{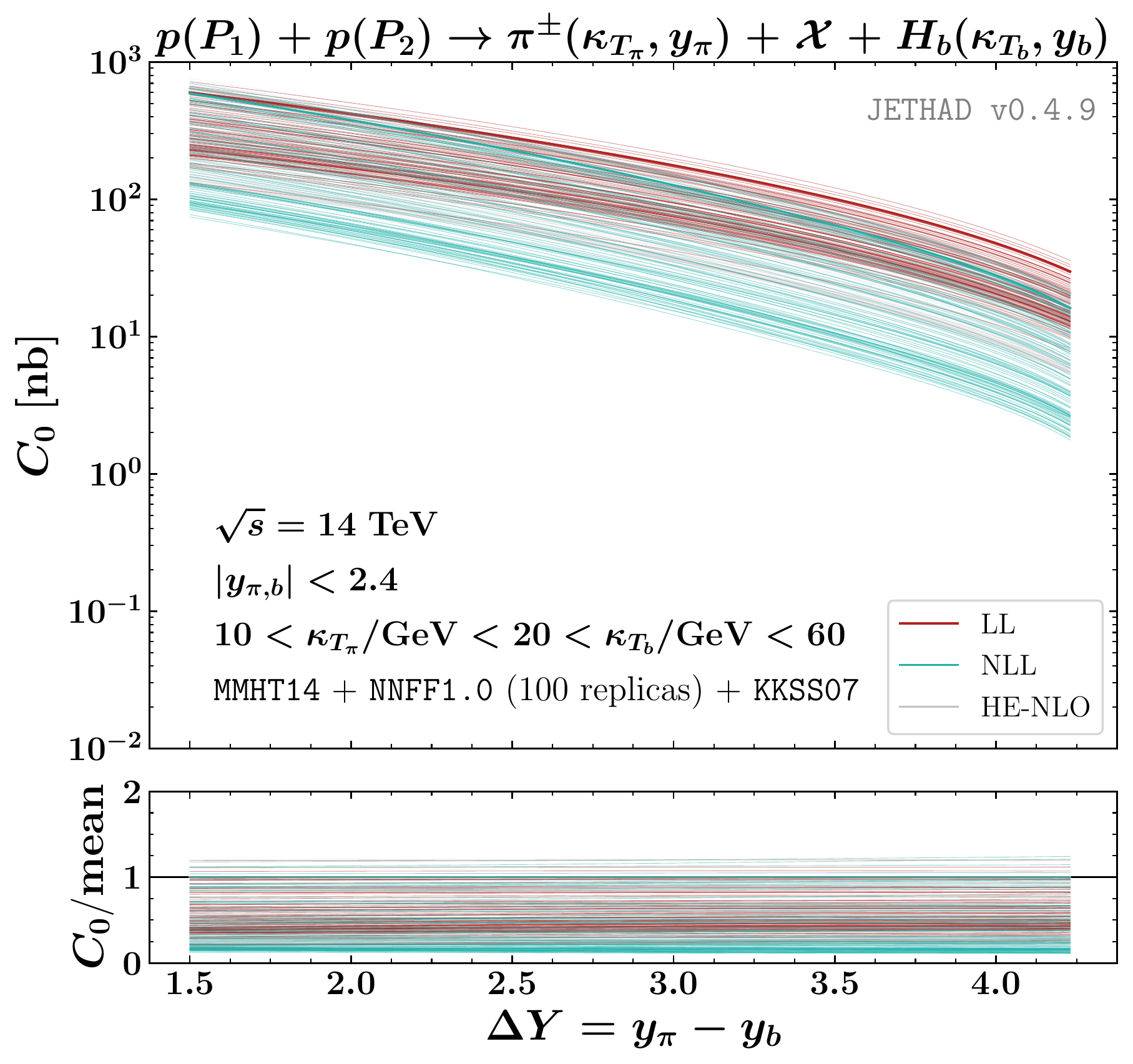}

\caption{$\DY$-distribution for the inclusive $\pi^\pm$~$+$~$b$~hadron production at the LHC for three different choices of the pion collinear FF set, and for $\sqrt{s} = 14$ TeV. Text boxes inside main plots show final-state kinematic ranges. 
The envelope of main results is built in terms of a replica-driven study on pion collinear FFs.
Ancillary panels below primary plots exhibit the reduced $\DY$-distribution, namely the envelope of replicas’ predictions divided by the mean value.}
\label{fig:C0_Pb_LHC_rep}
\end{figure*}

\begin{figure*}[t]

   \includegraphics[scale=0.48,clip]{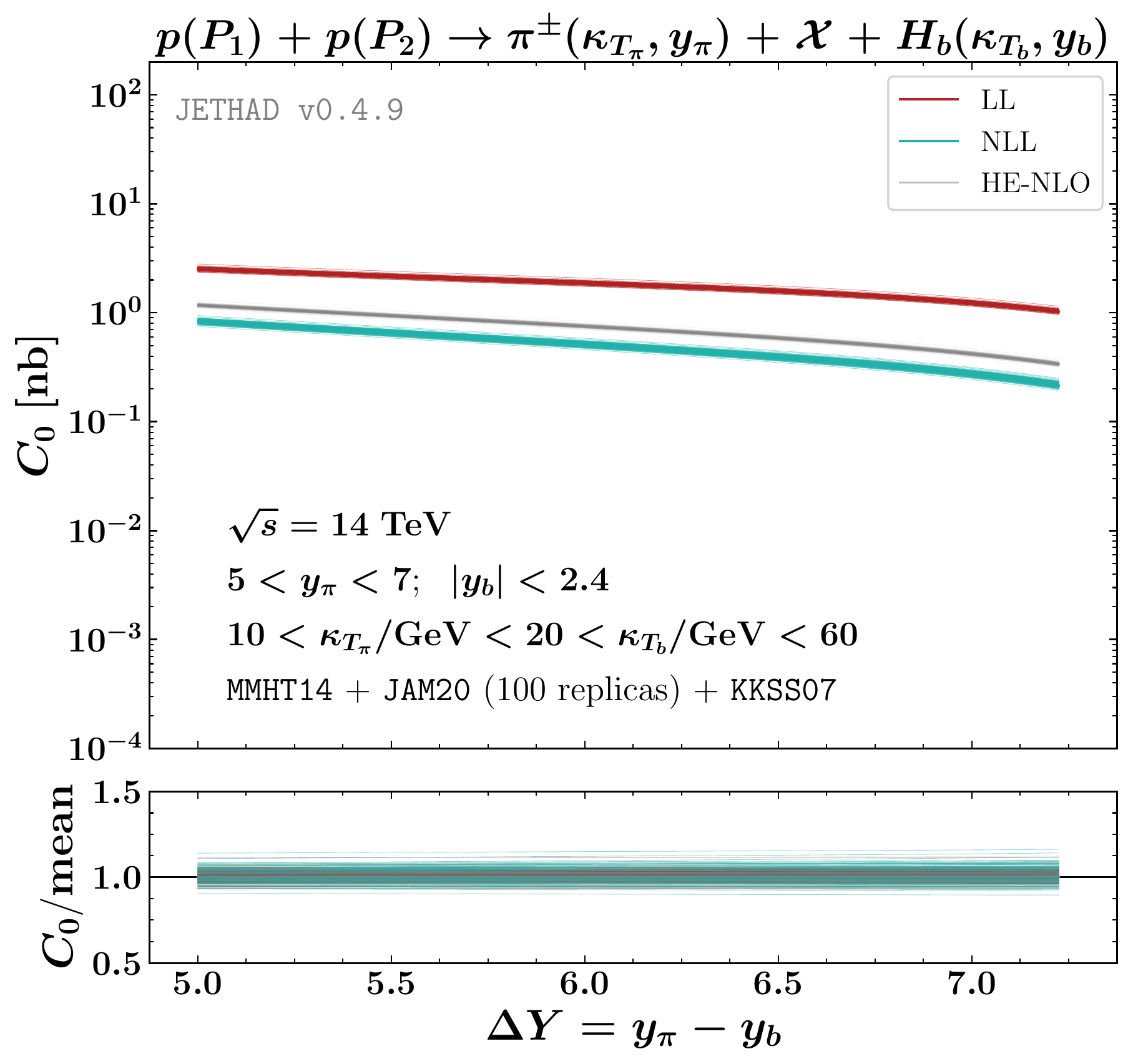}
   \includegraphics[scale=0.48,clip]{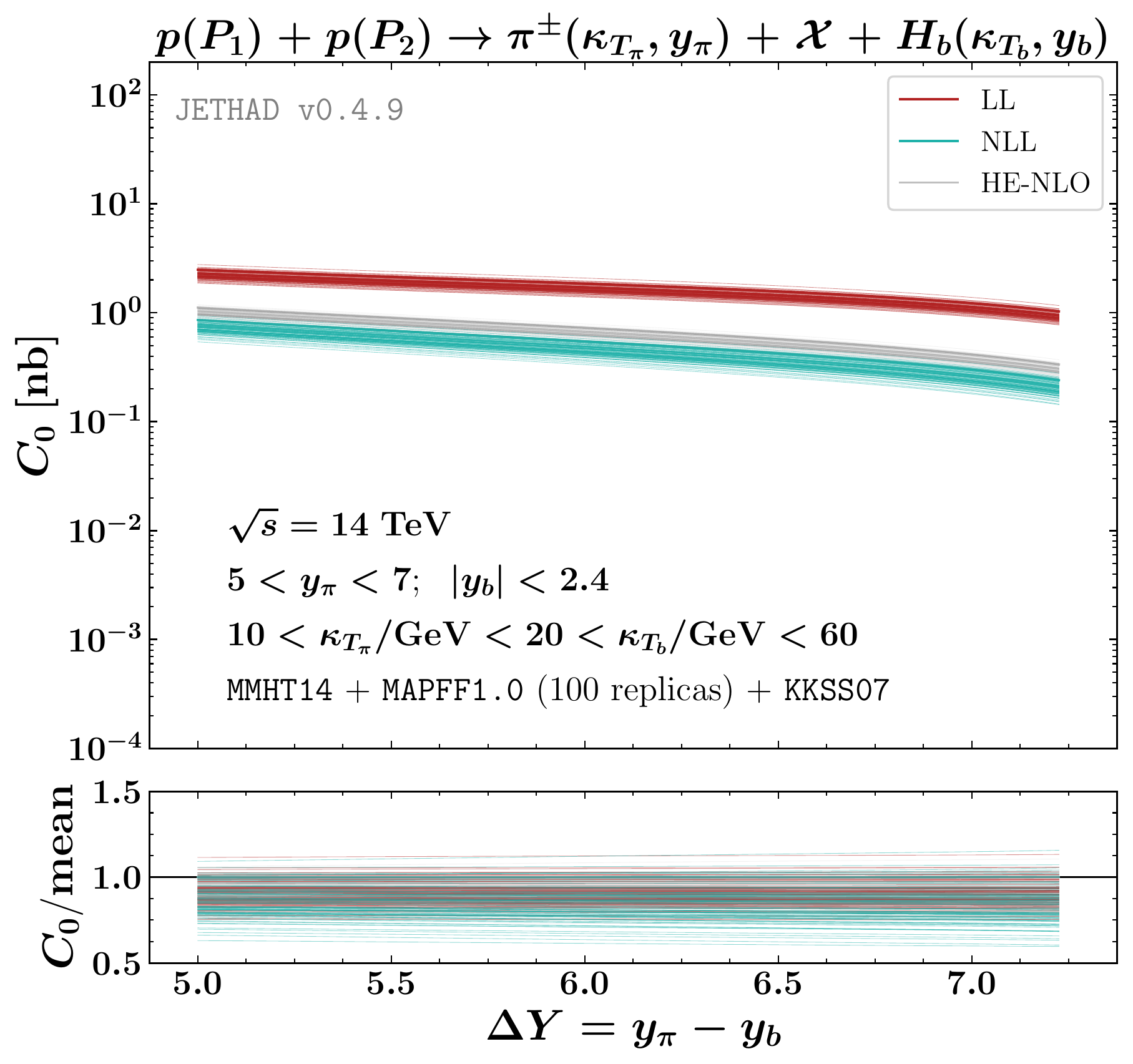}
   \includegraphics[scale=0.48,clip]{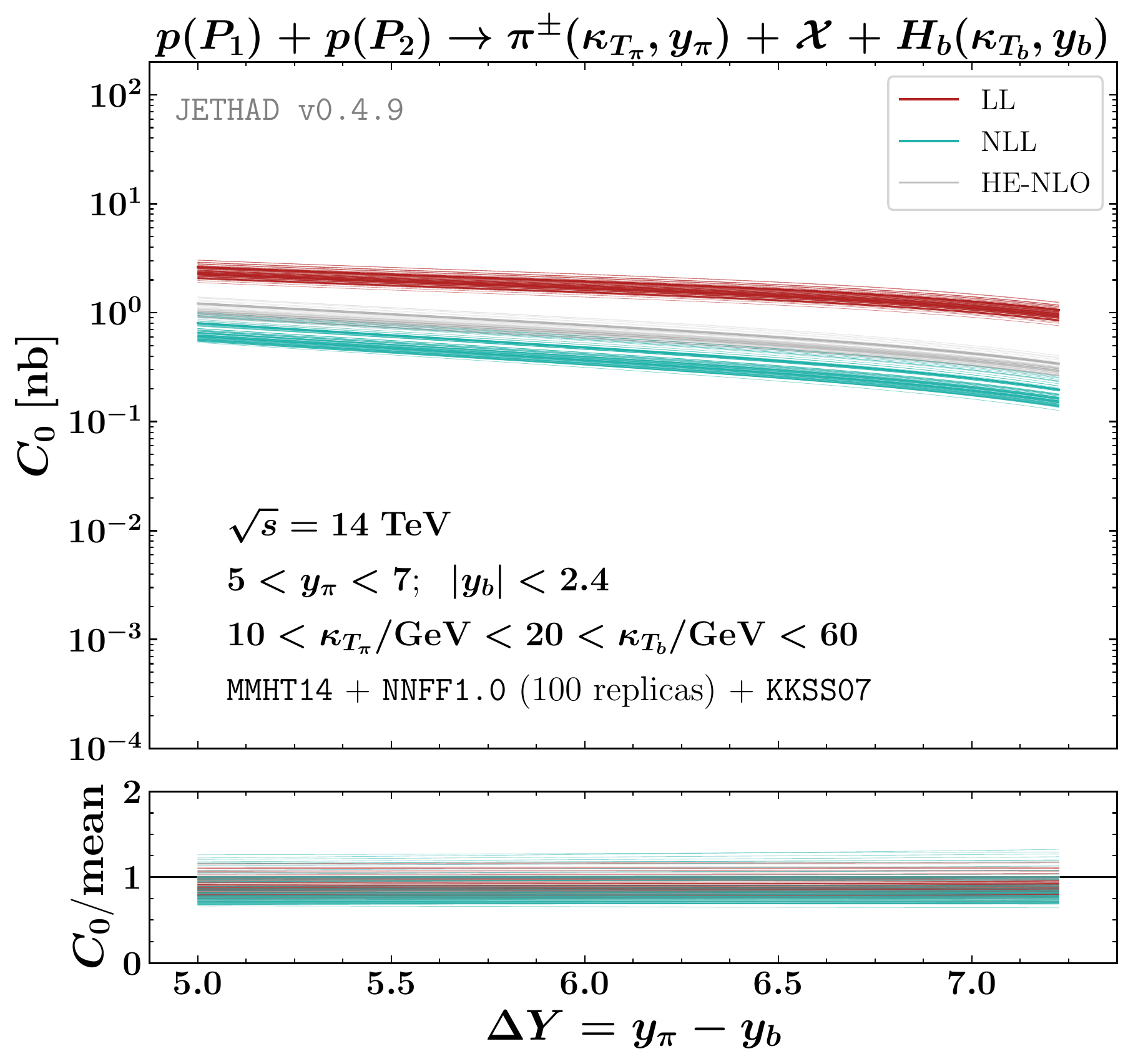}

\caption{$\DY$-distribution for the inclusive $\pi^\pm$~$+$~$b$~hadron production at the FPF~$+$~ATLAS for three different choices of the pion collinear FF set, and for $\sqrt{s} = 14$ TeV. Text boxes inside main plots show final-state kinematic ranges.
The envelope of main results is built in terms of a replica-driven study on pion collinear FFs.
Ancillary panels below primary plots exhibit the reduced $\DY$-distribution, namely the envelope of replicas’ predictions divided by the mean value.}
\label{fig:C0_Pb_FPF_rep}
\end{figure*}

\begin{figure*}[t]

   \includegraphics[scale=0.53,clip]{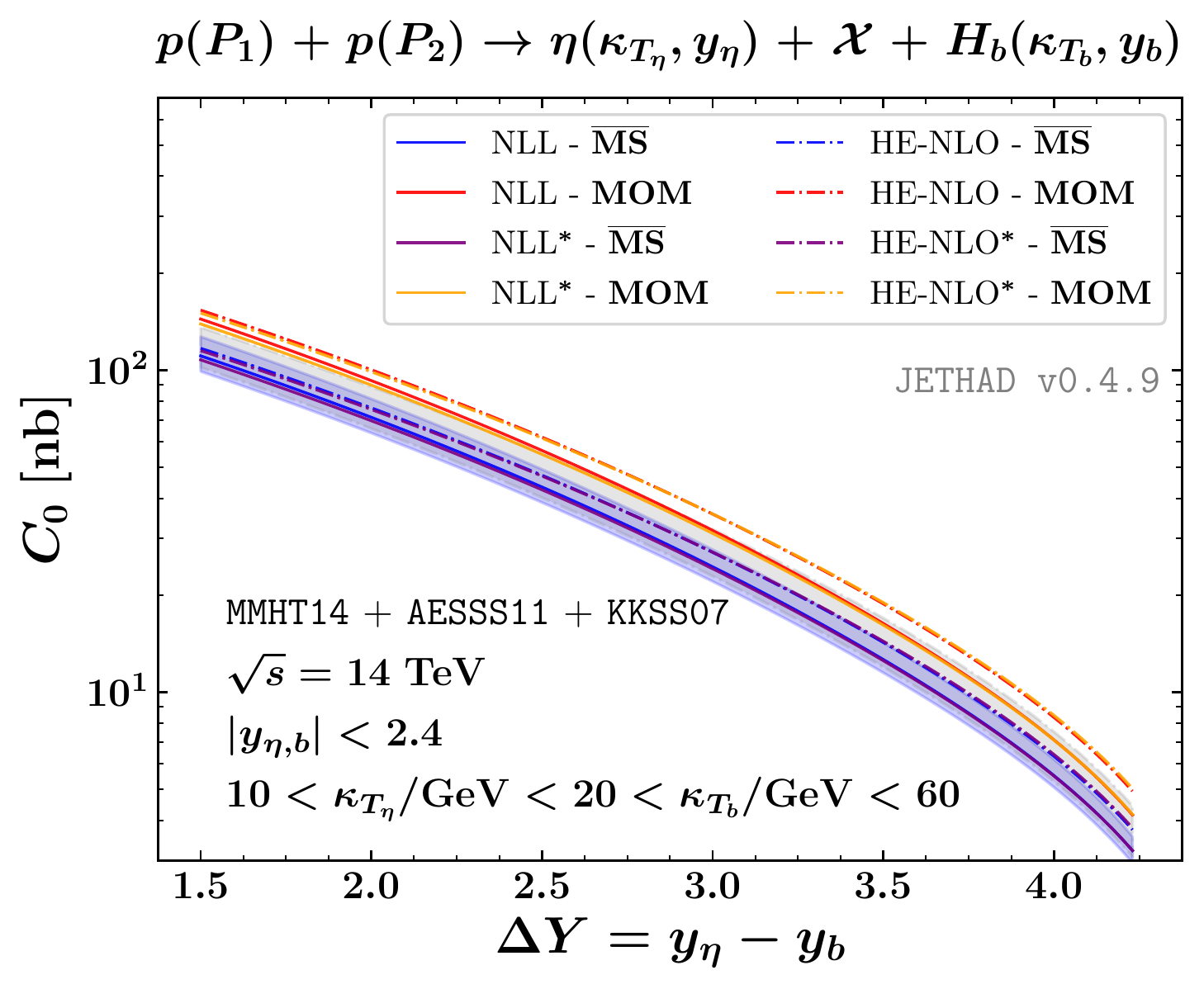}
   \includegraphics[scale=0.53,clip]{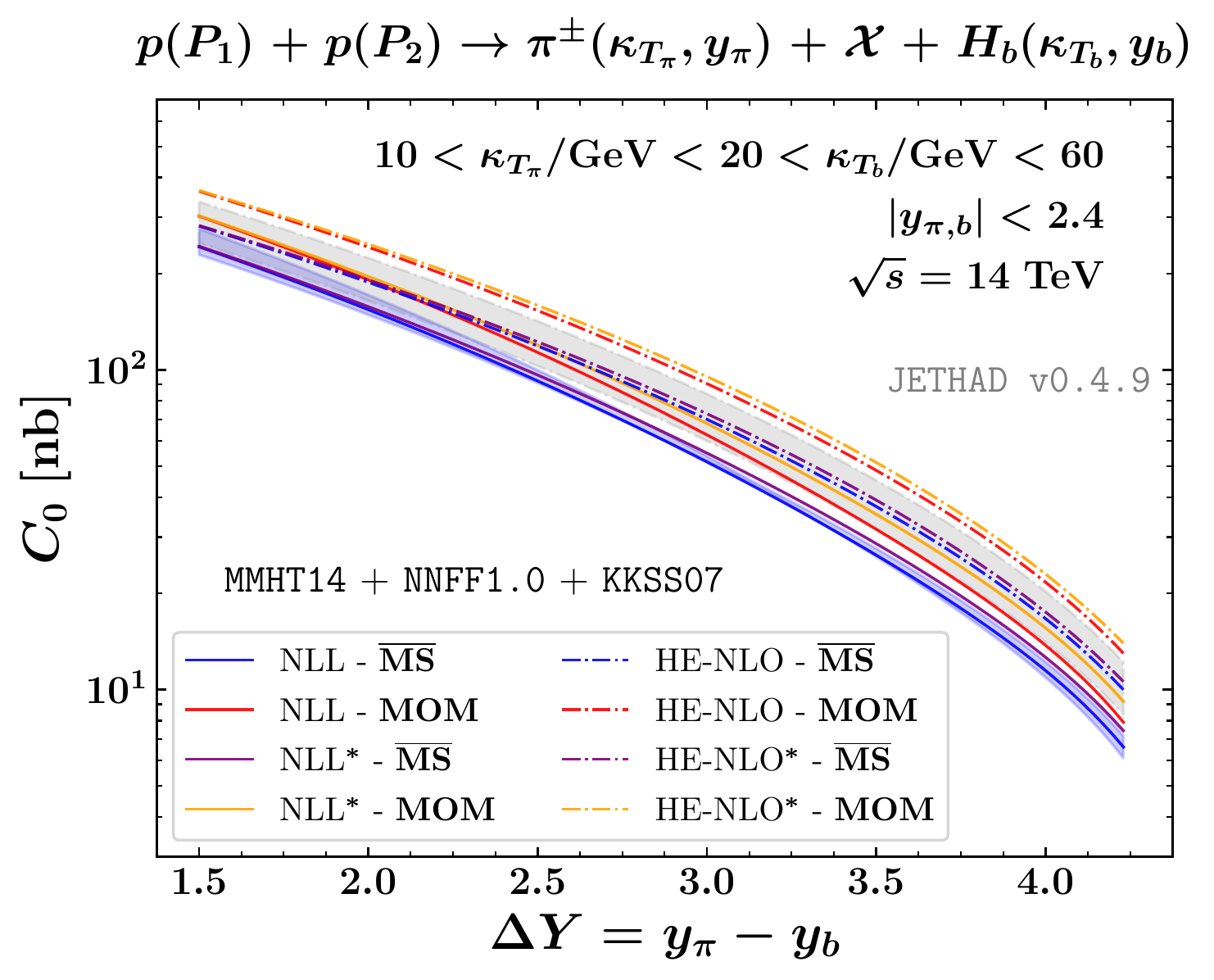}

   \includegraphics[scale=0.53,clip]{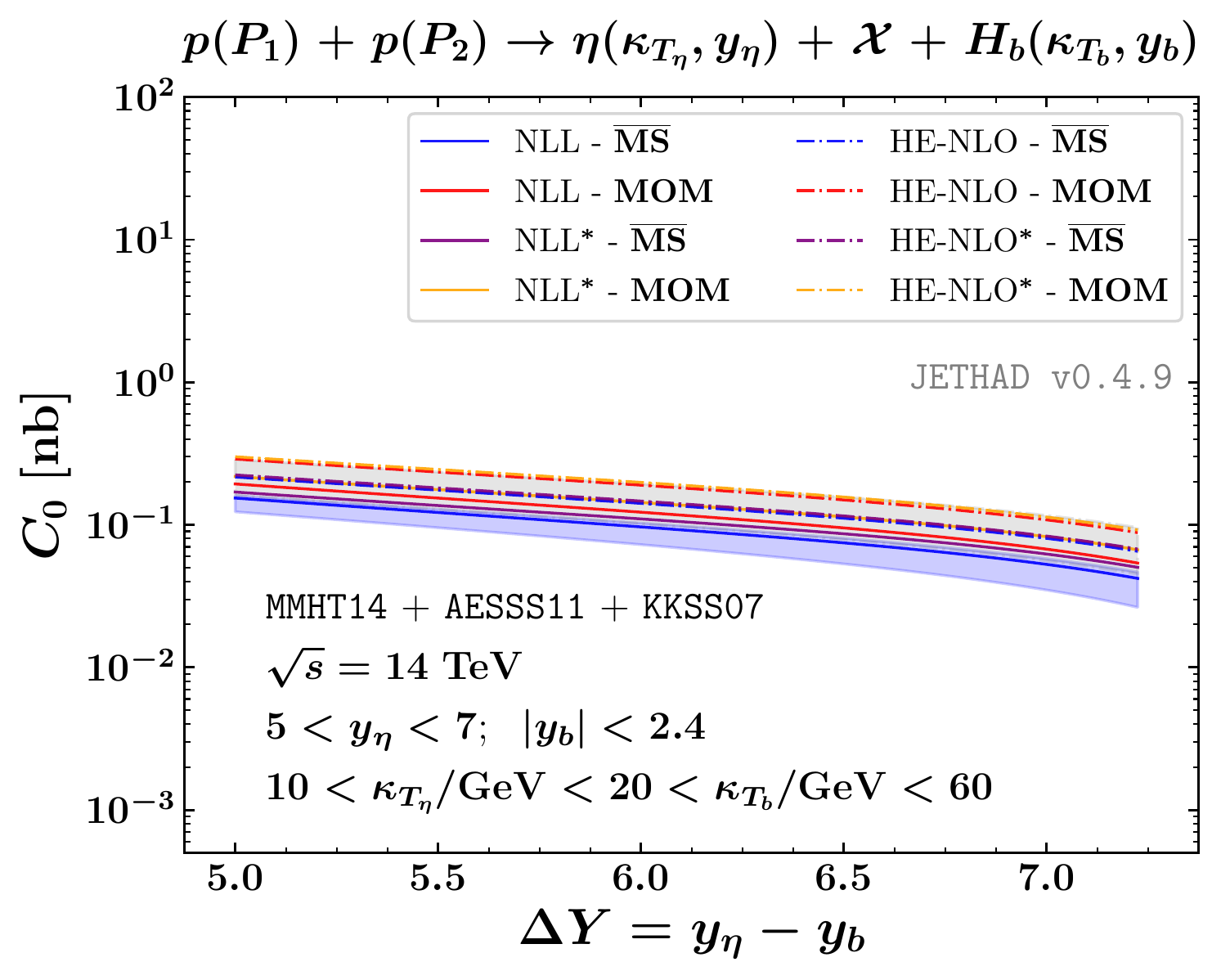}
   \includegraphics[scale=0.53,clip]{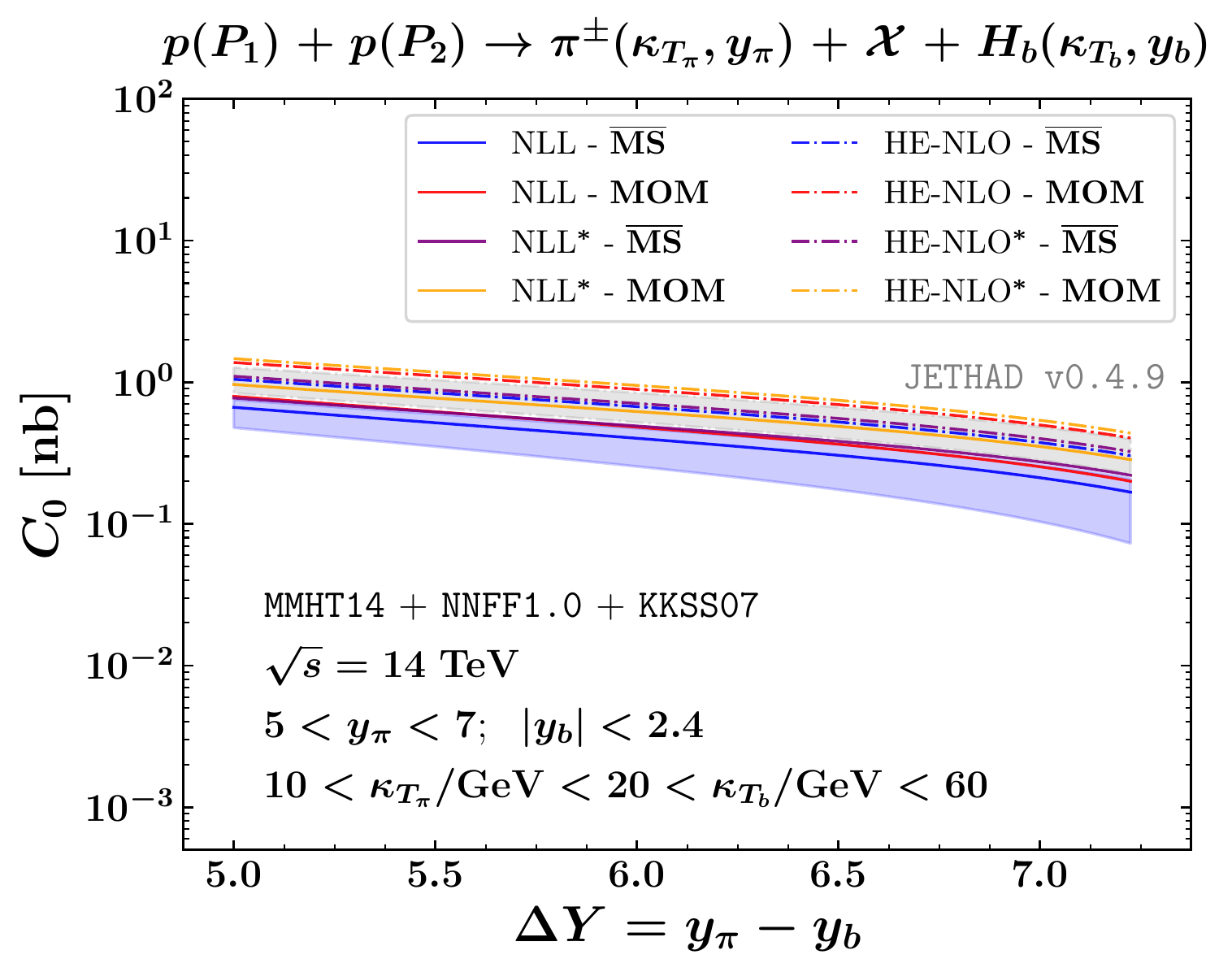}

\caption{$\DY$-distribution for the inclusive $\eta$~$+$~$b$~hadron (left) and $\pi^\pm$~$+$~$b$~hadron (right) channels at the LHC (upper) and at FPF~$+$~ATLAS (lower), for $\sqrt{s} = 14$ TeV.
Text boxes inside plots show final-state kinematic ranges. Predictions for two different representations of the NLL-resummed cross section, and for two renormalization-scheme selections are shown. Shaded bands refer to standard NLL calculations with the combined scale-variation and numerical-integration uncertainty.}
\label{fig:C0_sys-unc}
\end{figure*}

\begin{figure*}[t]

   \includegraphics[scale=0.53,clip]{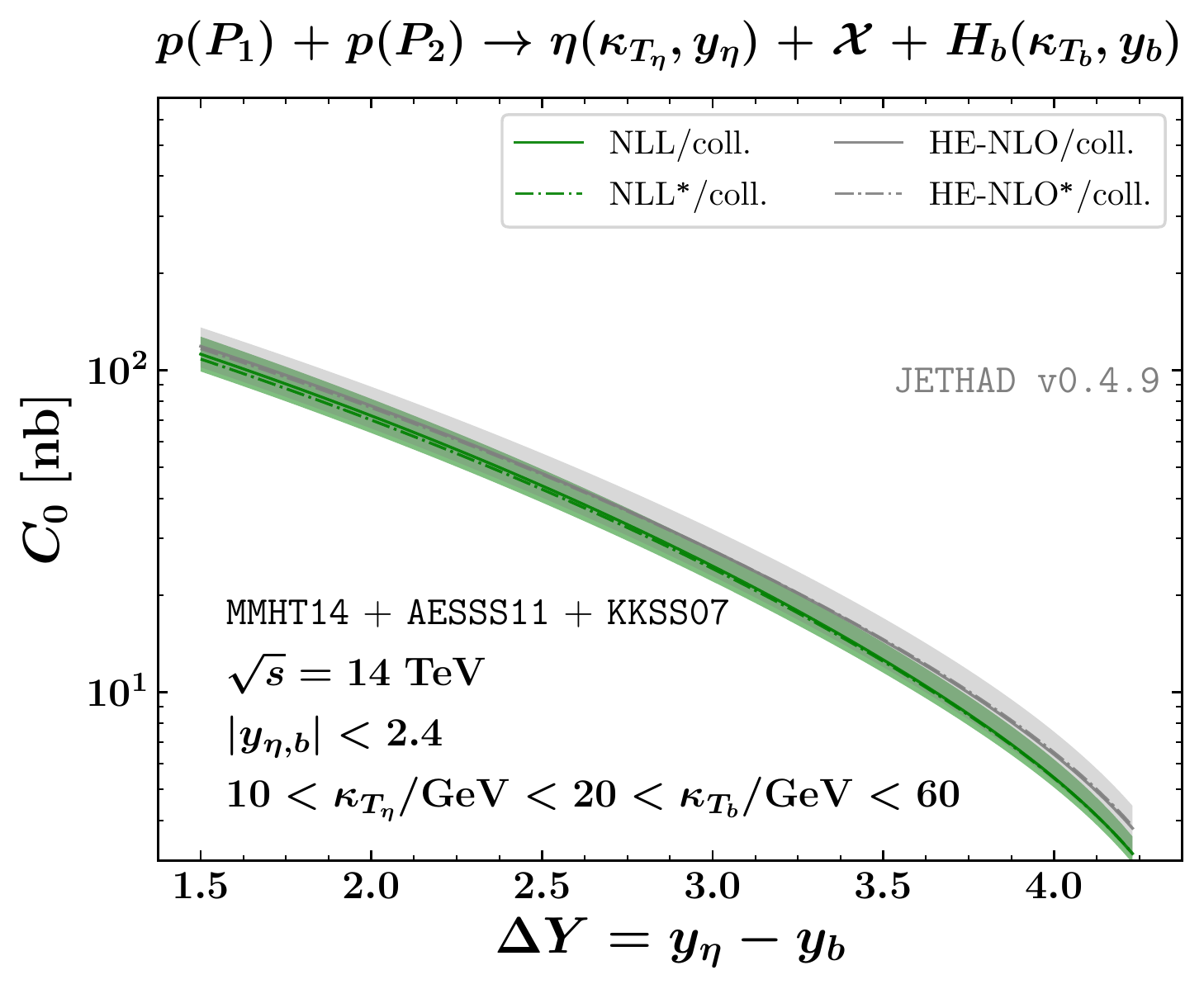}
   \includegraphics[scale=0.53,clip]{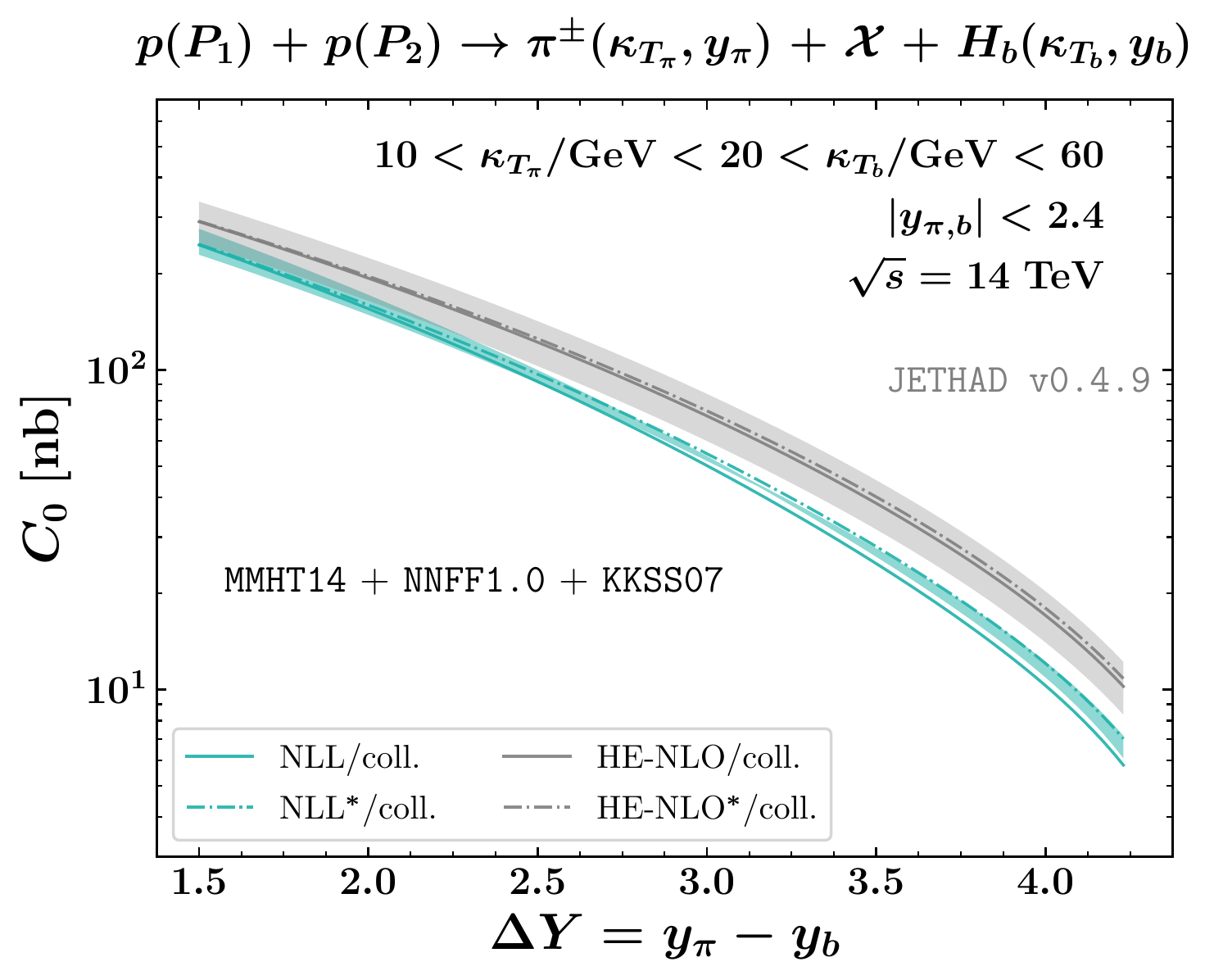}

   \includegraphics[scale=0.53,clip]{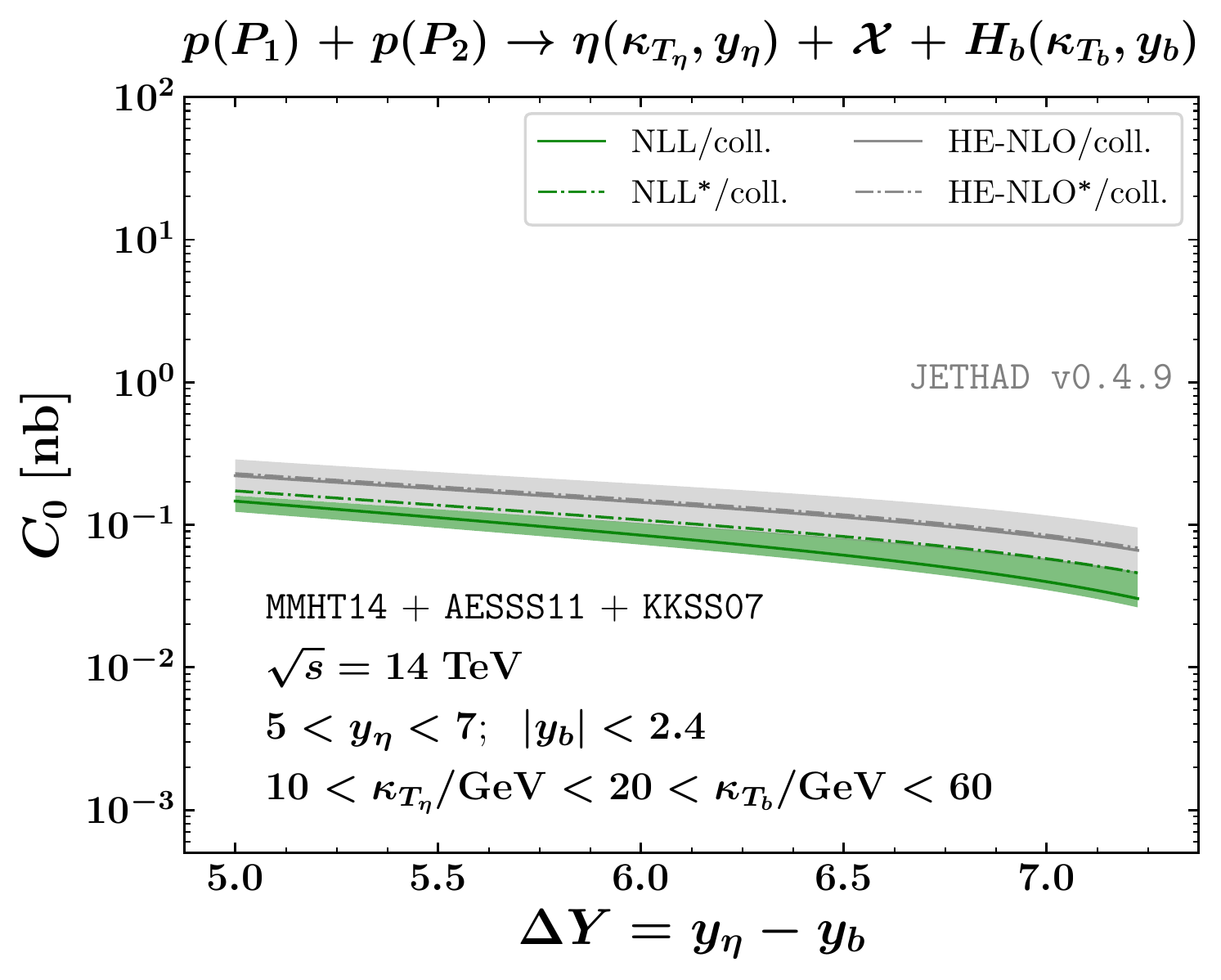}
   \includegraphics[scale=0.53,clip]{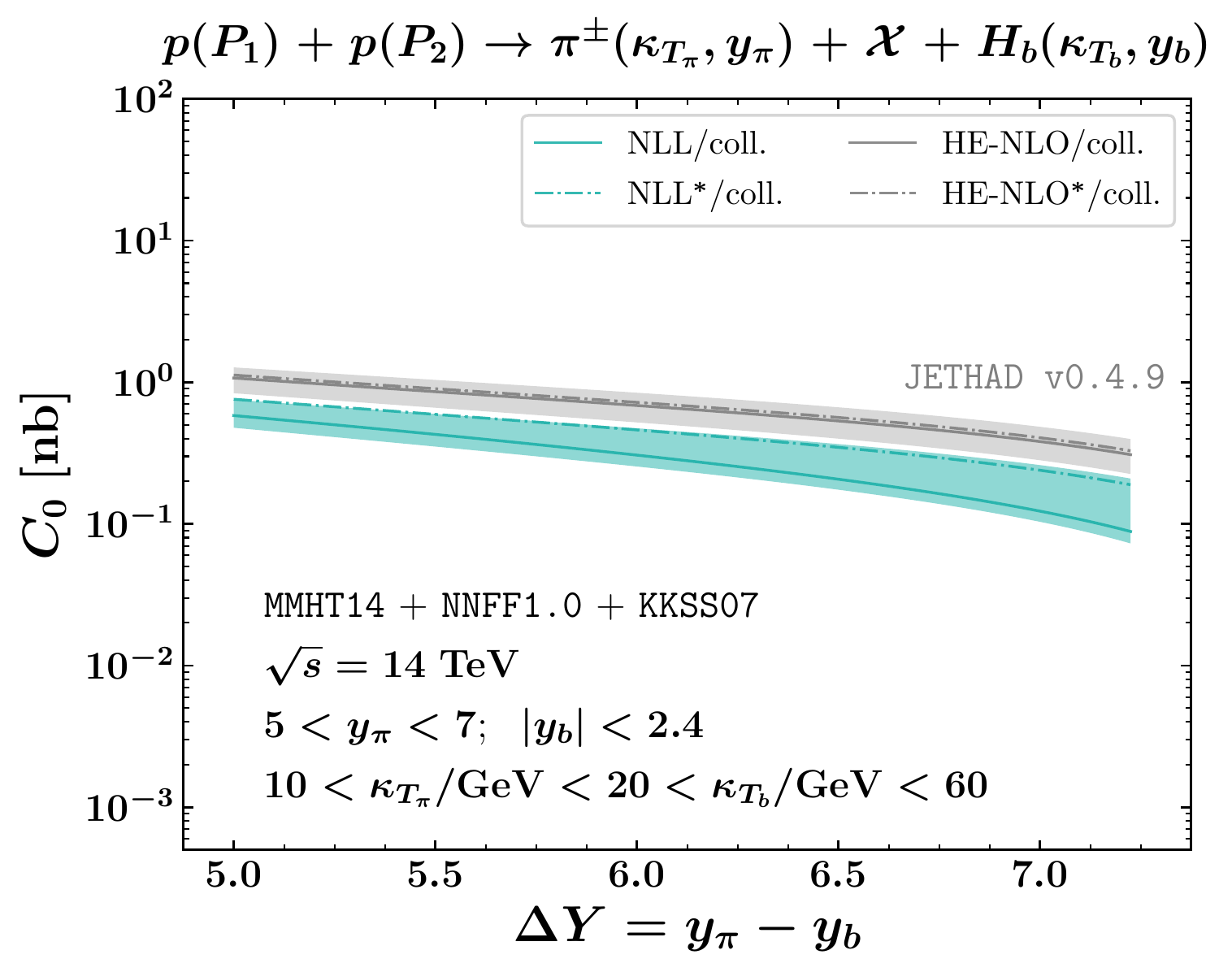}

\caption{$\DY$-distribution for the inclusive $\eta$~$+$~$b$~hadron (left) and $\pi^\pm$~$+$~$b$~hadron (right) channels at the LHC (upper) and at FPF~$+$~ATLAS (lower), for $\sqrt{s} = 14$ TeV.
Text boxes inside plots show final-state kinematic ranges. Predictions for the NLL kernel and the RG-improved one are compared in two different representations of the NLL-resummed cross section. Shaded bands refer to standard $\MSb$ NLL calculations with the combined scale-variation and numerical-integration uncertainty.}
\label{fig:C0_CI}
\end{figure*}

Plots of Fig.\tref{fig:C0_eta} show the behavior of $\DY$-distributions for the inclusive $\eta$-meson~$+$~heavy-flavor production at the LHC (upper panels) and at FPF~$+$~ATLAS (lower panel).
The $\DY$~shape of $C_0$ for the inclusive $\pi^\pm$~$+$~$\Lambda_c^\pm$ production at the LHC is presented in Fig.\tref{fig:C0_PL_LHC}.
Figs.\tref{fig:C0_Pb_LHC} and\tref{fig:C0_Pb_FPF} are for the cross section related to the inclusive $\pi^\pm$~$+$~$b$~hadron production at the LHC and at FPF~$+$~ATLAS, respectively.
Uncertainty bands are build in terms of the combined effect coming from energy-scale variation and numerical multi-dimensional integration over the final-state phase space, the former being sharply dominant.
From the inspection of results we generally observe a very favorable statistics, with our $\DY$-distributions lying in the range  $10^{-1}$ to $10^2$ nb.

The trend of bands in Figs.\tref{fig:C0_eta} to\tref{fig:C0_Pb_FPF} is a clear reflection of the usual dynamics of our hybrid factorization. Indeed, although the BFKL resummation predicts an increase with energy of the partonic hard-scattering cross section, its convolution with collinear PDFs and FFs brings, as an overall effect, to a lowering with $\DY$ of LL, NLL and HE-NLO results.
This drop-off is steeper when LHC kinematic configurations are considered, while it assumes a smoother shape in the FPF~$+$~ATLAS case. It could be due to the fact that, since the rapidity ranges covered by the FPF and ATLAS are not contiguous (see Section\tref{ssec:kinematics}), the increment with $\DY$ of the available phase space is slightly compensated by the absence of detected events in the interval between $y_{\rm FPF}^{\rm min}$ and $y_{\rm ATLAS}^{\rm max}$.

A similar pattern was also observed in a specular central~$+$~ultra-backward configuration, namely the CMS~$+$~CASTOR setup (see Fig.~10 of Ref.\tcite{Celiberto:2020wpk}).
Results for $C_0$ obtained with different pion FF parametrizations (see Figs.\tref{fig:C0_PL_LHC} and\tref{fig:C0_Pb_LHC}) are qualitatively similar, their mutual distance staying beyond a factor three.
This further motivates our dedicated analysis on FF uncertainty through the replica method (see Figs.\tref{fig:C0_PL_LHC_rep} to\tref{fig:C0_Pb_FPF_rep} and a related discussion in this Section).

We report the emergence of clear and natural stabilization effects of the high-energy series when standard LHC cuts are imposed (see upper panels of Fig.\tref{fig:C0_eta}, then Figs.\tref{fig:C0_PL_LHC} and\tref{fig:C0_Pb_LHC}). Indeed, $C_0$-distributions for all the channels of Fig.\tref{fig:process} feature NLL bands partially or even entirely nested in LL ones for small and moderate values of $\DY$. Then, when the rapidity interval grows, NLL BFKL corrections become more and more negative, thus making NLL predictions smaller than pure LL ones. NLL uncertainty bands are visibly narrower than LL and HE-NLO ones. Furthermore, their width generally decreases in the large $\DY$ range, namely where high-energy effects heavily dominate on pure DGLAP ones. This is in line with observations made in previous analyses on heavy-flavored emissions at CMS\tcite{Celiberto:2021dzy,Celiberto:2021fdp,Celiberto:2022dyf}. It reflects the fact that the energy-resummed series has reached a fair convergence thanks to the natural stabilizing effect of VFNS FFs depicting the hadronization mechanisms of the detected heavy-flavor species.

The high-energy stabilizing pattern is also present in the FPF~$+$~ATLAS coincidence setup, but its effects are milder (see lower panel of Figs.\tref{fig:C0_eta} and\tref{fig:C0_Pb_FPF}).
Indeed, albeit our required condition to assert evidence of stability is fulfilled, since cross sections can be fairly studied around the natural energy scales provided by kinematics, it turns out that FPF~$+$~ATLAS NLL predictions stay constantly below LL results. NLL uncertainty bands are narrower than LL ones, but slightly larger than NLL ones for the same channels investigated in the standard LHC configurations.
Moreover, LL results are always larger than HE-NLO ones, while NLL ones are smaller.
Although further, dedicated studies are needed to determine if the found natural-stability signals become worse when the FPF rapidity acceptances are pushed over the ones imposed in our analysis, an explanation for the increased sensitivity of $C_0$ to the resummation accuracy can be provided on the basis of our current knowledge about the dynamics behind other resummation mechanisms.

As a starting point, we remark that the semi-hard nature of the considered final states lead to high energies but not necessarily to $\mbox{small }x$.
This is particularly true for the FPF~$+$~ATLAS coincidence setup, where the strongly asymmetric final-state rapidity ranges make one of the two parton longitudinal fractions be always large, while the other one takes more moderate values.
As pointed out in Section\tref{sssec:FPF_ATLAS}, large-$x$ logarithms are not caught by our approach. They need to be accounted for via an adequate resummation mechanism, \emph{i.e.} the \emph{threshold} one.
A major outcome of a study conducted in Ref.\tcite{Almeida:2009jt} on inclusive di-hadron detections in hadronic collisions is that the inclusion of the NLL threshold resummation on top of pure NLO calculations leads to a substantial increase of cross sections.
Notably, this increment is of the same order of the gap between our LL and NLL high-energy predictions for $C_0$ at FPF~$+$~ATLAS configurations.
In Refs.\tcite{Liu:2020mpy,Shi:2021hwx} it was shown how the very large NLL instabilities emerging from forward hadron hadroproductions described by the hands of the \emph{saturation} framework\tcite{Stasto:2013cha} (see also Refs.\tcite{Altinoluk:2014eka,Ducloue:2017dit,Ducloue:2018lil}) and leading to negative values of cross sections can be sensibly reduced when threshold logarithms are included in those calculations.
In view of these results, we believe that the high-energy natural stability coming up from our studies is not worsened by the adoption of FPF~$+$~ATLAS coincidence setups. It is present and leads to a fair description of $C_0$ at natural scales.
The discrepancy between LL and NLL predictions is explained by the emergence of large-$x$, threshold logarithms, which are genuinely neglected by the hybrid high-energy and collinear factorization.
The inclusion of this large-$x$ resummation represents a key ingredient to improve the description of the considered observables and needs to be carried out as a next step to assess the feasibility of precision studies of $\DY$-distributions at FPF~$+$~ATLAS.

As discussed in Section\tref{ssec:jethad_settings}, previous analyses on semi-hard reactions have shown that the choice of different collinear PDF sets as well as of different members inside the same set does not have a sizable impact on rapidity-differential cross sections.
However, as already mentioned, for our considered final-state kinematic cuts (see Section\tref{ssec:kinematics}) we access the so-called \emph{threshold} sectors, where PDFs could not be well constrained. This is particularly true in the FPF~$+$~ATLAS coincidence setup.
Therefore, in Fig.\tref{fig:C0_pion_PDF_rep} we present a study on the sensitivity of $C_0$ for a limited selection of reactions in Fig.\tref{fig:process}, namely the $\pi^\pm$~$+$~heavy-flavor channel at the LHC (upper panels) and at FPF~$+$~ATLAS (lower panel).

The envelope of predictions in main panels is given as a replica-driven study on {\tt NNPDF4.0} PDFs, whereas the central member of {\tt NNFF1.0} pion FFs is considered.
Each of these plots is accompanied by an ancillary stripe panel showing the reduced $\DY$-distribution, namely the envelope of replicas’ predictions divided by its mean value.
It emerges that, for all the considered channels, the uncertainty related to PDF replicas stays by far below $5\%$ and do not lead to any overlap between LL, NLL and HE-NLO predictions. Thus, its effect turns out to have a little relevance when compared with the scale-variation uncertainty shown in Figs.\tref{fig:C0_eta} to\tref{fig:C0_Pb_FPF}.

In Figs.\tref{fig:C0_PL_LHC_rep} to\tref{fig:C0_Pb_FPF_rep} we present the behavior of $\DY$-distributions for the inclusive pion~$+$~heavy-flavor production, with the energy scales being taken at their natural values provided by kinematics, and the envelope of main results built in terms of a replica-driven study on {\tt JAM20}, {\tt MAPFF1.0}, and {\tt NNFF1.0} pion collinear FFs.
Panels of Figs.\tref{fig:C0_PL_LHC_rep} and\tref{fig:C0_Pb_LHC_rep} refer to final states where the pion is emitted at LHC configurations in association with a $\Lambda_c^\pm$ baryon or a $b$~hadron, respectively. Panels of\tref{fig:C0_Pb_FPF_rep} are for $\pi^\pm$~$+$~$b$~hadron detections at FPF~$+$~ATLAS.
The overall trend is a spread of our replica results, which is generally wider in the NLL case with respect to LL and HE-NLO ones.

The envelope of predictions obtained by making use {\tt JAM20} FFs is the narrowest one, around $10\div20\%$. It increases in the {\tt MAPFF1.0} case, around $25\div50\%$, and the broadest one is for {\tt NNFF1.0} functions,  around $25\div75\%$.
This noticeable difference is not surprising and has been already observed in a direct comparison among among different FF parametrizations, see, \emph{e.g.}, Ref.\tcite{Khalek:2021gxf}.
As pointed out in Section~V.B.1 of that work, the simultaneous determination of FFs and PDFs genuinely leads to a reduction of the spread of replicas for the {\tt JAM20} set. Conversely, {\tt MAPFF1.0} and {\tt NNFF1.0} FFs are determined alone.

Therefore, the information about gluon fragmentation is less constrained, since gluon-initiated channels are active starting from NLO only.
This brings us to a larger uncertainty of the gluon FF, whose contribution is heightened by the the gluon PDF in the collinear convolution encoded in the LO hadron impact factor (Eq.\eref{LOHIF}).
According to Ref.\tcite{Khalek:2021gxf}, the different assumptions made on the (partial) $SU(2)$ isospin symmetry mentioned in Section\tref{ssec:PDFs_FFs} of this article turn out not to have a relevant impact.
It emerges that the size of uncertainties coming from our FF-replica study in Figs.\tref{fig:C0_PL_LHC_rep} to\tref{fig:C0_Pb_FPF_rep} is of the same order and in some case larger than the one related to energy-scale variations presented in Figs.\tref{fig:C0_eta} to\tref{fig:C0_Pb_FPF}.

In Fig.\tref{fig:C0_sys-unc} we compare results for two different choices of the renormalization scheme, namely the $\MSb$ and the MOM one, as discussed in point~(\emph{iiii}) of Section\tref{ssec:jethad_settings}.
Furthermore, we gauge the impact of varying the representation of the NLL-resummed cross section, namely passing from the pure NLL to the NLL$^*$ one (see Section\tref{ssec:sigma} and point~(\emph{v}) of Section\tref{ssec:jethad_settings}).
We apply the same procedure to the high-energy fixed order case, \emph{i.e.} moving from the HE-NLO to the HE-NLO$^*$ representation.

For the sake of simplicity, we consider a restricted selection of processes in Fig.\tref{fig:process}.
Panels of Fig.\tref{fig:C0_sys-unc} show the behavior of $\DY$-distributions for $\eta$~$+$~$b$~hadron (left) and $\pi^\pm$~$+$~$b$~hadron (right) channels at the LHC (upper) and FPF~$+$~ATLAS (lower) kinematic setups. Solid (dashed) lines refer to resummed (high-energy fixed-order) calculations done at natural $\mu_{R,F}$ scales. Shaded bands embody the combined scale-variation and numerical-integration uncertainty of standard $\MSb$ NLL and HE-NLO predictions, taken as reference results.

We observe that both $\MSb$ NLL$^*$ and HE-NLO$^*$ selections do not bring to a relevant change of the general trend. They are entirely contained inside the corresponding $\MSb$ NLL and HE-NLO band.
Conversely, MOM predictions are systematically larger, and they often stay above the corresponding $\MSb$ bands.
As already mentioned (see Section\tref{ssec:pert} and discussion in point~(\emph{iiii}) of Section\tref{ssec:jethad_settings}),
a consistent MOM analysis on $C_0$-distributions would rely on MOM-determined PDFs and FFs. Therefore, results presented in Fig.\tref{fig:C0_sys-unc} can be interpreted as a guess of the upper limit, whose completed effect still needs to be assessed.

We complement our analysis on $C_0$ by comparing next-to-leading predictions obtained with the standard NLL kernel and the RG-improved one (see discussion in point~(\emph{vi}) of Section\tref{ssec:jethad_settings}, and Appendix~\hyperlink{app:NLL_kernel}{A}).
The two resummed representations are also considered, as well as the corresponding high-energy fixed order cases.
To recognize the RG-improved distributions we add the suffix ``/coll." to labels employed so far.

Plots in Fig.\tref{fig:C0_CI} show the $\DY$~shape of $C_0$ for $\eta$~$+$~$b$~hadron (left) and $\pi^\pm$~$+$~$b$~hadron (right) channels at the LHC (upper) and FPF~$+$~ATLAS (lower) kinematic setups. Solid (dashed) lines refer to $\MSb$ NLL resummed and fixed-order predictions (with the inclusion of next-to-NLL factors) done at natural $\mu_{R,F}$ scales, and embodying the RG-improvement of the BFKL kernel. Shaded bands refer to the combined scale-variation and numerical-integration uncertainty of standard $\MSb$ NLL and  results, used as reference distributions.

In all the presented plots we note that the RG improvement of the kernel produces an effect which is visible, but almost entirely contained inside bands for corresponding nonimproved predictions. As expected, its weight is almost irrelevant in the HE-NLO case, since the truncation of the exponentiated kernel in Eq.\eref{Cn_HENLOstar_MSbar} brings to a loss of sensitivity of such kind of improvement on the whole calculation.

The overall result coming out from the discussion of predictions shown in this Section is that light-meson plus heavy-flavor production processes allow for a fair stabilization of the high-energy resummation, as expected. $\DY$-distributions are promising observables where to hunt for signals of the onset of high-energy dynamics as well as possible candidates to discriminate among BFKL-driven and fixed-order computations. 
Their sensitivity to collinear FFs permits us to assess the weight of uncertainties coming from the hadronization mechanisms of different hadron species in ranges complementary to the currently accessible ones.

Further studies of these distributions will offer us an intriguing chance to explore the interplay between the high-energy QCD dynamics and other resummations, in particular the large-$x$ threshold one.

\subsection{Azimuthal distributions}
\label{ssec:phi}

Semi-hard phenomenology has always been characterized by the definition and study of observables more and more sensitive to final-state rapidity intervals.
When these observables are taken to be differential also in the azimuthal angles, a breach in the core of high-energy QCD is made. In our case of two-particle hadroproduction reactions, distinctive clues of the onset of the BFKL dynamics emerge when large $\DY$ values heighten the number of undetected gluons strongly ordered in rapidity. These gluon emissions are accounted for by the resummation of energy logarithms and they lead to a \emph{decorrelation} on the azimuthal plane of the final-state objects, which grows with $\DY$.

The azimuthal decorrelation was first observed in the $\DY$ behavior of cross-section azimuthal moments\tcite{Cerci:2008xv,Colferai:2010wu,Ducloue:2013bva,Ducloue:2013hia,Caporale:2014gpa}, operationally defined as the $R_{n0} \equiv C_n/C_0$ ratios between a given azimuthal coefficient, $C_{n \ge 1}$, and the $\phi$-summed $\DY$-distribution, $C_0$, studied in Section\tref{ssec:C0}.
The $R_{n0}$ ratio can be thought as the mean value of the cosine $\langle \cos \phi \rangle$ which directly measures the azimuthal decorrelation between the two outgoing particles. The $R_{n0}$ ratios are the higher moments $\langle \cos (n \phi) \rangle$. In Refs.\tcite{Vera:2006un,Vera:2007kn} ratios between azimuthal moments, $R_{nm} \equiv C_n/C_m = \langle \cos (n \phi) \rangle / \langle \cos (m \phi) \rangle$, were proposed as further probes of BFKL. 

NLL-resummed predictions for azimuthal correlations of two Mueller--Navelet jet turned out to be in a satisfactory agreement with LHC data at $\sqrt{s} = 7$ TeV and for symmetric $\kappa_T$ windows\tcite{Ducloue:2013hia,Ducloue:2013bva,Caporale:2014gpa}.
However, due to the aforementioned instabilities arising in the NLL series for processes featuring the emission of two light objects, a theory-versus-experiment comparison could not be done at natural energy scales, but at the much larger ones provided by different scale-optimization procedures\tcite{Ducloue:2013hia,Caporale:2014gpa,Caporale:2015uva}.

A side outcome of recent studies on inclusive hadroproductions of heavy-flavored hadrons is that the stabilizing effects coming out from the use of heavy-flavor VFNS FFs is less effective when a heavy hadron is emitted together with a light jet instead of another heavy object\tcite{Celiberto:2021dzy,Celiberto:2021fdp}. This brings to just a partial reduction of instabilities for azimuthal moments.

As pointed out in Ref.\tcite{Celiberto:2022dyf}, starting from the azimuthal coefficients it is possible to construct a more stable observable that contains signals of high-energy dynamics coming from all azimuthal modes. We refer to the azimuthal distribution, defined as
\begin{widetext}
\begin{equation}
\label{azimuthal_distribution}
 \frac{1}{\sigma}
 \frac{\drv \sigma}{\drv \phi} (\phi, \DY; s) = \frac{1}{2 \pi} \left\{ 1 + 2 \sum_{l = 1}^{\infty} \cos(l \phi) \langle \cos(l \phi) \rangle \right\}
 = \frac{1}{2 \pi} \left\{ 1 + 2 \sum_{l = 1}^{\infty} \cos(l \phi) \frac{C_l (\DY; s)}{C_0 (\DY; s)} \right\} \; .
\end{equation}
\end{widetext}
This observable was originally proposed in Ref.\tcite{Marquet:2007xx} in the context of Mueller--Navelet analyses and then studied with NLL accuracy in Ref.\tcite{Ducloue:2013hia} for the same process. Investigating its behavior is advantageous both from a theoretical and an experimental viewpoint. On the one hand, as already mentioned, it collects the whole spectrum of high-energy signals coming from all the $C_n$ coefficients, thus making it one of the most robust observables to hunt for BFKL effects.

Being a normalized distribution, namely a \emph{multiplicity}, the effect of uncertainties coming from the choice of different PDFs and/or FFs as well as the of ones stemming from distinct replicas inside the same PDF and/or FF set are strongly suppressed. 
This permits us to focus on the uncertainties which intrinsically come from the high-energy resummation and thus to conduct stringent BFKL tests.
On the other hand, detector acceptances hardly cover the whole ($2 \pi$) azimuthal-angle range. Therefore, the comparison with data of a $\phi$-dependent observable, such as our azimuthal distribution, is much easier than for a $R_{nm}$ correlation moment. 
From a numerical perspective, in order to get reliable predictions for our $\phi$-distribution, a large number of $C_n$ coefficients is needed. We checked the numerical stability of calculations by gradually increasing the effective upper bound of the $l$~sum in Eq.\eref{azimuthal_distribution}. A fair numerical convergence was got by reaching $l_{\rm [bound]} = 20$.

\begin{figure*}[!pht]

   \includegraphics[scale=0.53,clip]{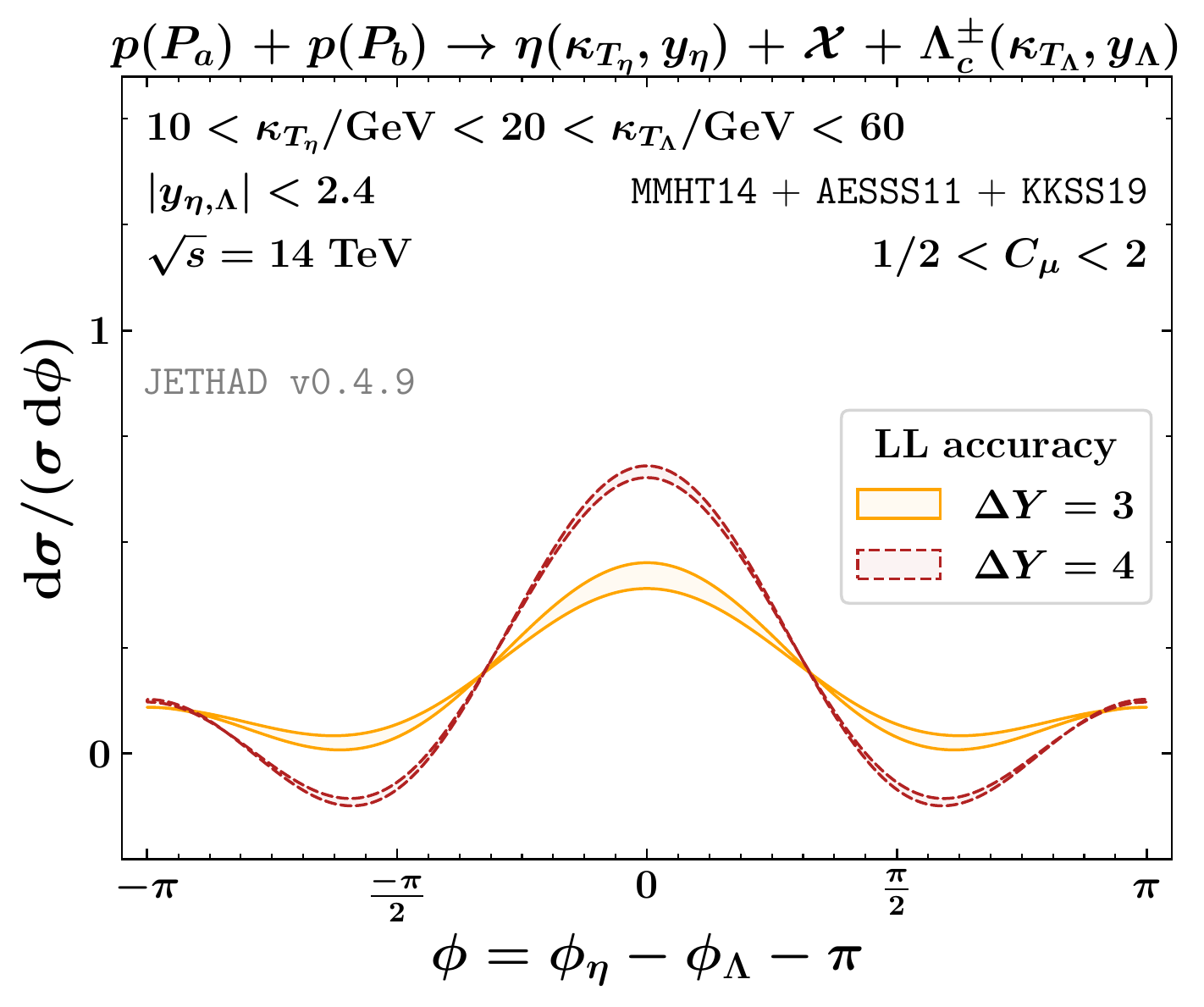}
   \includegraphics[scale=0.53,clip]{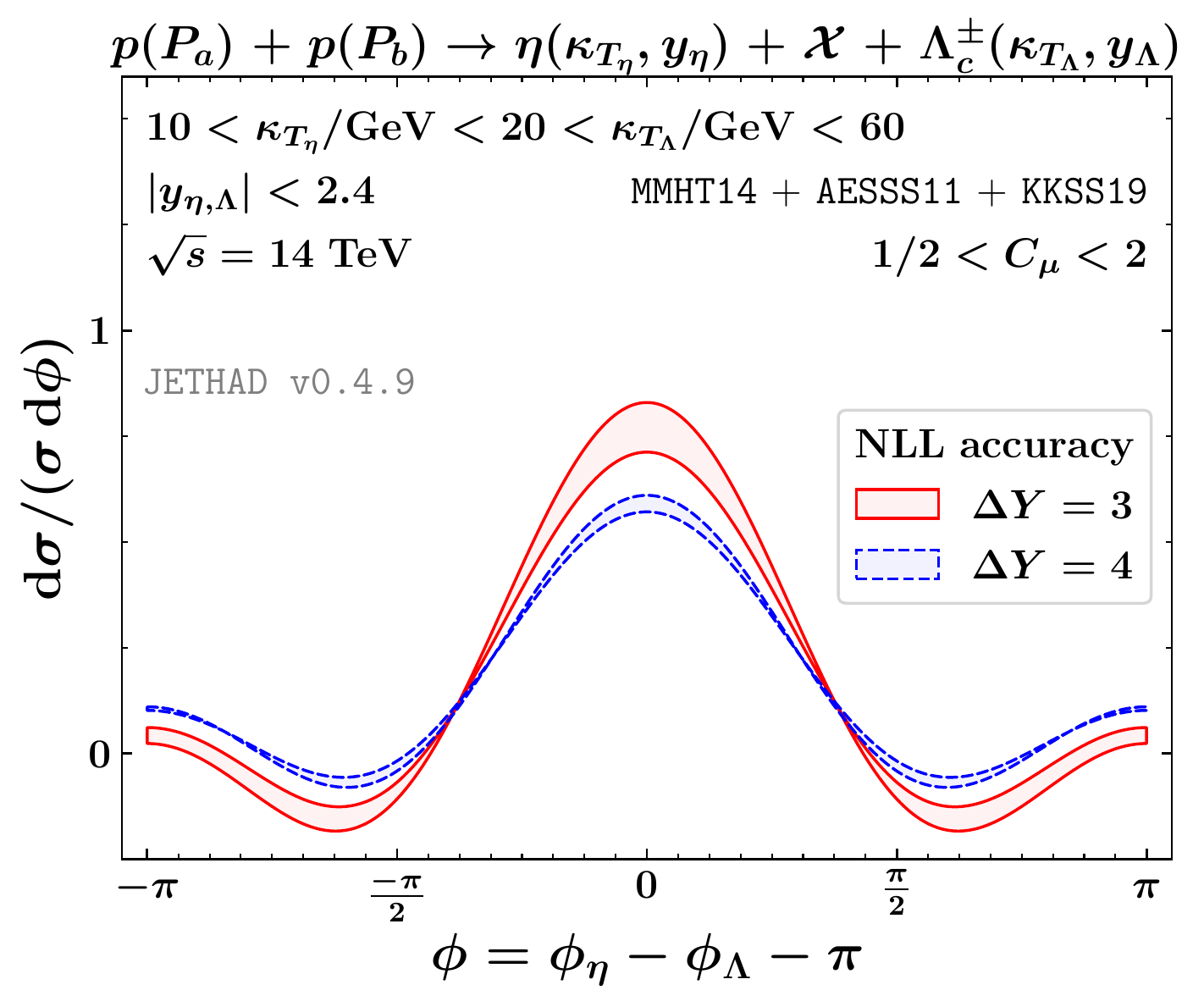}

   \includegraphics[scale=0.53,clip]{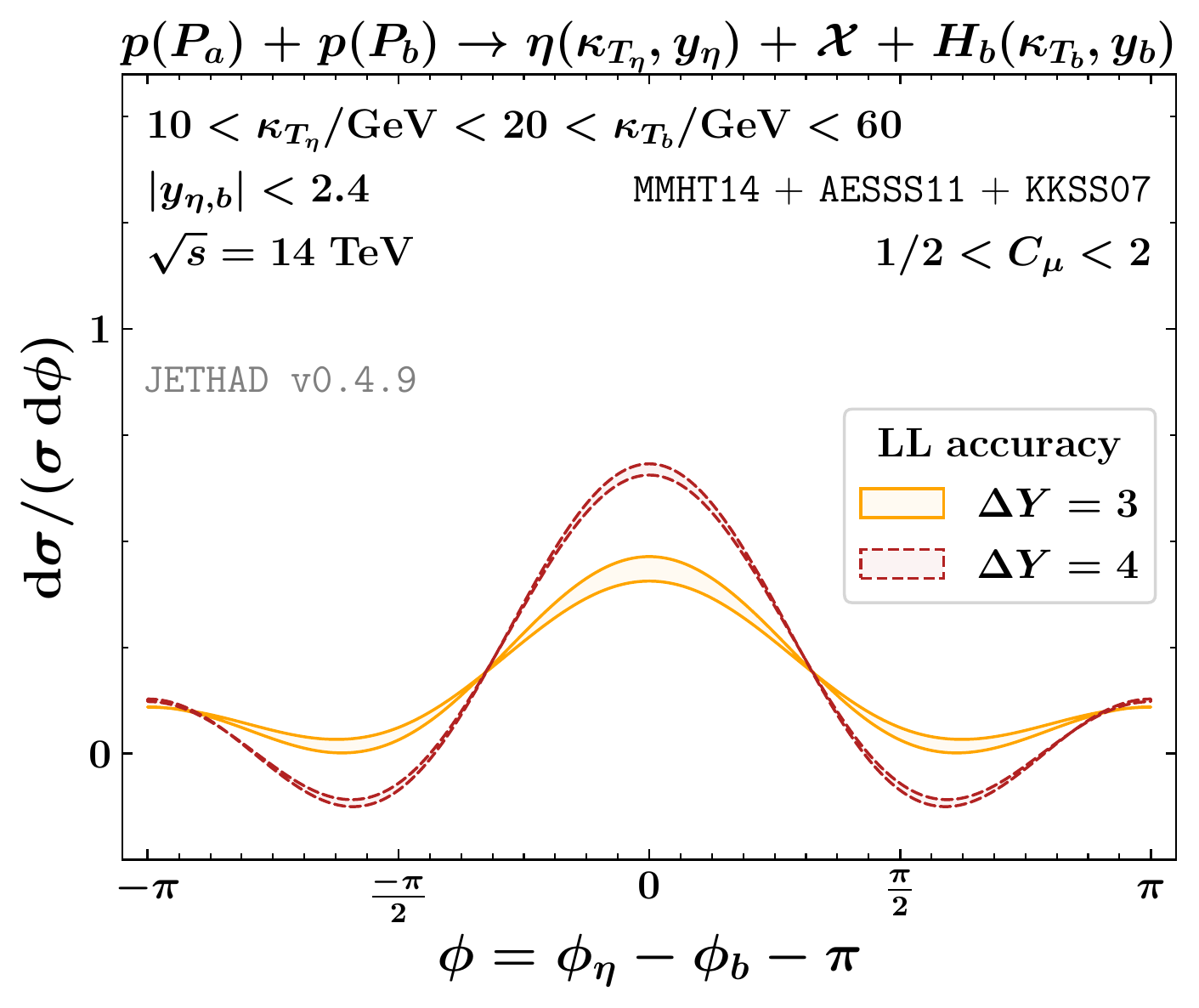}
   \includegraphics[scale=0.53,clip]{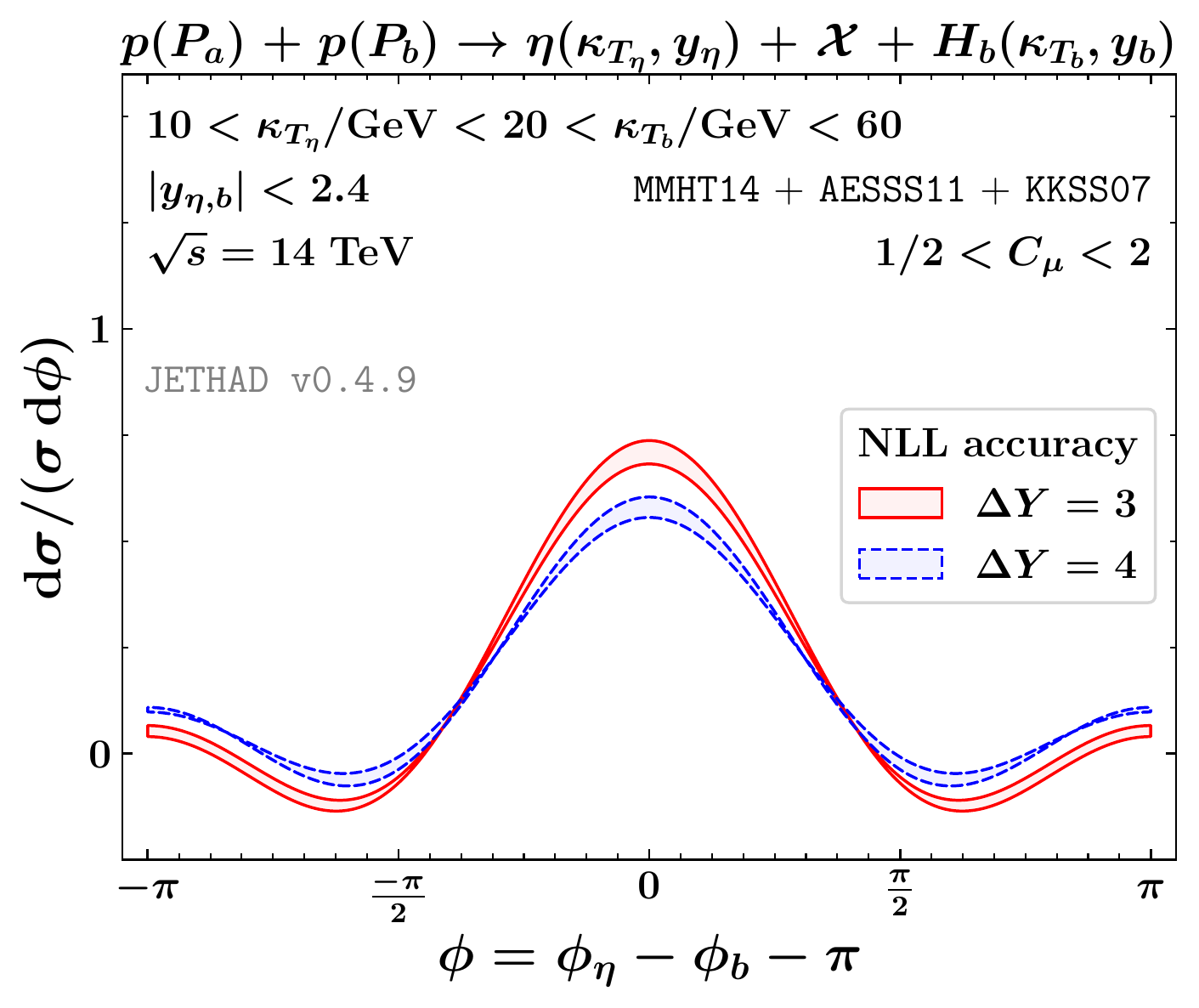}

   \includegraphics[scale=0.53,clip]{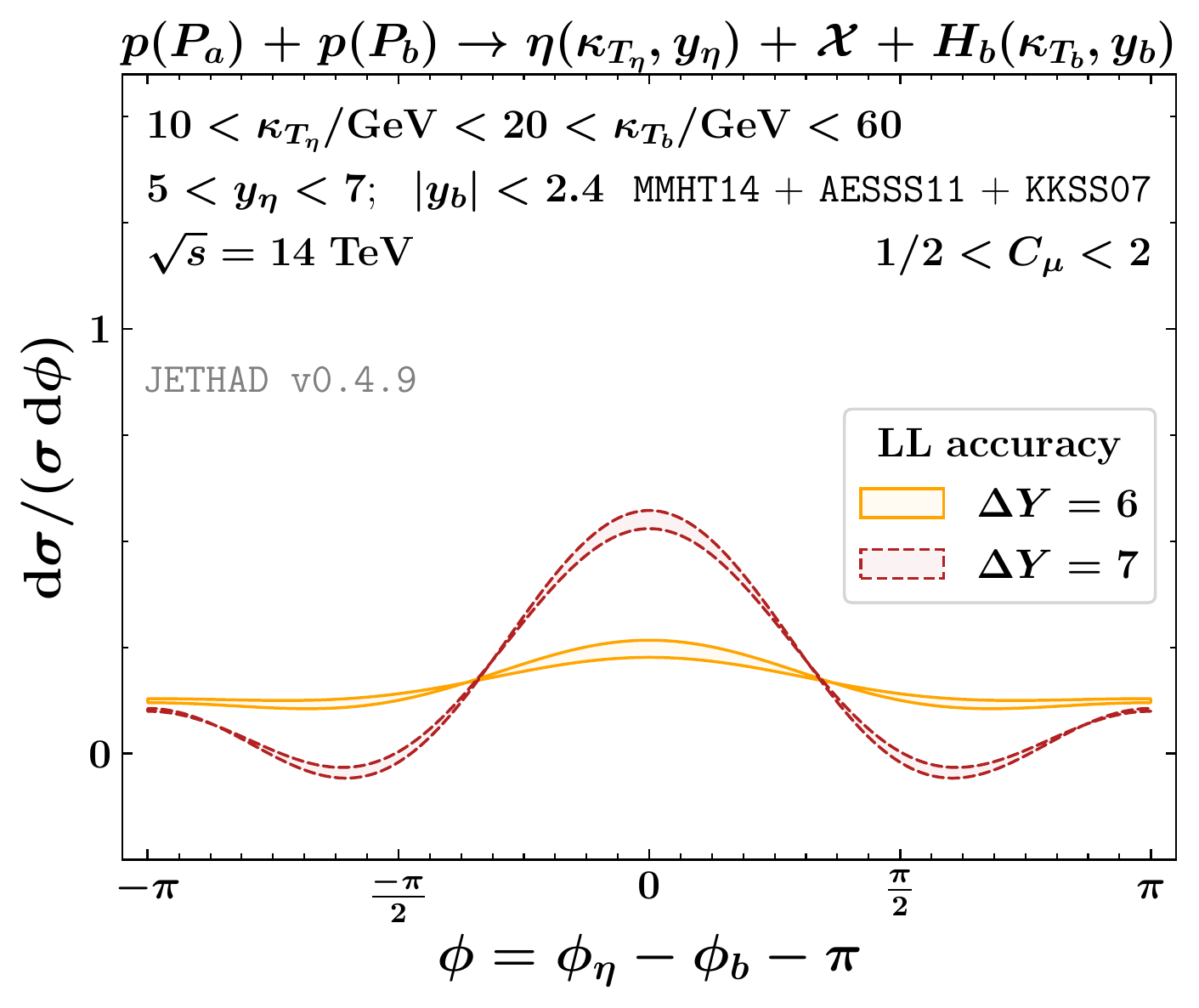}
   \includegraphics[scale=0.53,clip]{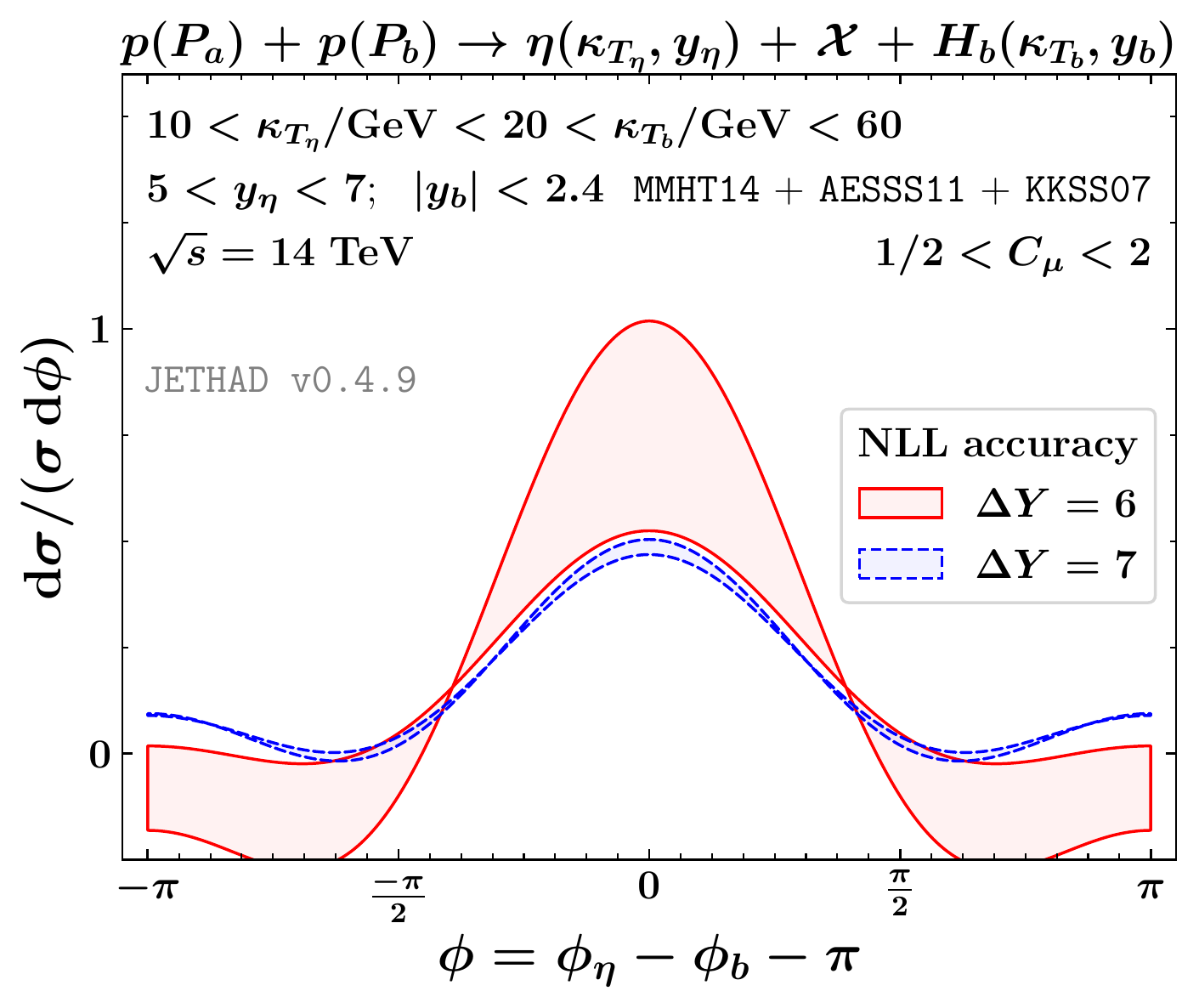}

\caption{Azimuthal distribution for the inclusive $\eta$-meson~$+$~heavy-flavor production at the LHC (upper panels) and at FPF~$+$~ATLAS (lower panels), for $\sqrt{s} = 14$ TeV. Text boxes inside plots show final-state kinematic ranges. Uncertainty bands encode the net effect of the scale variation and the multi-dimensional integration over the final-state phase space.}
\label{fig:phi_eta}
\end{figure*}

\begin{figure*}[!pht]

   \includegraphics[scale=0.53,clip]{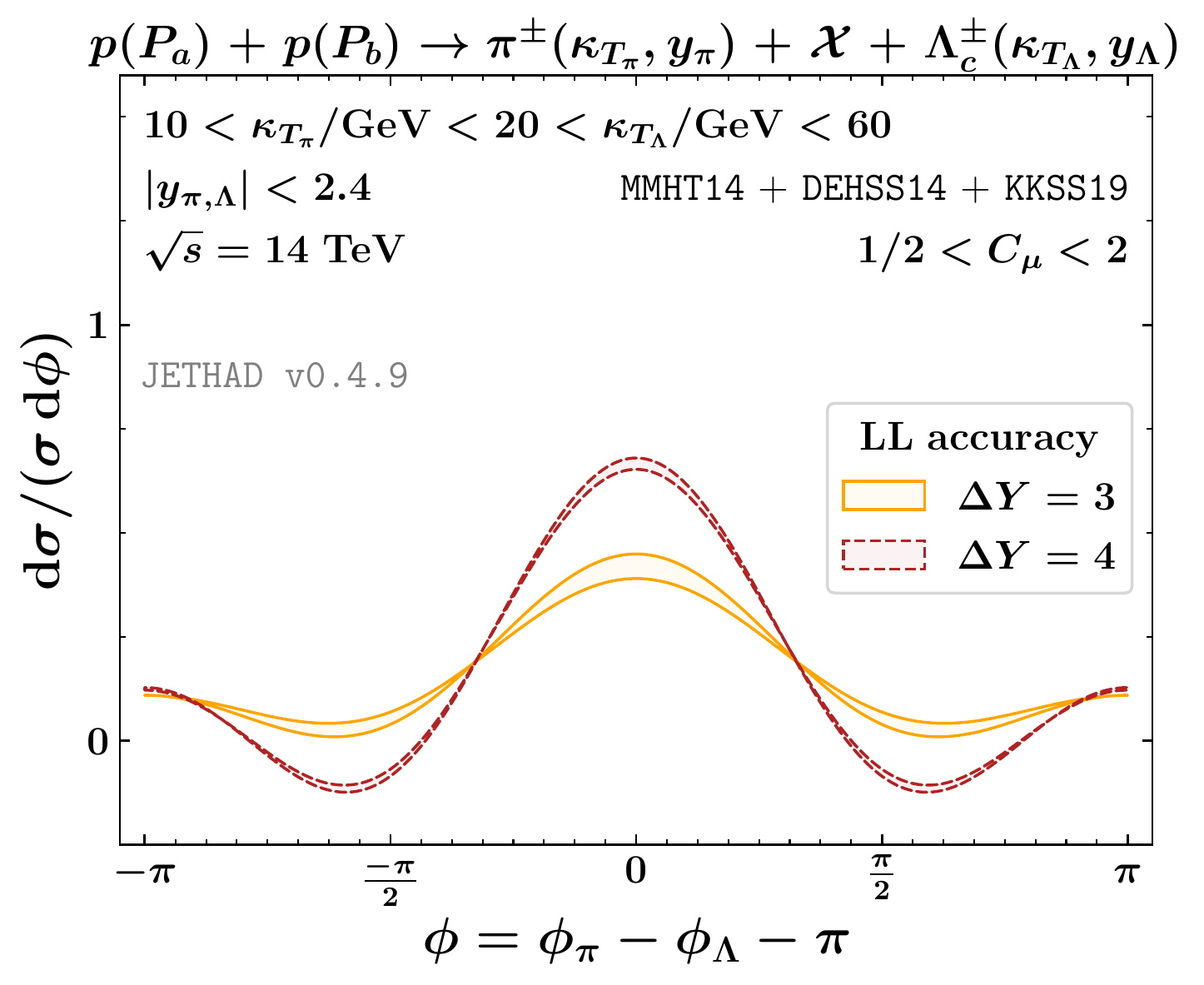}
   \includegraphics[scale=0.53,clip]{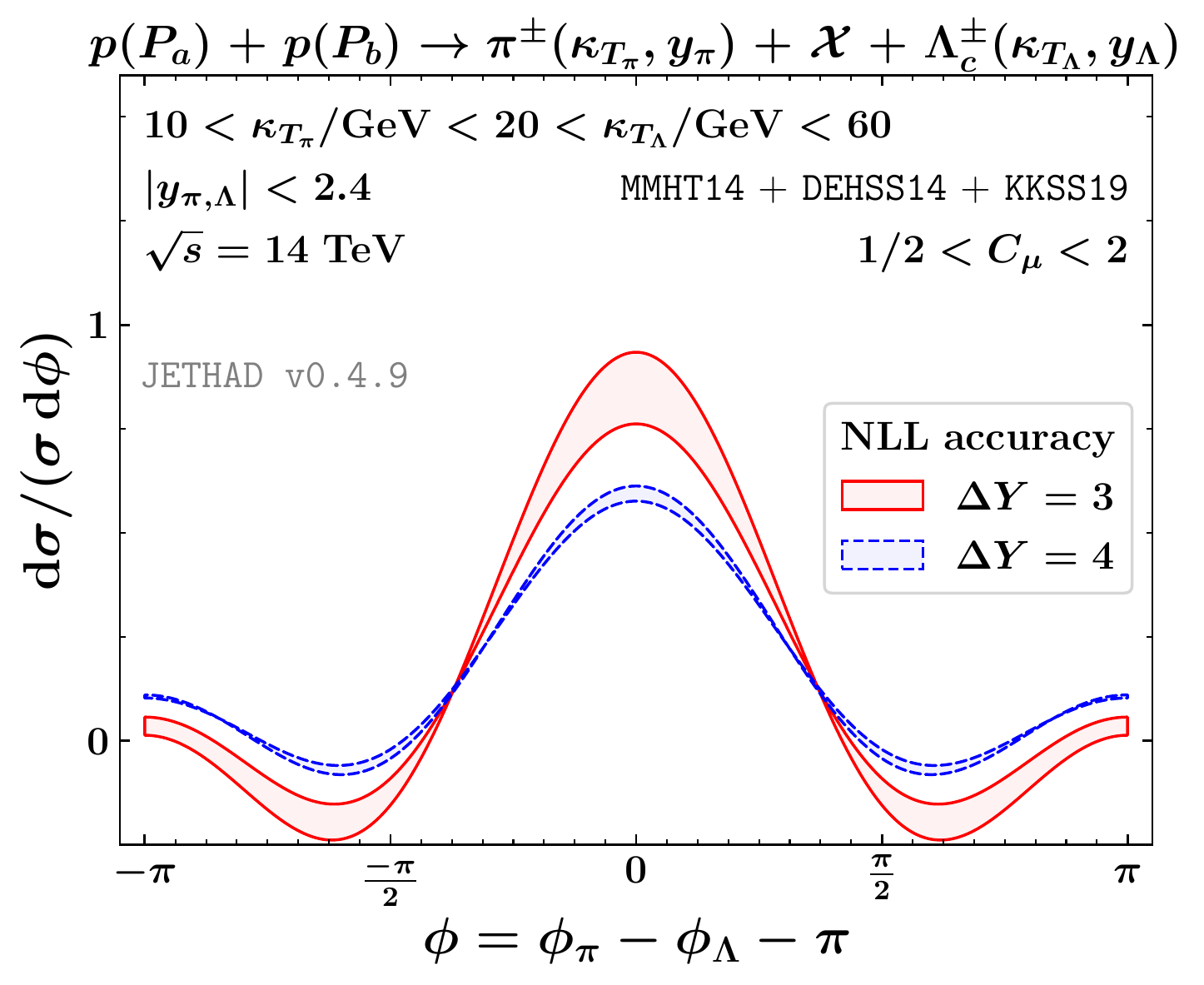}

   \includegraphics[scale=0.53,clip]{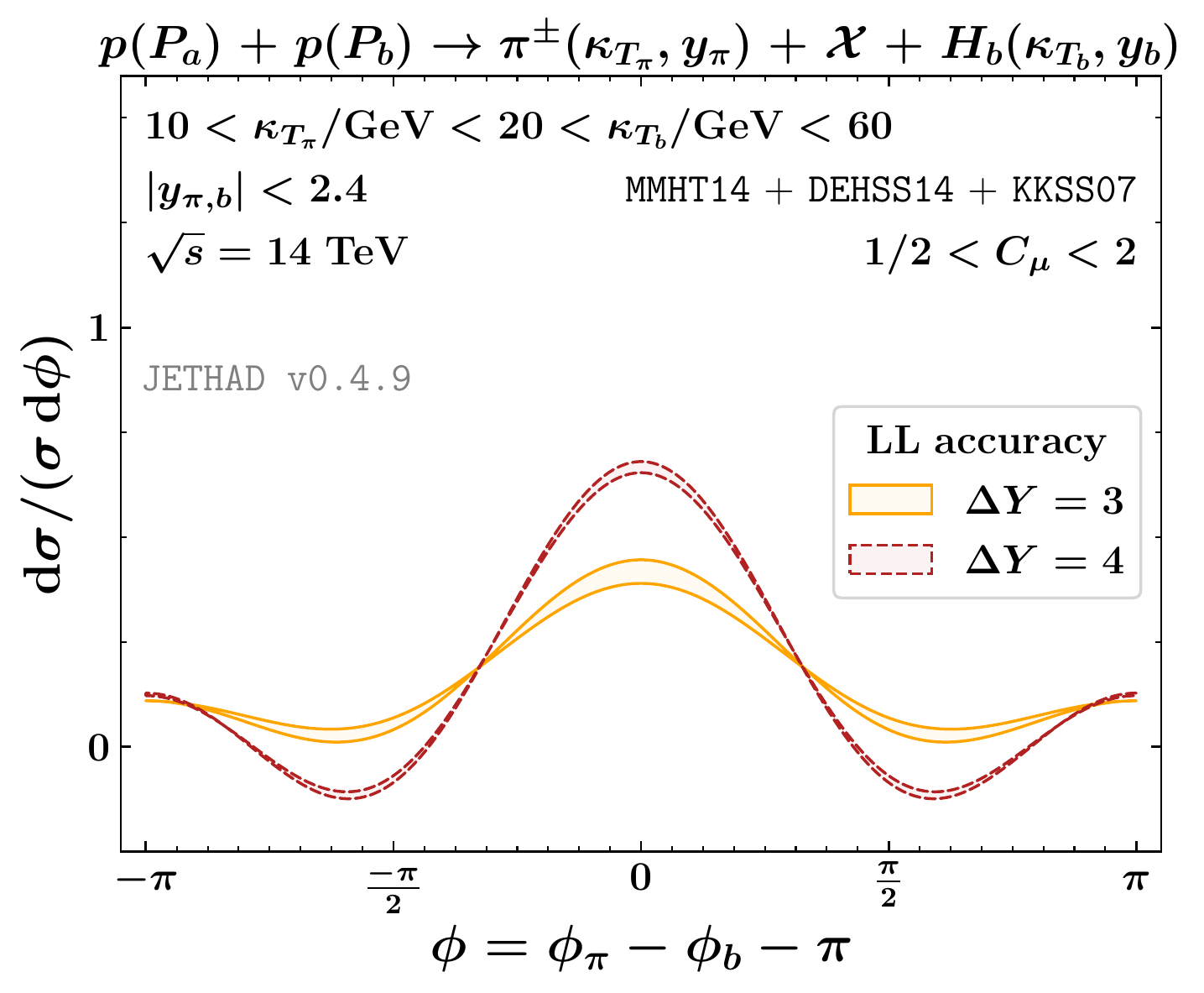}
   \includegraphics[scale=0.53,clip]{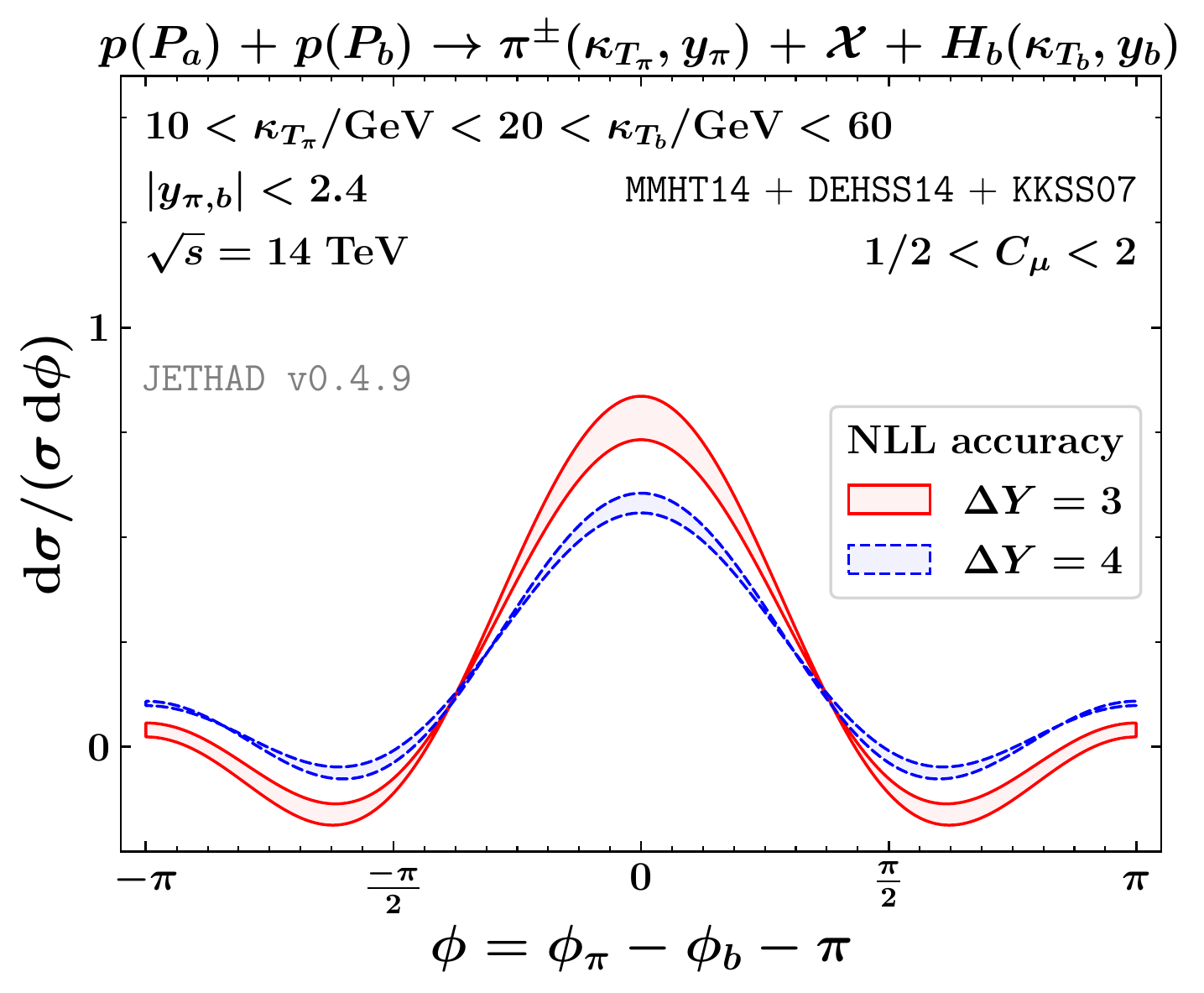}

   \includegraphics[scale=0.53,clip]{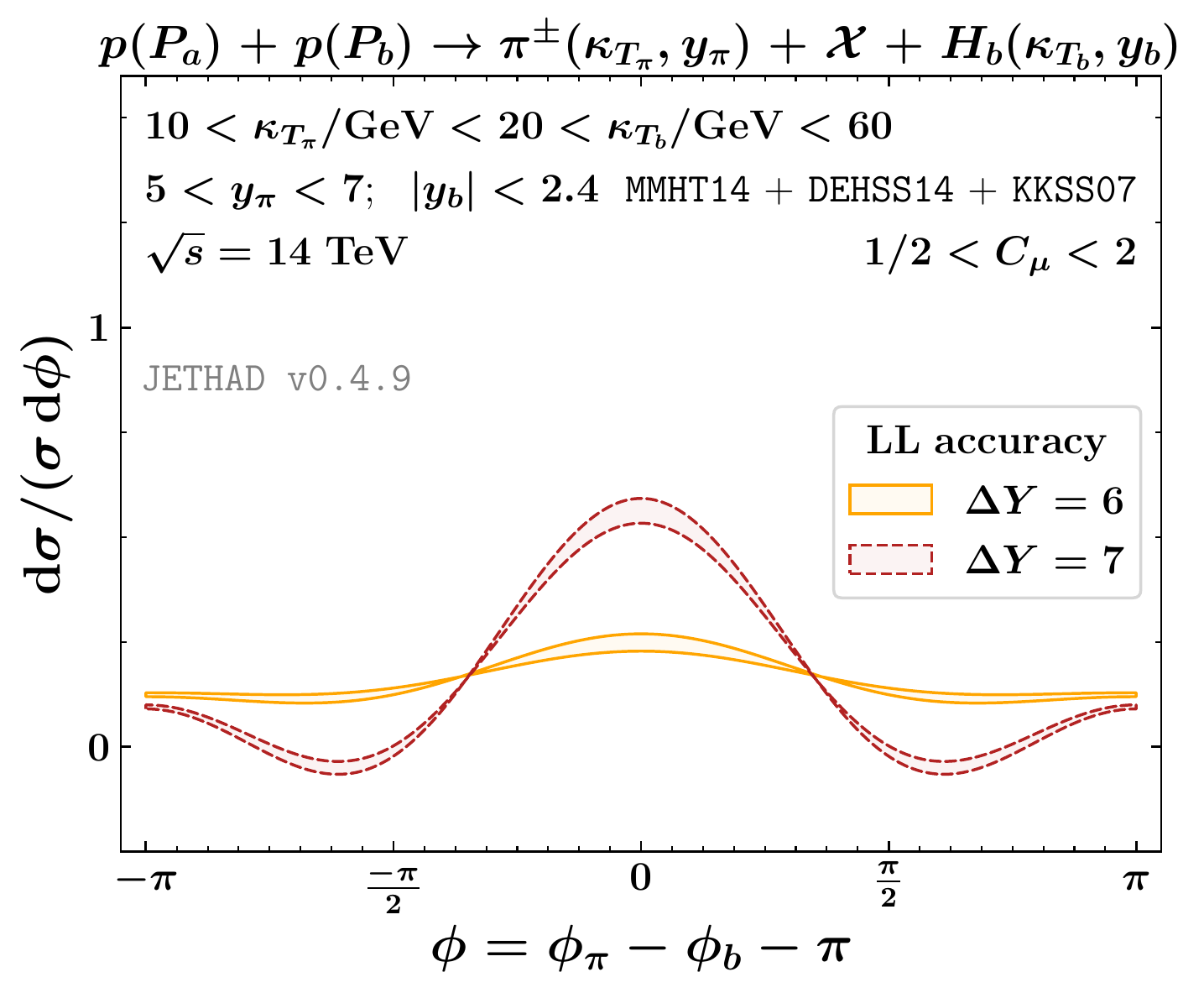}
   \includegraphics[scale=0.53,clip]{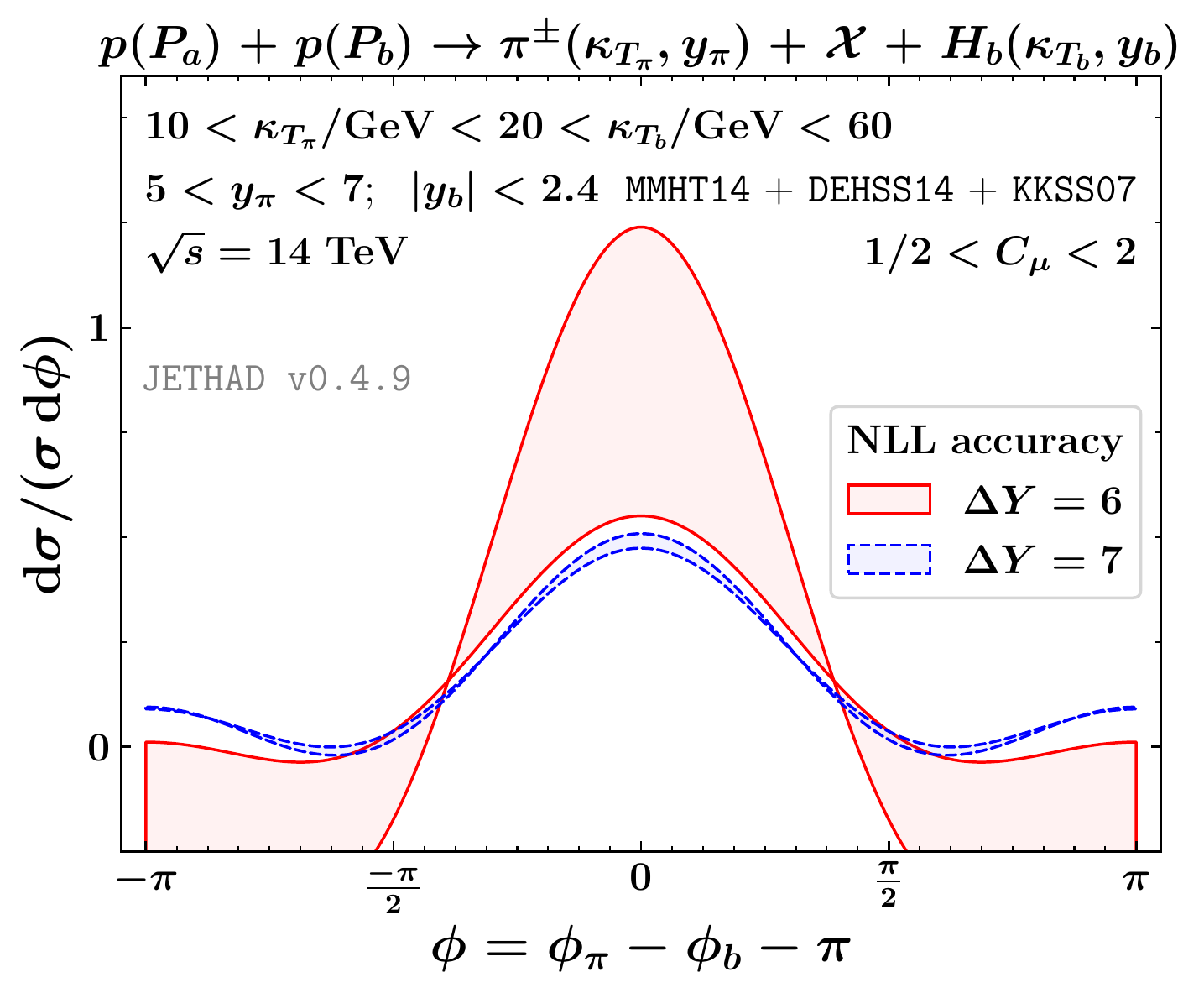}

\caption{Azimuthal distribution for the inclusive $\pi^\pm$~$+$~heavy-flavor production at the LHC, for $\sqrt{s} = 14$ TeV. Predictions for the $\pi^\pm$~$+$~$\Lambda_c^\pm$ (LHC), the $\pi^\pm$~$+$~$b$~hadron (LHC) and the $\pi^\pm$~$+$~$b$~hadron (FPF~$+$~ATLAS) channels are, respectively, presented in upper, central and lower panels. {\tt DEHSS14} collinear FFs are employed in the description of $\pi^\pm$ production. Text boxes inside plots show final-state kinematic ranges. Uncertainty bands encode the net effect of the scale variation and the multi-dimensional integration over the final-state phase space.}
\label{fig:phi_pi_DKh10}
\end{figure*}

\begin{figure*}[!pht]

   \includegraphics[scale=0.53,clip]{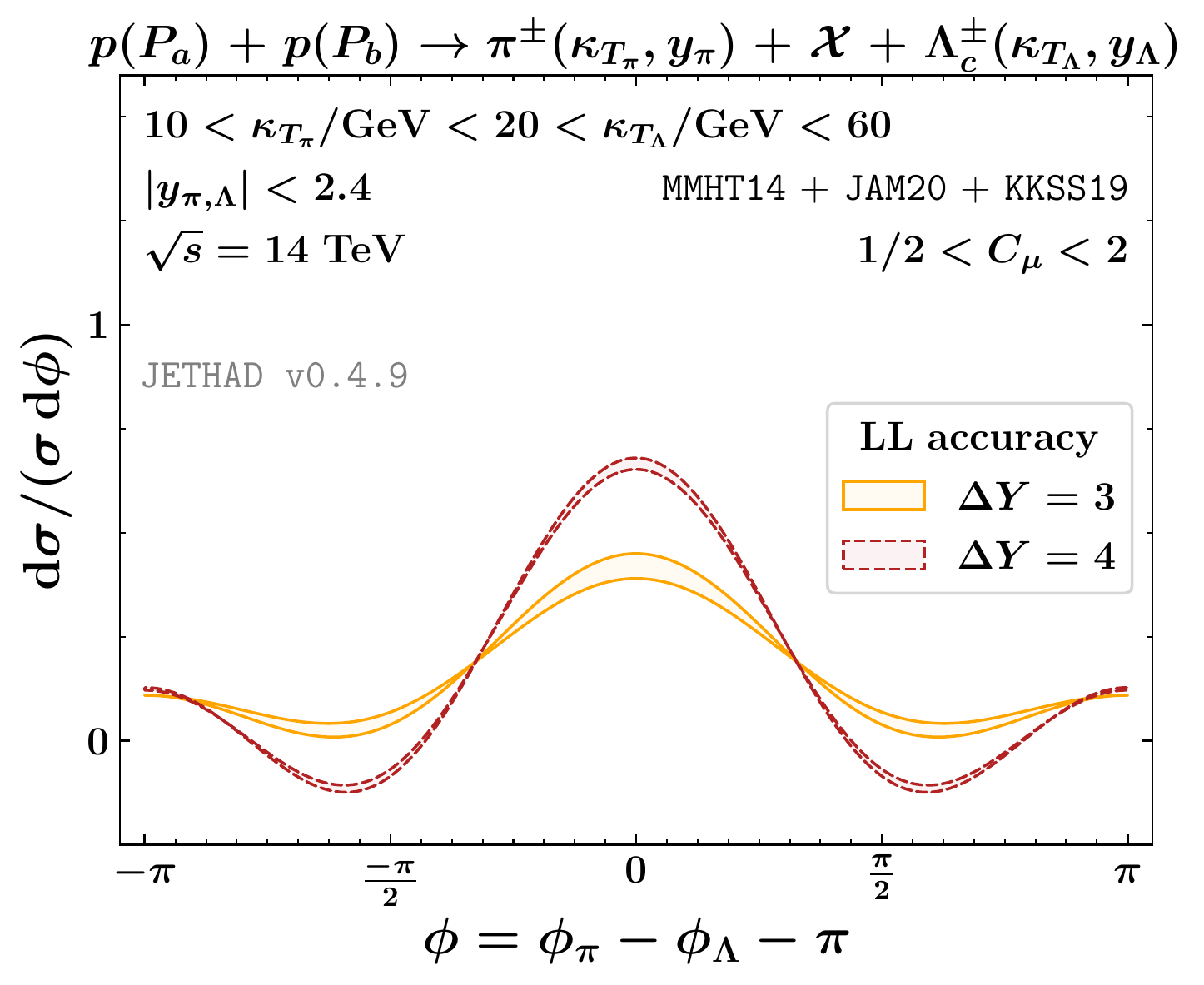}
   \includegraphics[scale=0.53,clip]{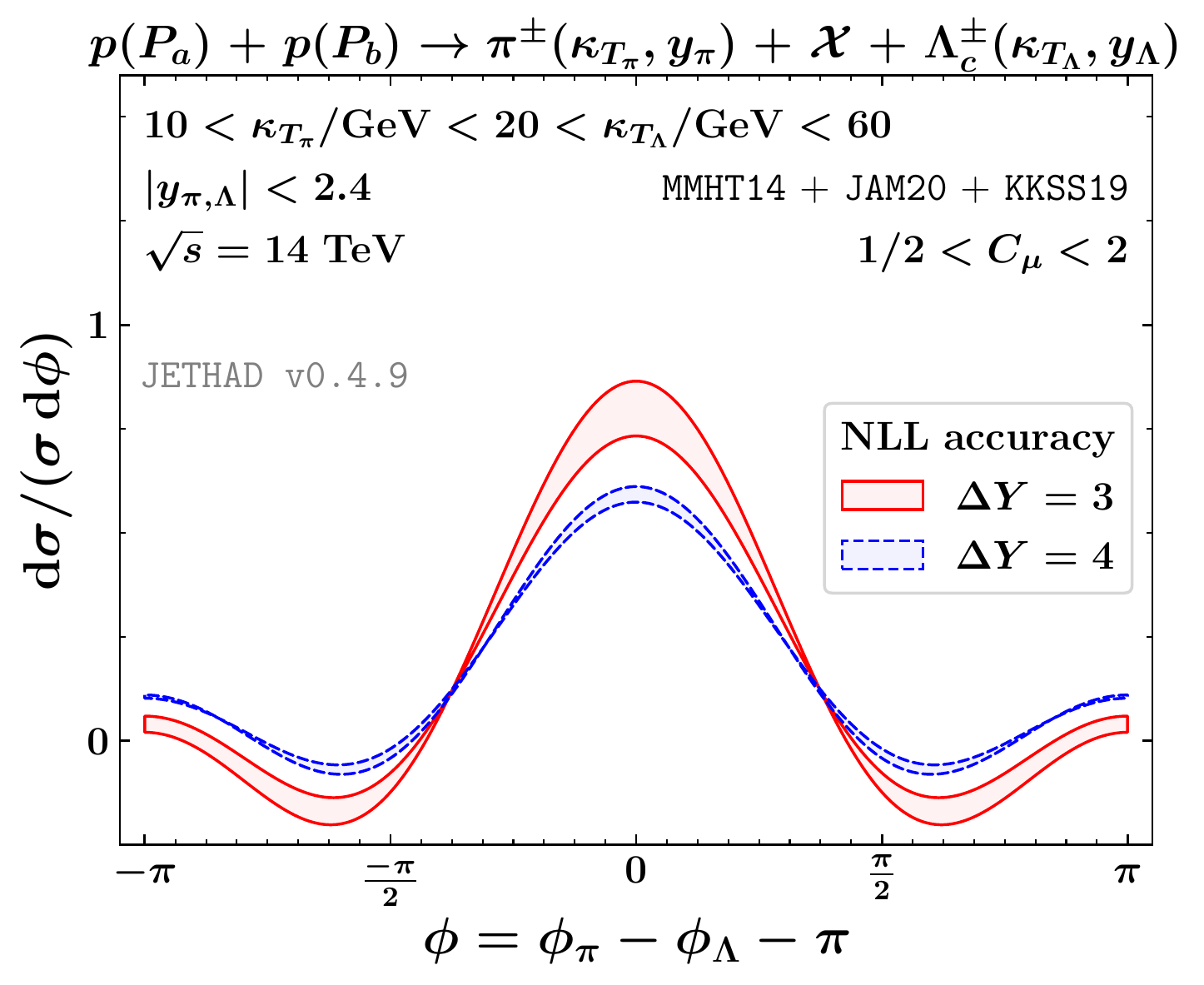}

   \includegraphics[scale=0.53,clip]{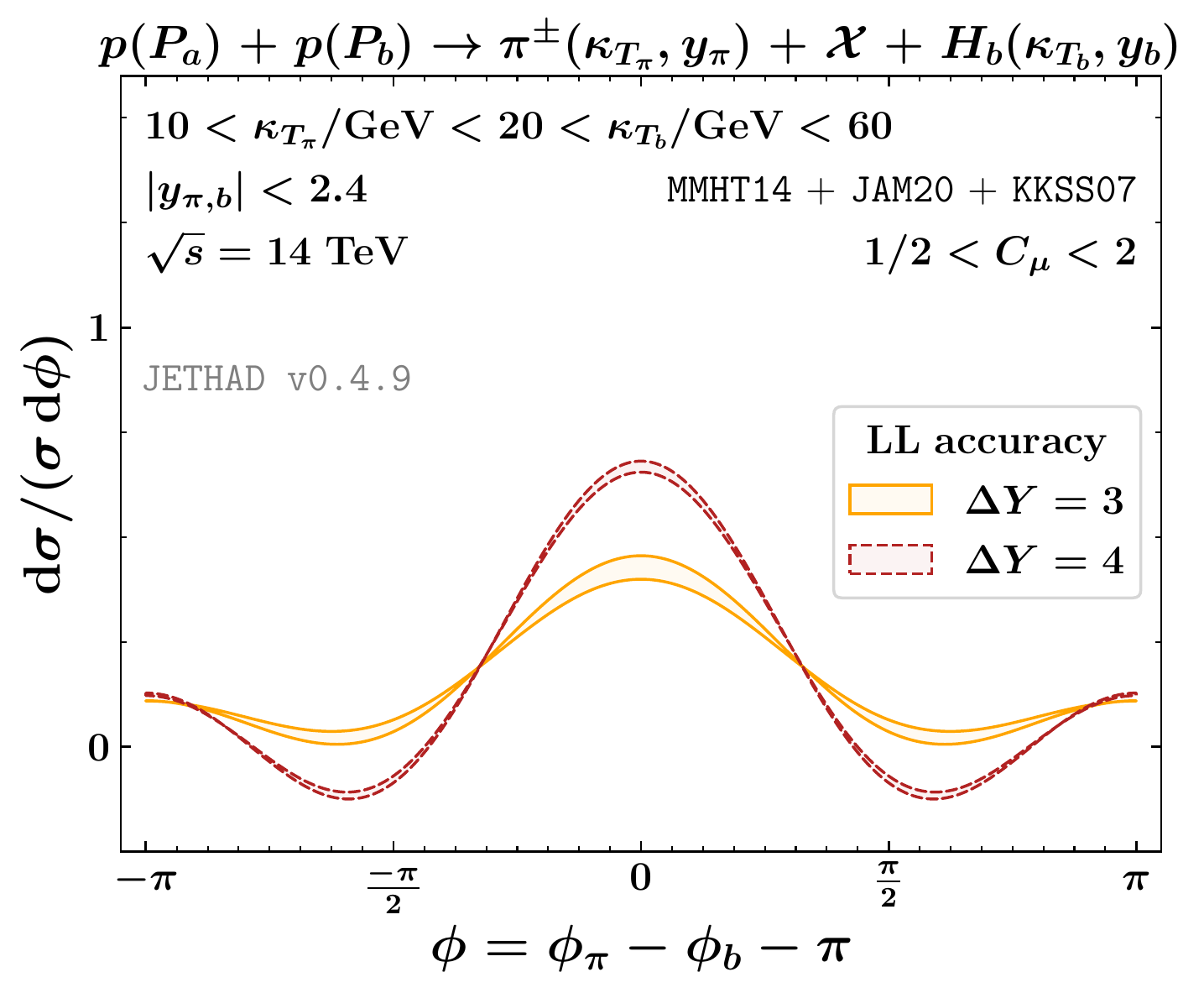}
   \includegraphics[scale=0.53,clip]{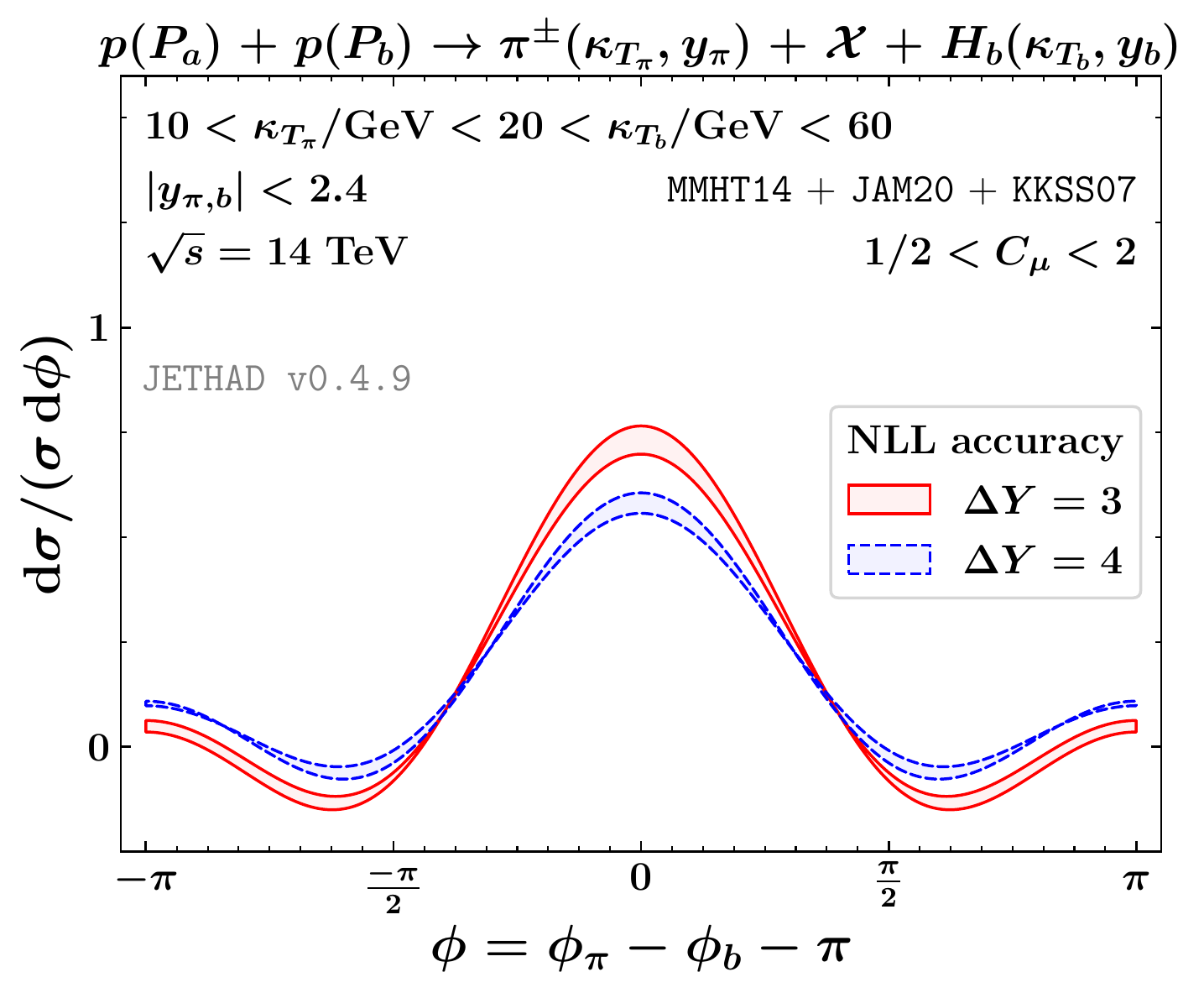}

   \includegraphics[scale=0.53,clip]{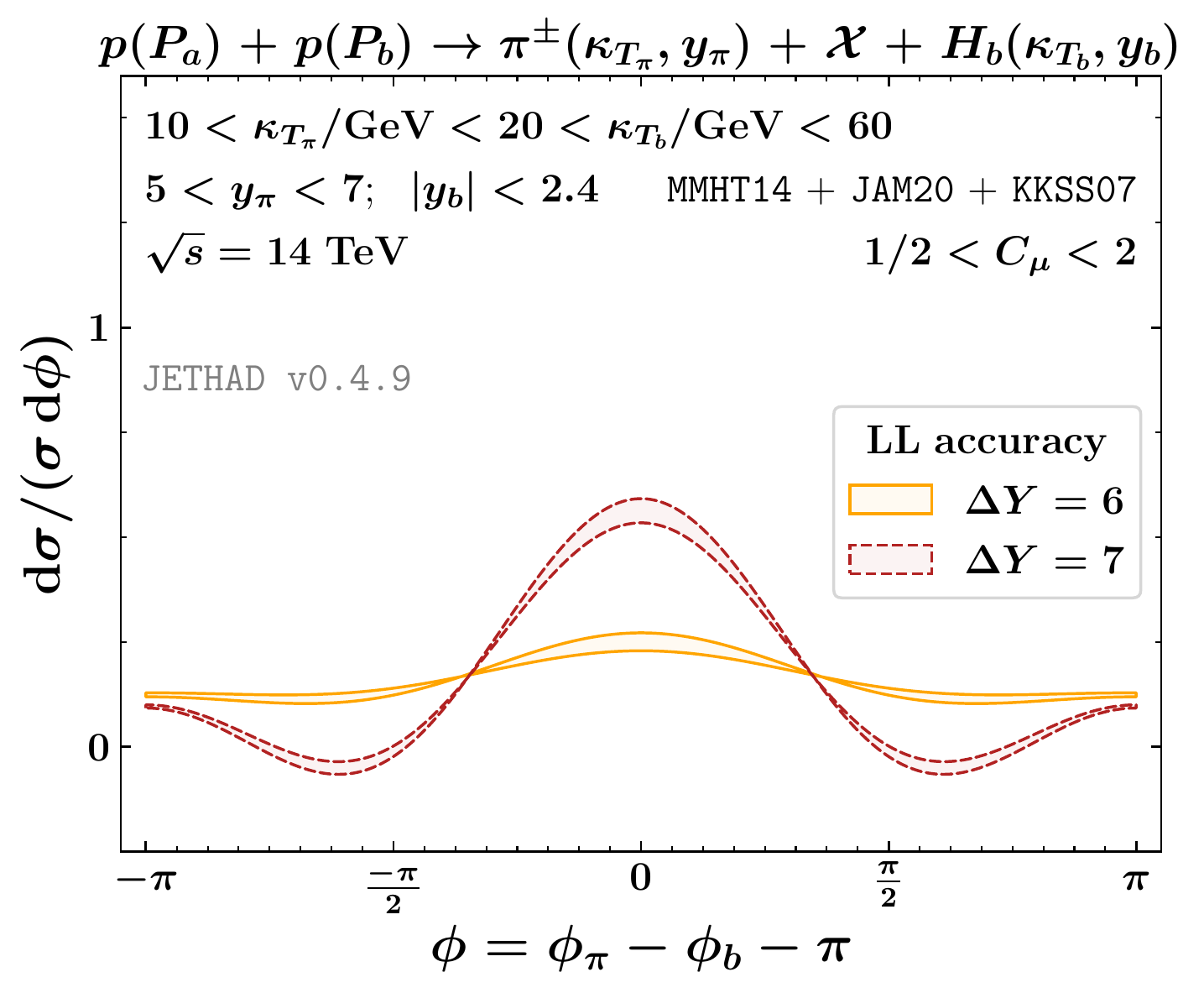}
   \includegraphics[scale=0.53,clip]{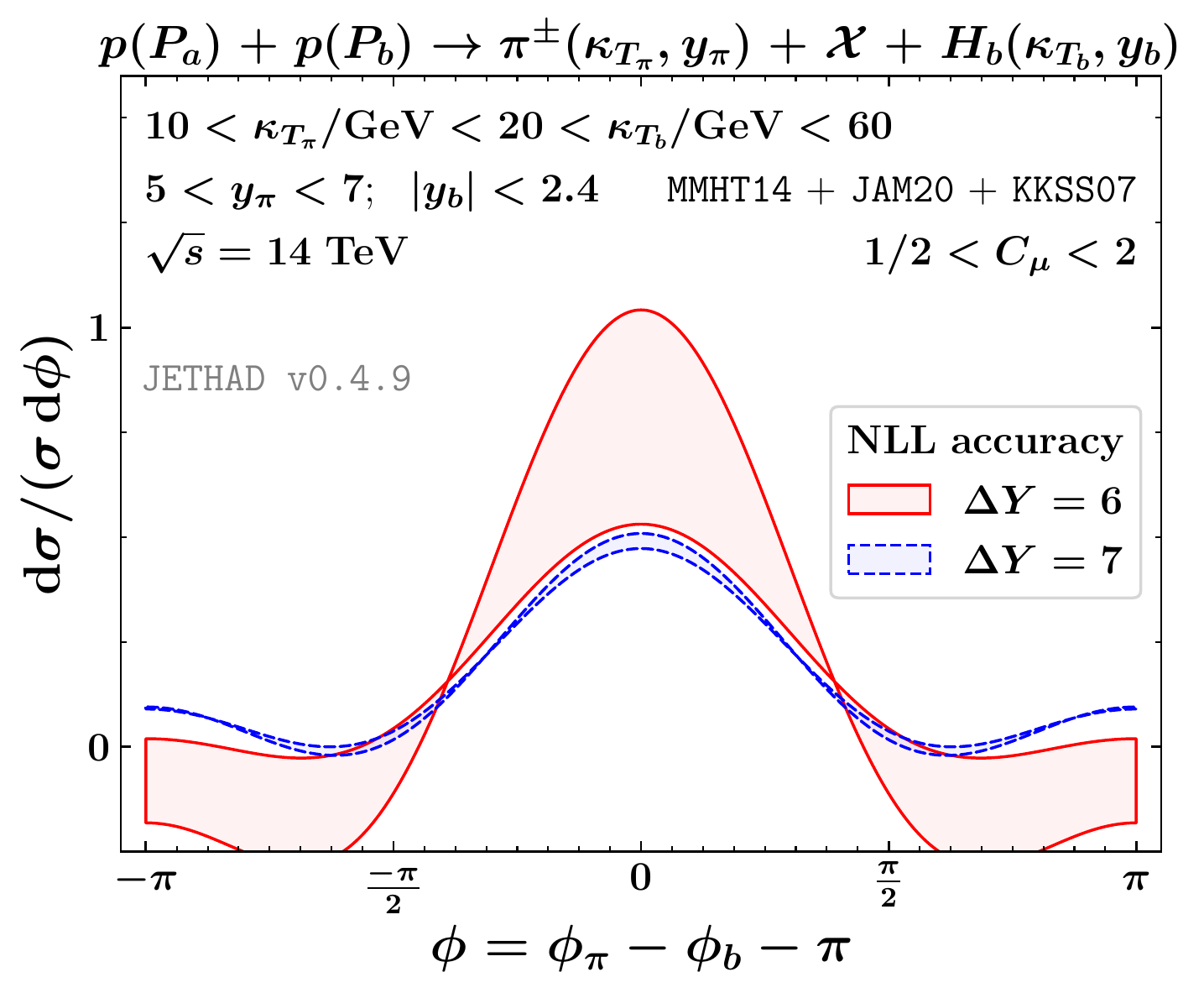}

\caption{Azimuthal distribution for the inclusive $\pi^\pm$~$+$~heavy-flavor production at the LHC, for $\sqrt{s} = 14$ TeV. Predictions for the $\pi^\pm$~$+$~$\Lambda_c^\pm$ (LHC), the $\pi^\pm$~$+$~$b$~hadron (LHC) and the $\pi^\pm$~$+$~$b$~hadron (FPF~$+$~ATLAS) channels are, respectively, presented in upper, central and lower panels. {\tt JAM20} collinear FFs are employed in the description of $\pi^\pm$ production. Text boxes inside plots show final-state kinematic ranges. Uncertainty bands encode the net effect of the scale variation and the multi-dimensional integration over the final-state phase space.}
\label{fig:phi_pi_JKh10}
\end{figure*}

\begin{figure*}[!pht]

   \includegraphics[scale=0.53,clip]{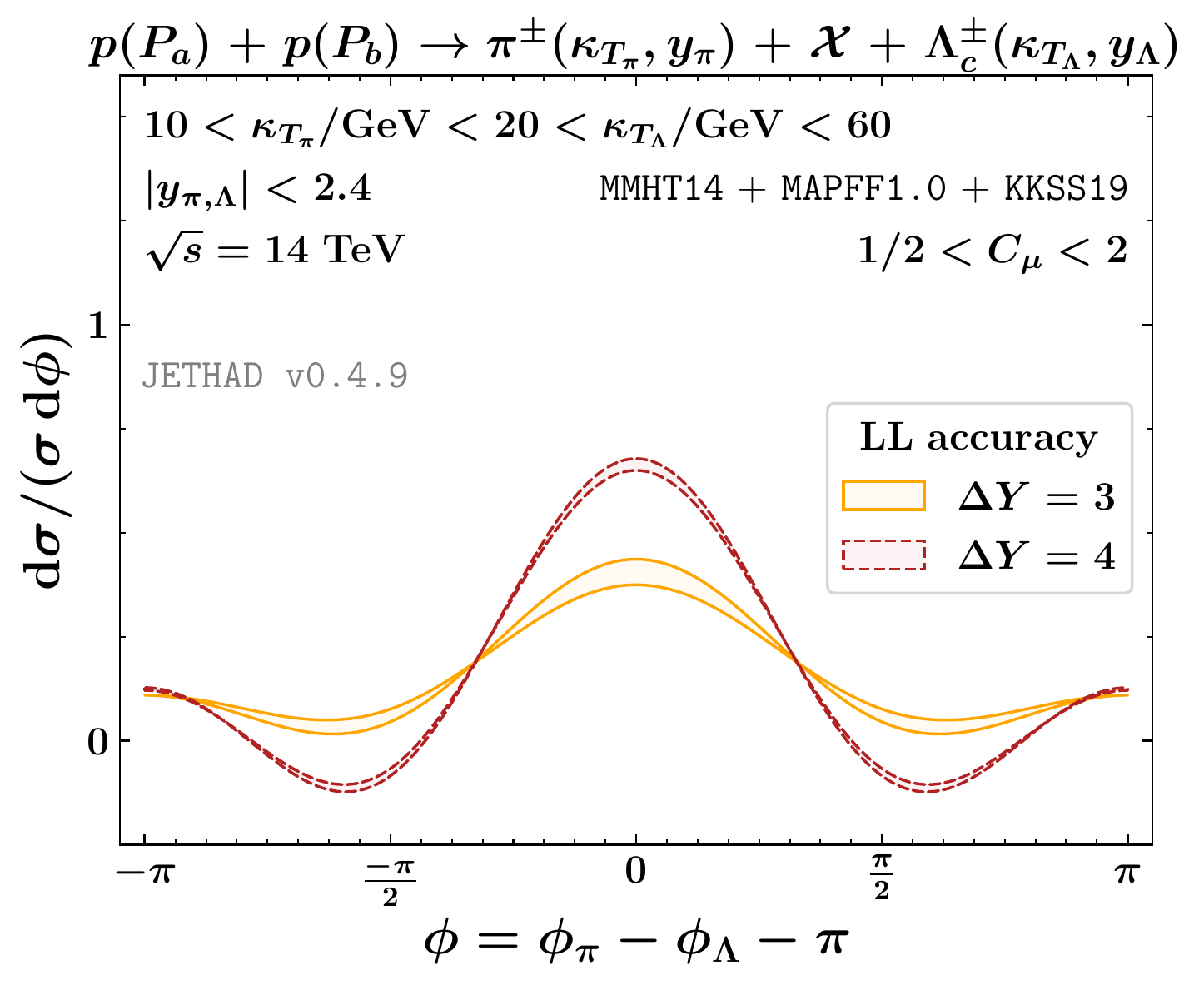}
   \includegraphics[scale=0.53,clip]{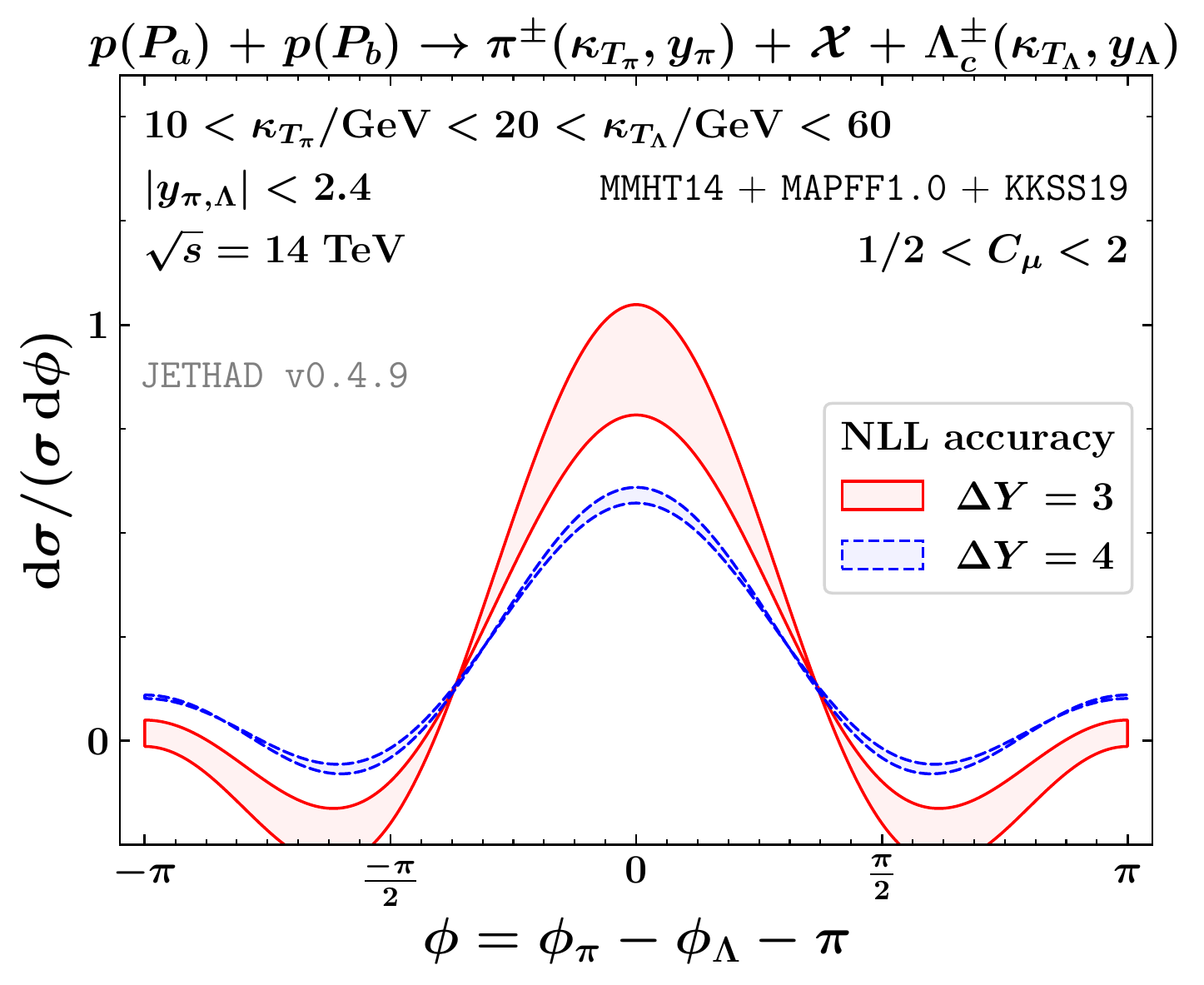}

   \includegraphics[scale=0.53,clip]{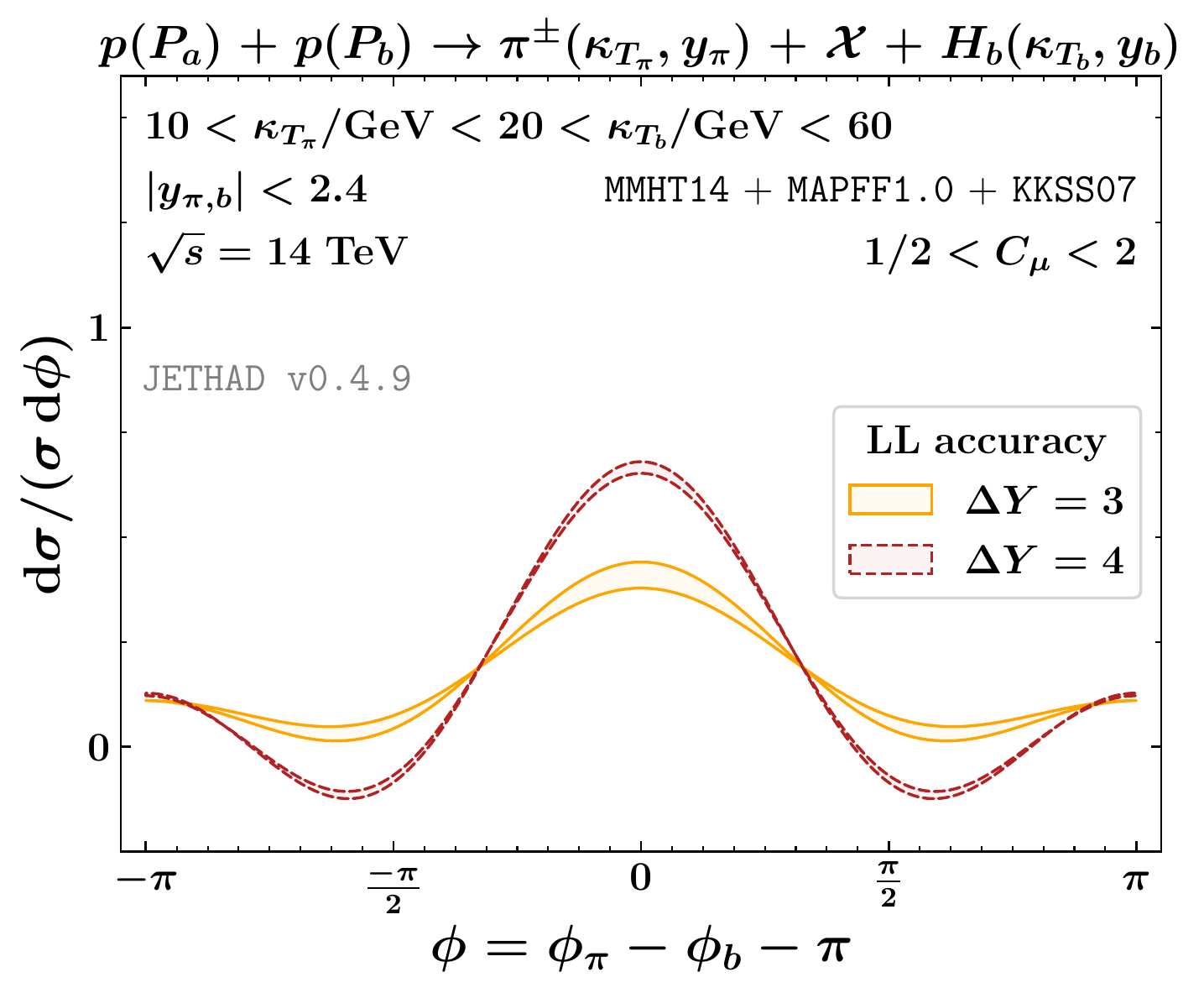}
   \includegraphics[scale=0.53,clip]{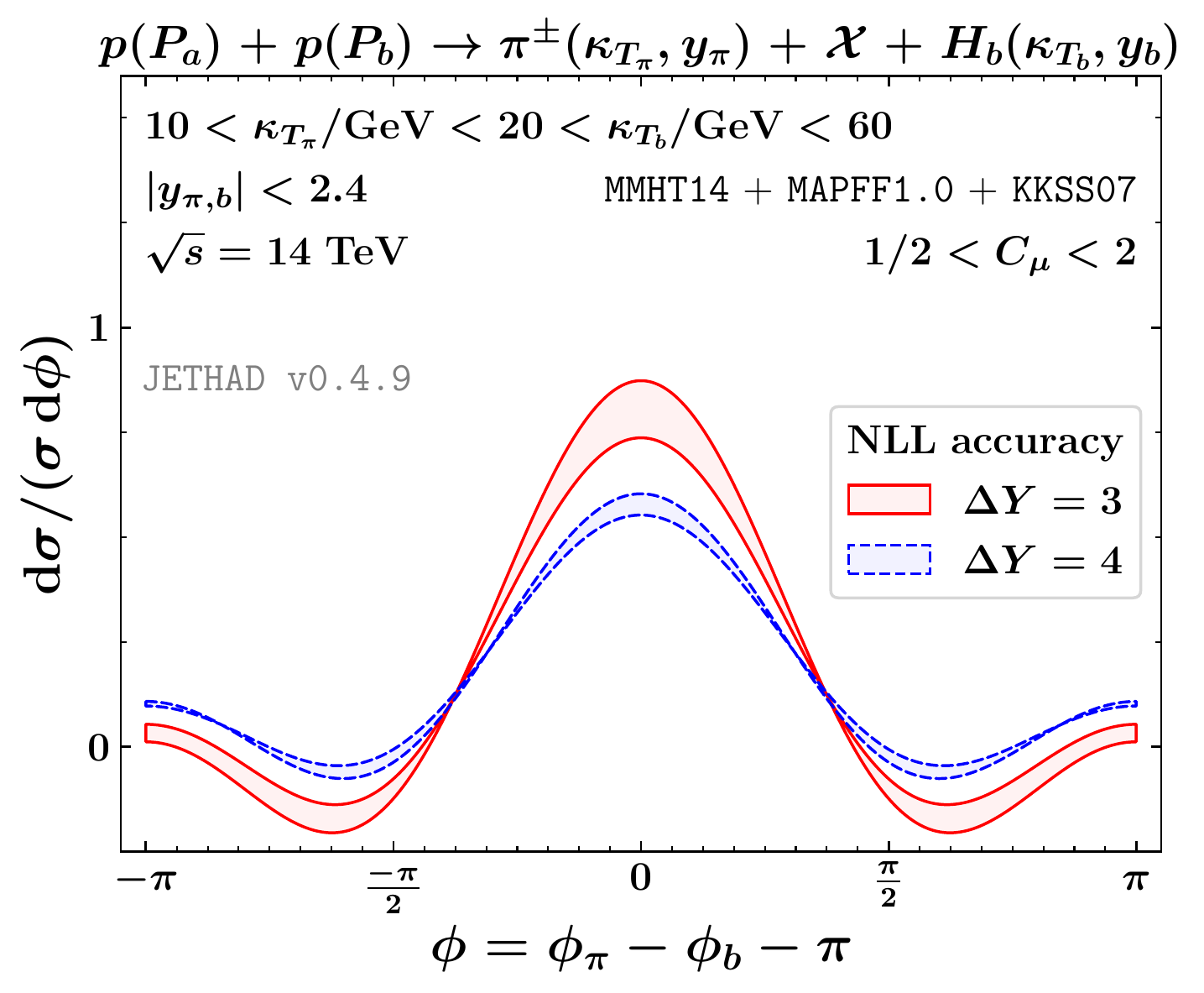}

   \includegraphics[scale=0.53,clip]{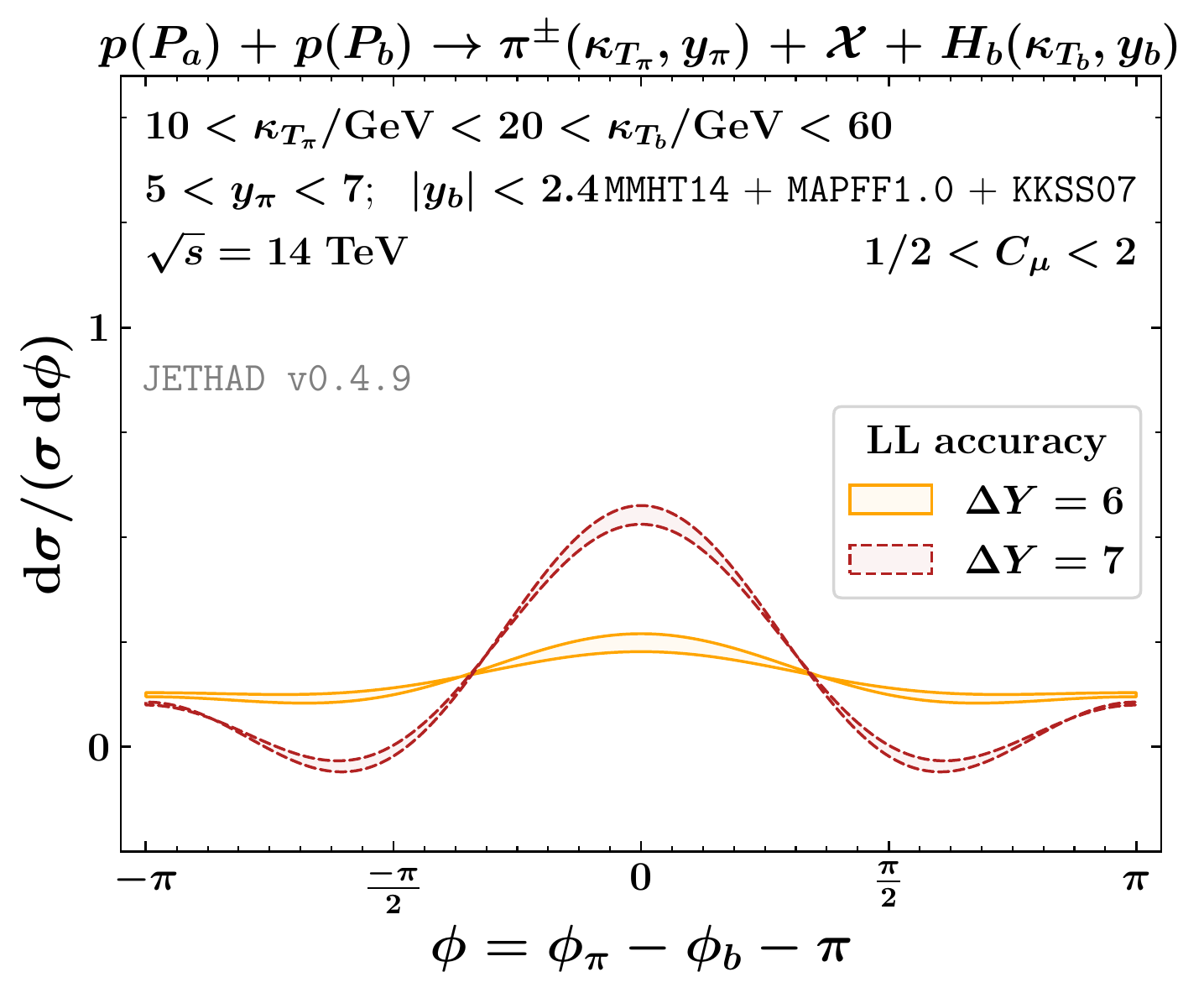}
   \includegraphics[scale=0.53,clip]{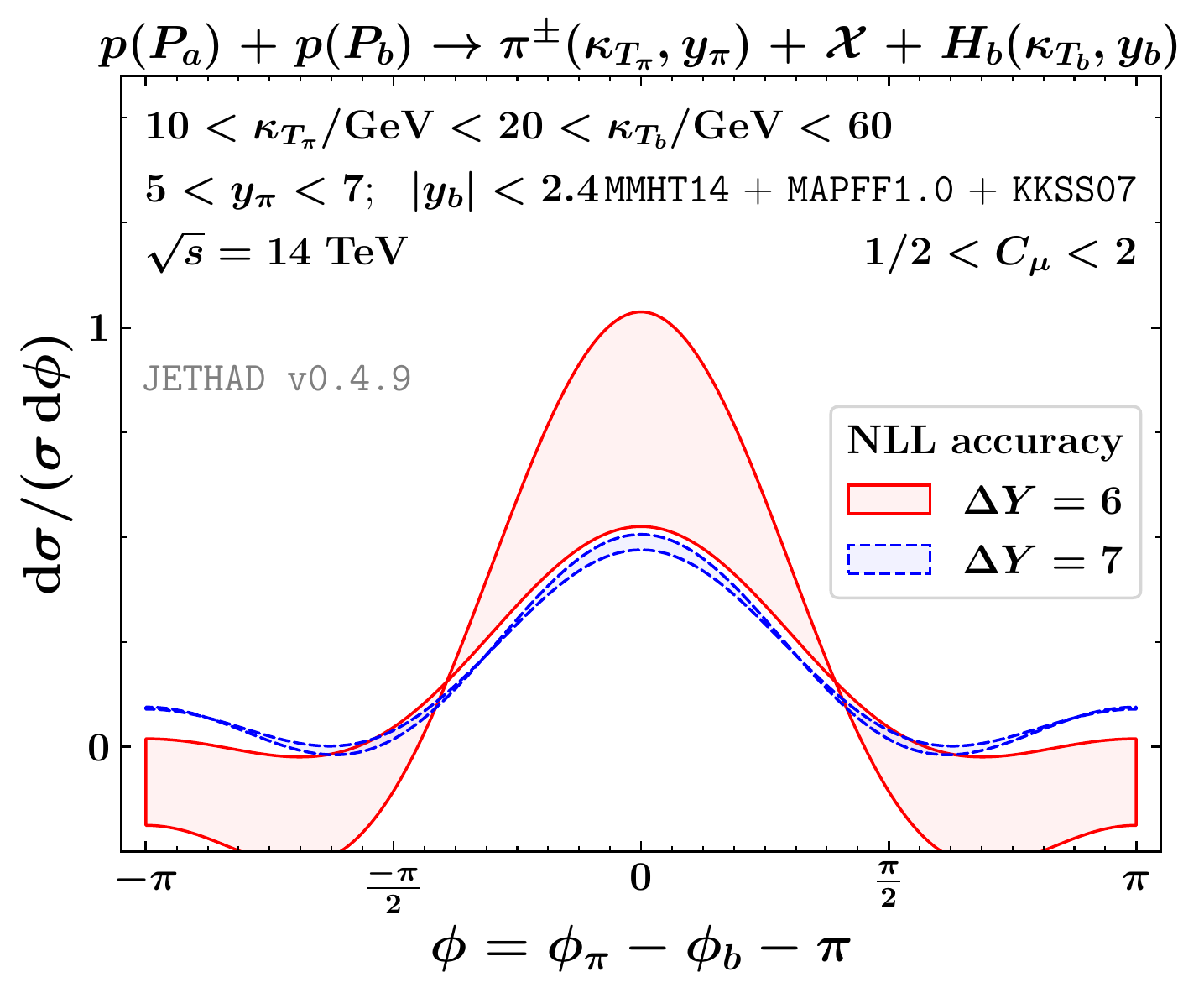}

\caption{Azimuthal distribution for the inclusive $\pi^\pm$~$+$~heavy-flavor production at the LHC, for $\sqrt{s} = 14$ TeV. Predictions for the $\pi^\pm$~$+$~$\Lambda_c^\pm$ (LHC), the $\pi^\pm$~$+$~$b$~hadron (LHC) and the $\pi^\pm$~$+$~$b$~hadron (FPF~$+$~ATLAS) channels are, respectively, presented in upper, central and lower panels. {\tt MAPFF1.0} collinear FFs are employed in the description of $\pi^\pm$ production. Text boxes inside plots show final-state kinematic ranges. Uncertainty bands encode the net effect of the scale variation and the multi-dimensional integration over the final-state phase space.}
\label{fig:phi_pi_MKh10}
\end{figure*}

\begin{figure*}[!pht]

   \includegraphics[scale=0.53,clip]{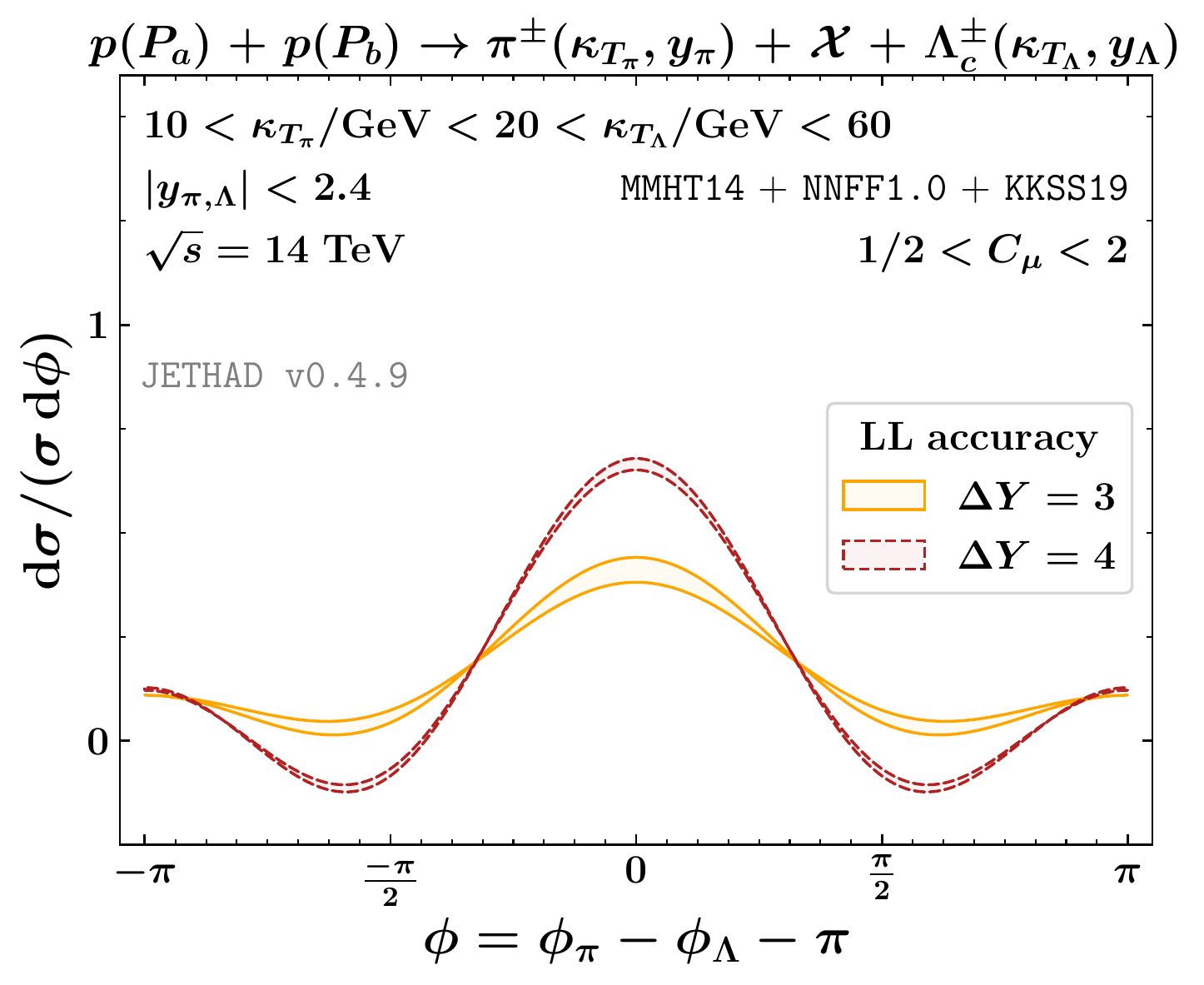}
   \includegraphics[scale=0.53,clip]{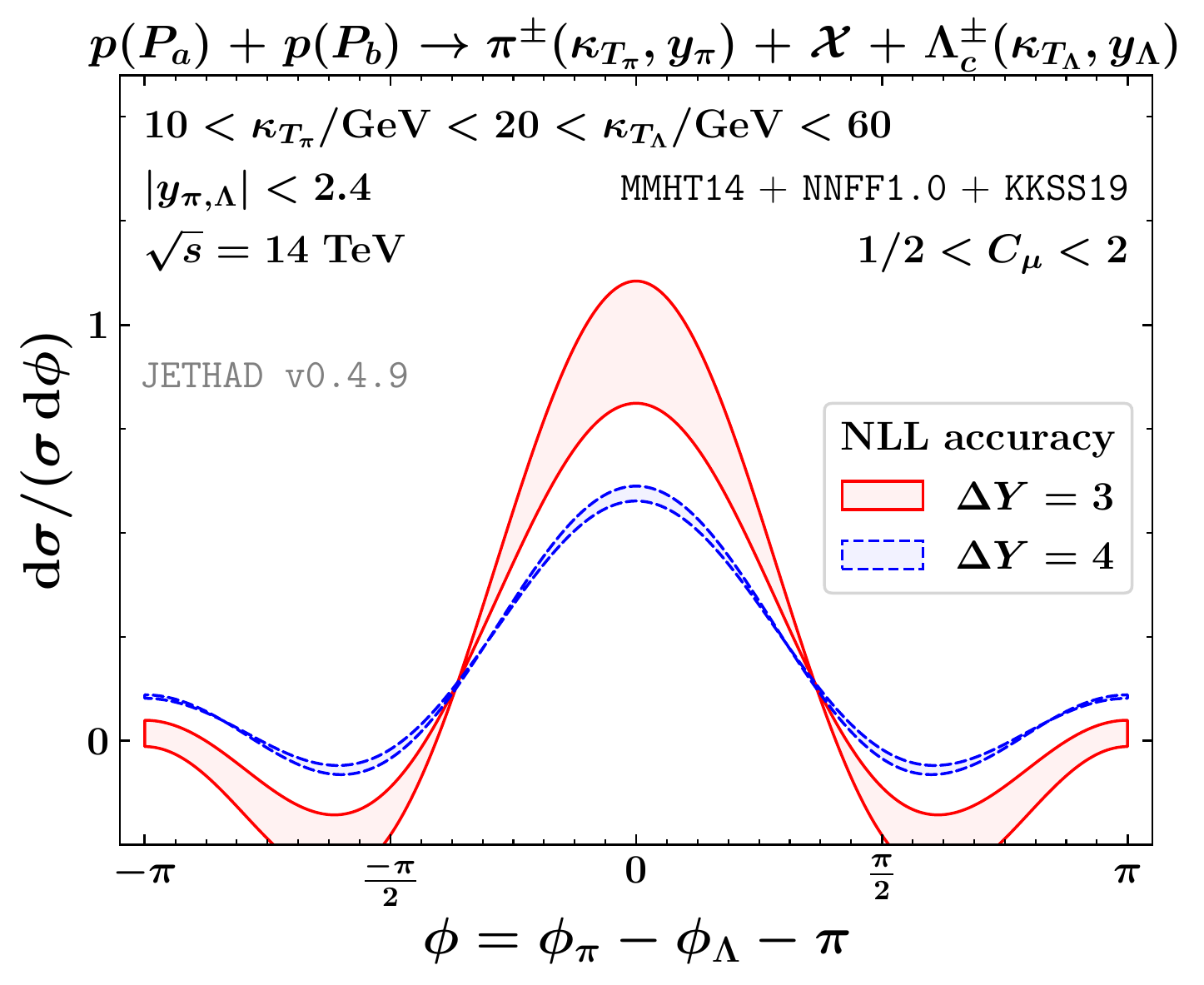}

   \includegraphics[scale=0.53,clip]{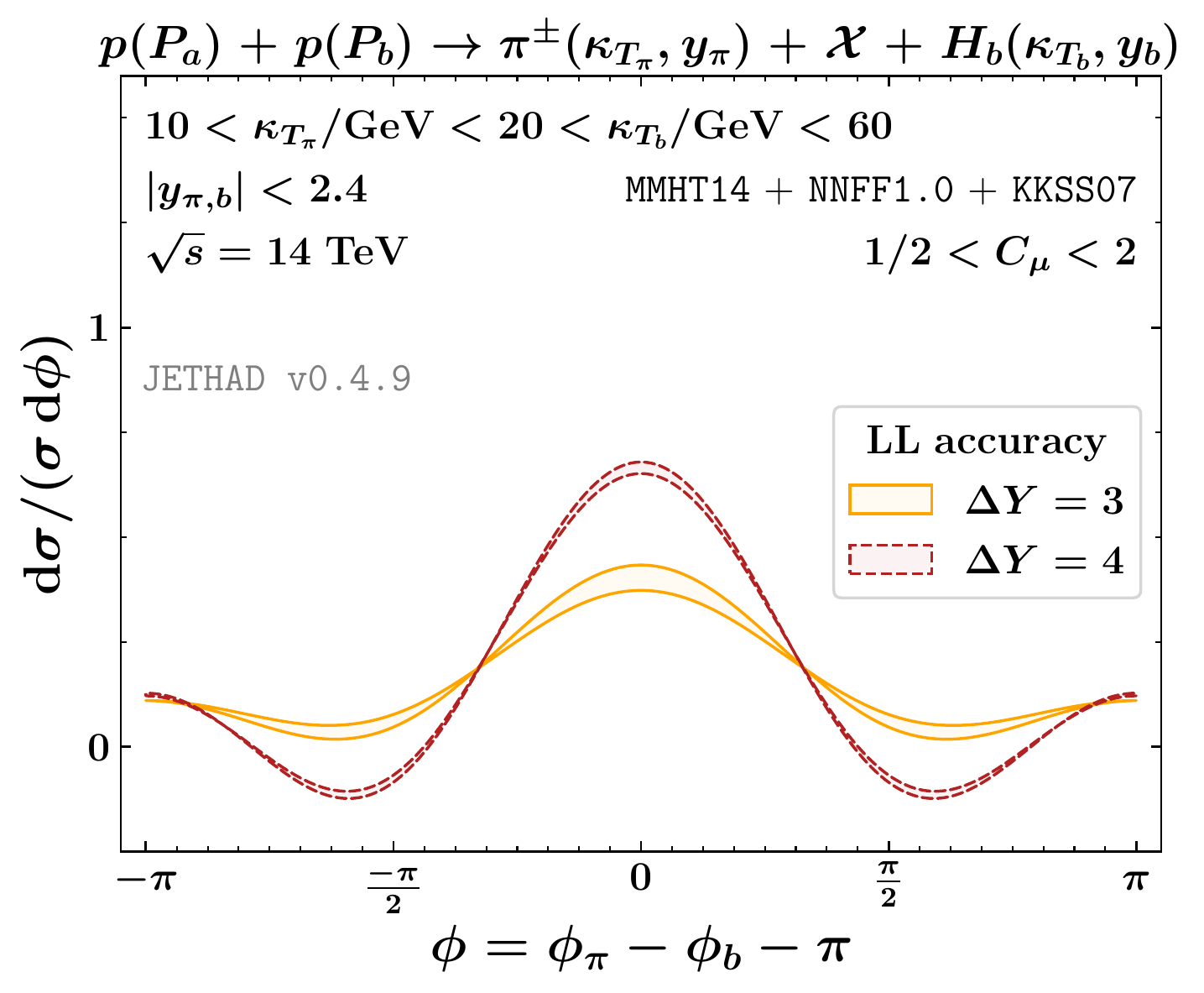}
   \includegraphics[scale=0.53,clip]{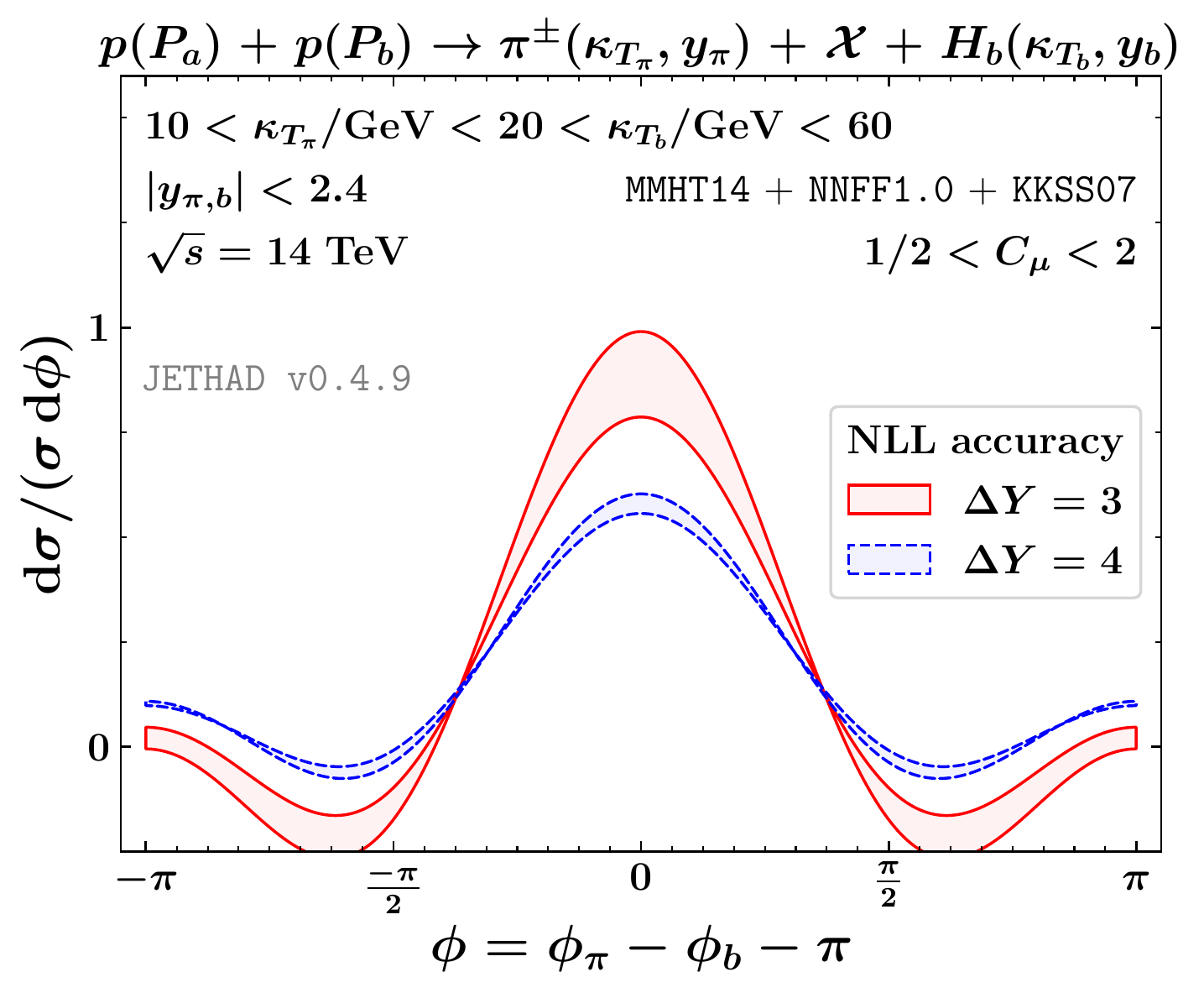}

   \includegraphics[scale=0.53,clip]{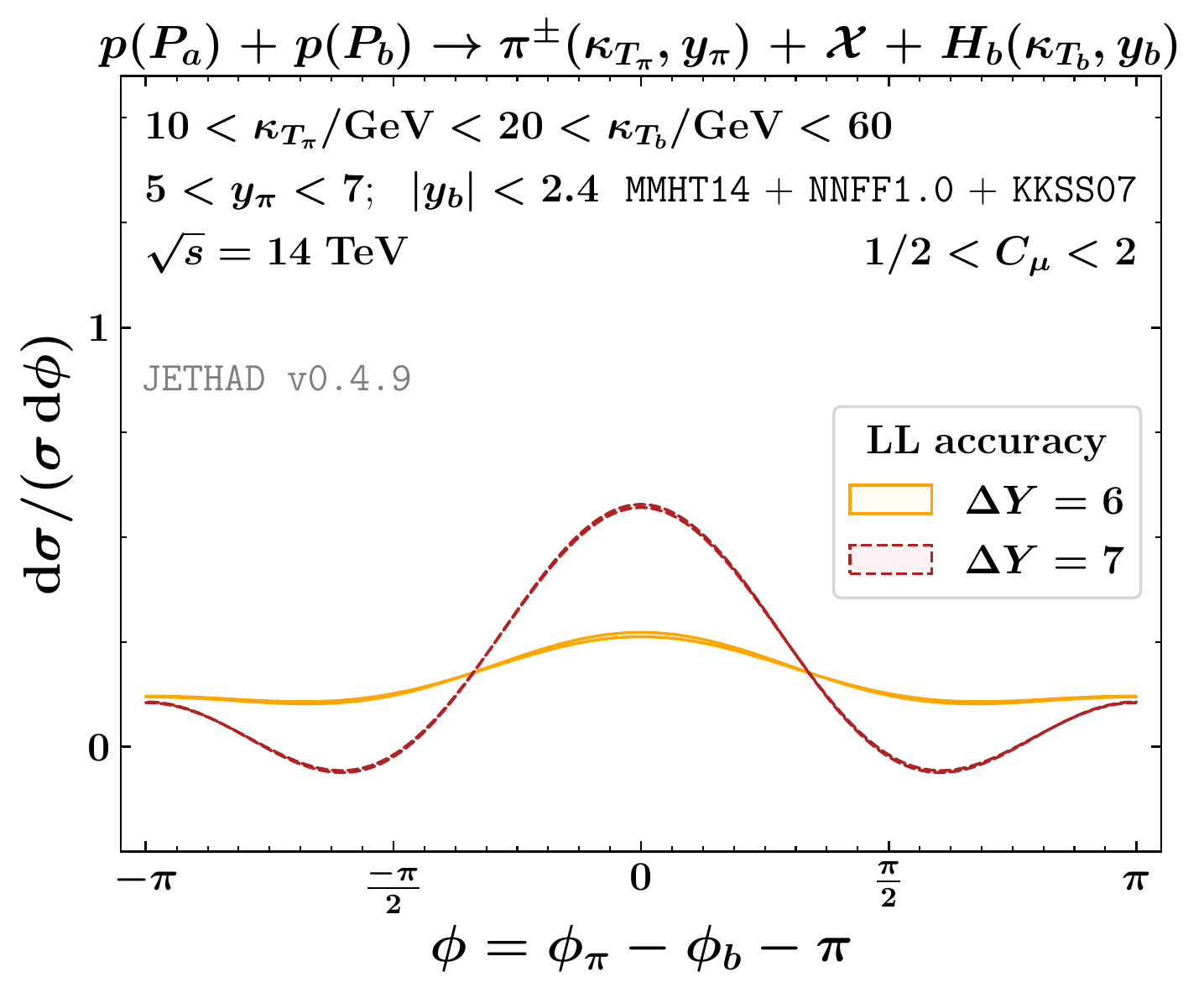}
   \includegraphics[scale=0.53,clip]{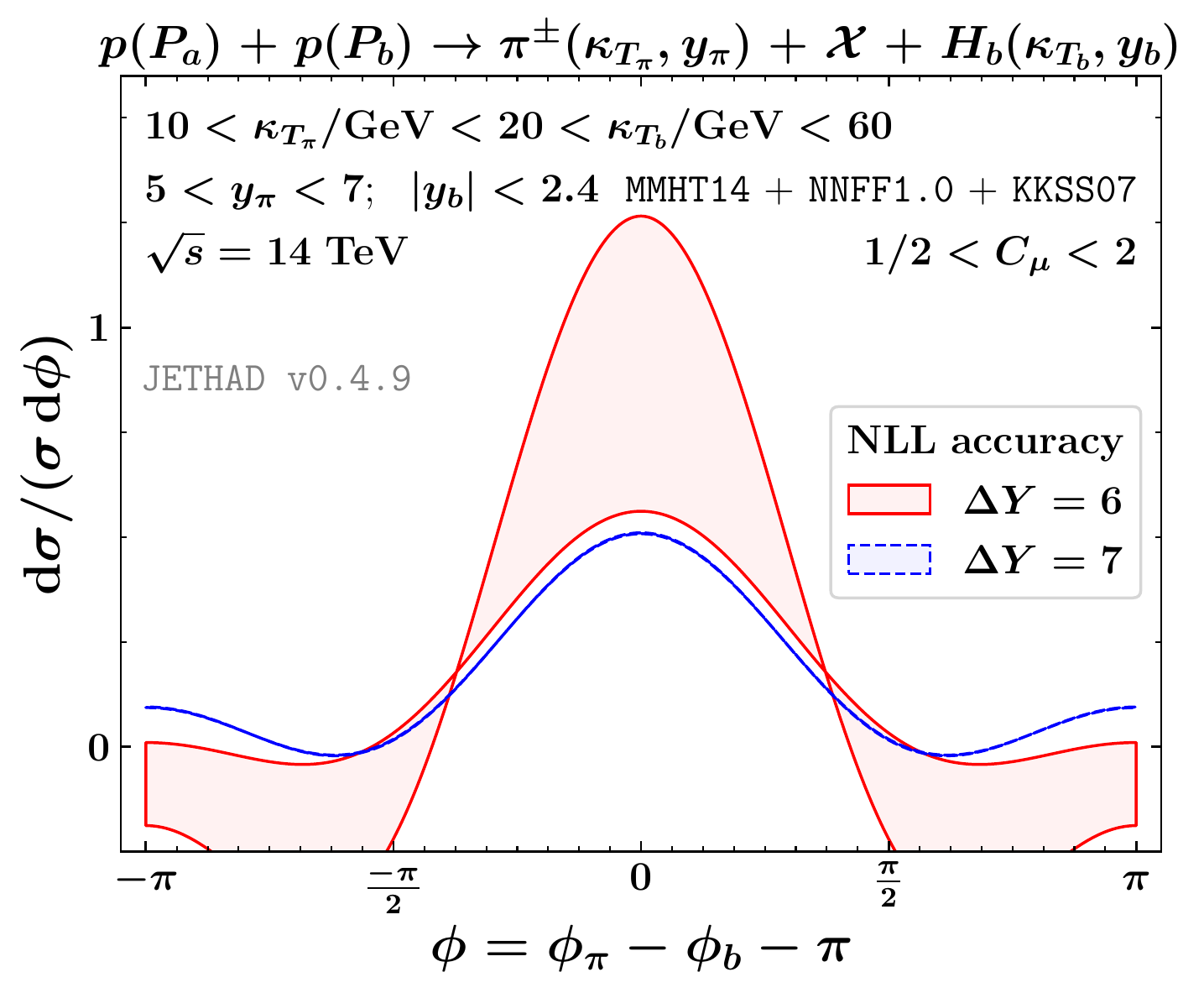}

\caption{Azimuthal distribution for the inclusive $\pi^\pm$~$+$~heavy-flavor production at the LHC, for $\sqrt{s} = 14$ TeV. Predictions for the $\pi^\pm$~$+$~$\Lambda_c^\pm$ (LHC), the $\pi^\pm$~$+$~$b$~hadron (LHC), and the $\pi^\pm$~$+$~$b$~hadron (FPF~$+$~ATLAS) channels are respectively presented in upper, central, and lower panels. {\tt NNFF1.0} collinear FFs are employed in the description of $\pi^\pm$ production. Text boxes inside plots show final-state kinematic ranges. Uncertainty bands encode the net effect of the scale variation and the multi-dimensional integration over the final-state phase space.}
\label{fig:phi_pi_NKh10}
\end{figure*}

Predictions for the azimuthal distribution as a function of $\phi$ and for two distinct sets of values for the rapidity interval, $\DY = 3, 4$ (LHC) or $\DY = 6, 7$ (FPF~$+$~ATLAS), are presented in Figs.\tref{fig:phi_eta} to\tref{fig:phi_pi_NKh10}.
Plots are organized as follows. Upper, central and lower panels of all figures refer to final states featuring the emission of a given light meson plus a $\Lambda_c^\pm$ at the LHC, a $b$~hadron at the LHC, and a $b$~hadron at FPF~$+~{\rm ATLAS}$, respectively. Left (right) panels contains LL (NLL) distributions. Figure\tref{fig:phi_eta} refers to $\eta$ channels, while the remnant Figs.\tref{fig:phi_pi_DKh10}~to\tref{fig:phi_pi_NKh10} show results for $\pi^\pm$ channels with {\tt DEHSS14}, {\tt JAM20}, {\tt MAPFF1.0}, and {\tt NNFF1.0} pion collinear FFs.

Preliminary studies on our distributions have shown no significant sensitivities to the uncertainties listed in points (\emph{ii}) to (\emph{vi}) of Section\tref{ssec:jethad_settings}. More in particular, the uncertainty coming from the choice of different members inside the same set of pion FFs is negligible,\footnote{A similar result was recently obtained in a preliminary analysis on azimuthal distributions for the inclusive light meson plus $D^{*\pm}$ at FPF~$+$~ATLAS, see Fig.~6.16 of Ref.\tcite{Feng:2022inv}.} while the one related to the selection of a distinct set has a more evident effect.
Therefore, shaded bands in our plots exhibit the combined effect energy-scale variation and numerical multi-dimensional integration over the final-state phase space. Results for different pion FF parametrizations are presented separately.

The common trend to all the presented azimuthal distributions is the emergence of a peak centered at $\phi = 0$, which corresponds to the physical configuration where the light meson and the heavy hadron are emitted (almost) back to back.
Due to high-energy dynamics, the NLL peak height decreases when $\DY$ grows, whereas its width expands. This stems from the increase with $\DY$ of the number of secondary gluons (the $\cal X$ system in Eq.\eref{process} and in labels of our plots) emitted with a strong separation in rapidity, as predicted by the BFKL equation. Thus, the correlation in the azimuthal plane between the two tagged particles lowers, and the number of almost back-to-back events diminishes.

Notably, LL distributions exhibit an opposite pattern, namely the peak grows and its width comes down with $\DY$.
Here, the connection between the strongly asymmetric transverse-momentum windows at which the light meson and the heavy particle are detected (see Section\tref{ssec:kinematics}) and the corresponding longitudinal-momentum fractions brings to a reduction of $y_{\eta,\pi}$ and $y_{\Lambda,b}$ combinations for the given $\DY$.
This translates to a recorrelation pattern in the azimuthal plane for LL distributions when $\DY$ grows, which turns out to be unphysical since, as previously mentioned, one of the main imprints of the high-energy resummation is the loss of correlation due to the weight of rapidity-ordered gluons forming the inclusive system $\cal X$.
The correct behavior is recovered when full NLL corrections are considered.

Furthermore, we note that the size of uncertainty bands due to scale variations shrinks when $\DY$ increases. This effect is particularly pronounced in the FPF~$+$~ATLAS configuration. Distributions taken at the lower reference values of the rapidity interval, $\DY = 3$ (LHC) or $\DY = 6$ (FPF~$+$~ATLAS), feature two symmetric minima at $|\phi| \gtrapprox \pi/2$ which reach unphysical values beyond zero, whereas no negative values are observed for larger $\DY$~values.
This represents a clear indication that the \emph{natural stability} of the high-energy series gets more and more evident in the large rapidity-interval regime, as expected.
A possible technical explanation for the origin of these negative values could rely on the fact that higher azimuthal modes in the sum of Eq.\eref{azimuthal_distribution} have a larger absolute size at lower $\DY$-values, thus enhancing the oscillating pattern of our distributions.

From a qualitative comparison of results in Figs.\tref{fig:phi_eta} to\tref{fig:phi_pi_NKh10} with corresponding ones previously studied in other semi-hard channels, we observe that novel features emerge. Light-meson plus heavy-flavor $\phi$-distributions are much less peaked than vector quarkonium plus jet ones (see Fig.~6 of Ref.\tcite{Celiberto:2022dyf}). They appear more similar to light-hadron or jet ones (see Figs.~17~and~18 of Ref.\tcite{Celiberto:2020wpk}). Furthermore, the symmetric minima observed in panels of this Section have no analogs in the mentioned studies.

The general message coming out from the inspection of results shown in this Section is the feasibility of studying azimuthal distributions for light-meson plus heavy-flavor production reactions by means of our hybrid factorization around natural values of $\mu_R$ and $\mu_F$ scales.
A \emph{natural stabilization} of the high-energy resummation emerges and its effect becomes significant in its expected applicability domain, namely the large $\DY$~sector.
Azimuthal distributions can be easily measured at current LHC experimental configurations as well as in future analyses doable via the FPF~$+$~ATLAS coincidence methods.
This will heighten our chances of making stringent tests of the high-energy QCD dynamics.

\section{Towards precision studies of high-energy QCD?}
\label{sec:conclusions}

By making use of the hybrid high-energy and collinear factorization, we have studied the inclusive detection of a light-flavored hadron, a $\eta$ meson or a pion, in association with a heavy-flavored particle, a $\Lambda_c$ baryon or a bottomed hadron, at current LHC energies and kinematics configurations, as well as at the ones accessible via a FPF~$+$~ATLAS tight timing-coincidence setup\tcite{Anchordoqui:2021ghd,Feng:2022inv}.

Our analysis on distributions differential in the observed rapidity interval ($\DY$) or in the azimuthal-angle distance ($\phi$) between the two tagged objects has confirmed that the remarkable property of \emph{natural stability} of the high-energy resummation, recently discovered in the context of heavy-flavor studies in forward directions\tcite{Celiberto:2021dzy,Celiberto:2021fdp,Celiberto:2022dyf}, is at work and allows for a fair description of the considered observables around the natural values of energy scales provided by process kinematics.
This represents a required condition and a first step towards precision studies of high-energy QCD via inclusive heavy-light di-hadron system emissions in proton collisions.

We have provided evidence that $\DY$-distributions are able to disengage the high-energy signal from the fixed-order background. Furthermore, they are valid probe channels to assess the weight of the uncertainty connected to collinear FFs in ranges that are complementary to the currently accessible ones.
Conversely, $\phi$-distributions have shown a solid stability in the large $\DY$ regime. Their study is helpful to hunt for novel and distinctive high-energy imprints.

We believe that the very promising statistics of our observables in the FPF~$+$~ATLAS configuration supports the interest of the FPF Community in exploring the intriguing possibility of making FPF and ATLAS detectors work in coincidence. It will require very precise timing procedures, whose technical feasibility needs to be encouraged and complemented by a positive feedback from the theoretical side.

The main outcome of our recent study on the interplay between BFKL and DGLAP in inclusive semi-hard emissions of light-jet and hadrons at the LHC\tcite{Celiberto:2020wpk} is the need for a \emph{multi-lateral} formalism in which several different resummation mechanisms are simultaneously and  consistently implemented, as core element for precision studies of high-energy QCD.
The sensitivity of FPF~$+$~ATLAS results presented in this work on both the high-energy and the threshold resummation is a clear message that the development of such a unified description should be carried with top priority in the medium-term future.
In addition to these perspectives on formal aspects, two main paths can be traced as phenomenological outlooks.

First, we plan to complement our investigation on light mesons emitted in FPF-like kinematic configurations in association with heavy particles detected by ATLAS detectors by considering an opposite configuration, namely when an ultra-forward heavy-flavored hadron is tagged at the FPF while another central object stays inside ATLAS cuts. 
Then, we will study the high-energy behavior of observables sensitive to single inclusive emissions of heavy hadrons reconstructed by FPF detectors (see diagram ($a$) of Fig.\tref{fig:HE-factorization}).
Our project is to access the proton content in ultra-forward (low-$x$) regimes provided by FPF cuts. 
Here, our hybrid factorization could serve as a theoretical common basis to explore production mechanisms and decays of heavy-flavored particles.

Mapping the proton structure in the very $\mbox{low-}x$ regime will rely on a profound exploration of the connections among different approaches. We particularly refer to the interplay between our hybrid factorization, which permits to describe cross sections for single forward emissions in terms of a $\kappa_T$-factorization between off-shell matrix elements and the UGD, and the ABF formalism which, as previously mentioned, allowed us to determine $\mbox{low-}x$ improved collinear PDFs (see also Section~6.1.2 of Ref.\tcite{Feng:2022inv}).

Furthermore, as outlined by model studies of leading-twist gluon TMDs\tcite{Bacchetta:2020vty} (see also Refs.\tcite{Celiberto:2021zww,Bacchetta:2021oht,Bacchetta:2021lvw,Bacchetta:2021twk,Bacchetta:2022esb}), the distribution of linearly-polarized gluons can give rise to spin effects even in collisions of unpolarized hadrons,\footnote{These features are collectively known as Boer--Mulders effect. It was observed first in the quark-polarization case\tcite{Boer:1997nt} (see also Refs.\tcite{Bacchetta:2008xw,Barone:2008tn,Barone:2009hw}).} and the weight of these effects is asymptotically similar to unpolarized-gluon one in the $\mbox{small-}x$ limit. The gluon Boer--Mulders density is easily accessed via inclusive emissions of heavy-flavored objects in hadron collisions, such as the ones that can be studied at the FPF (see Section~6.1.7 of Ref.\tcite{Feng:2022inv}).
In view of these considerations, we plan to employ FPF kinematic ranges as a tool to unveil the connection between the BFKL UGD and the (un)polarized gluon TMDs.

We believe that analyses presented in this work represent a robust step towards precision studies of high-energy QCD.
Our hybrid factorization, possibly enhanced via the inclusion of other resummations, is helpful to perform a systematic reduction of uncertainties coming both from perturbative calculations of high-energy scatterings and from parton densities. This will serve as a benchmark for SM measurements as well as a common basis for searches of New Physics beyond the SM.

\section*{Acknowledgements}

The author is grateful to members of the \textbf{FPF Collaboration} for the exciting conversations and the warm atmosphere of the related meetings which inspired him in realizing the analyses proposed in this article.

The author thanks Christine Aidala for providing native numerical grids of {\tt AESSS11} FFs, Valerio Bertone for recommendations on the use of pion collinear FFs, Ajjath A.H. for insightful discussions on the threshold resummation, and Marco Bonvini for fruitful conversations on challenges and prospects of the use of joint resummations in the formal description of processes at new-generation colliders.

The author would like to express his gratitude to Alessandro Papa, Mohammed M.A. Mohammed and Juan Rojo for a critical reading of the manuscript, for useful suggestions and for encouragement.

The author acknowledges support from the INFN/NINPHA project and thanks the Universit\`a degli Studi di Pavia for the warm hospitality.

\begin{widetext}

\setcounter{appcnt}{0}
\hypertarget{app:NLL_kernel}{
\section*{Appendix~A: NLL BFKL kernel}}
\label{app:NLL_kernel}

The characteristic function entering the expression for NLL correction to the BFKL kernel in Eq.\eref{chi_NLO} reads
\begin{equation}
 \label{kernel_NLO}
 \bar \chi(n,\nu)\,=\, - \frac{1}{4}\left\{\frac{\pi^2 - 4}{3}\chi(n,\nu) - 6\zeta(3) - \frac{\drv^2 \chi}{\drv\nu^2} + \,2\,\phi(n,\nu) + \,2\,\phi(n,-\nu)
 \right.
\end{equation}
\[
 \left.
 +\frac{\pi^2\sinh(\pi\nu)}{2\,\nu\, \cosh^2(\pi\nu)}
 \left[
 \left(3+\left(1+\frac{n_f}{N_c^3}\right)\frac{11+12\nu^2}{16(1+\nu^2)}\right)
 \delta_{n0}
 -\left(1+\frac{n_f}{N_c^3}\right)\frac{1+4\nu^2}{32(1+\nu^2)}\delta_{n2}
\right]\right\} \, ,
\]
with
\begin{equation}
\label{kernel_NLO_phi}
 \phi(n,\nu)\,=\,-\int\limits_0^1 \drv x\,\frac{x^{-1/2+i\nu+n/2}}{1+x}\left\{\frac{1}{2}\left(\psi^\prime\left(\frac{n+1}{2}\right)-\zeta(2)\right)+\mbox{Li}_2(x)+\mbox{Li}_2(-x)\right.
\end{equation}
\[
\left.
 +\ln x\left[\psi(n+1)-\psi(1)+\ln(1+x)+\sum_{k=1}^\infty\frac{(-x)^k}{k+n}\right]+\sum_{k=1}^\infty\frac{x^k}{(k+n)^2}\left[1-(-1)^k\right]\right\}
\]
\[
 =\sum_{k=0}^\infty\frac{(-1)^{k+1}}{k+(n+1)/2+i\nu}\left\{\psi^\prime(k+n+1)-\psi^\prime(k+1)\right.
\]
\[
 \left.
 +(-1)^{k+1}\left[\beta_{\psi}(k+n+1)+\beta_{\psi}(k+1)\right]-\frac{\psi(k+n+1)-\psi(k+1)}{k+(n+1)/2+i\nu}\right\} \; ,
\]
\begin{equation}
\label{kernel_NLO_phi_beta_psi}
 \beta_{\psi}(z)=\frac{1}{4}\left[\psi^\prime\left(\frac{z+1}{2}\right)
 -\psi^\prime\left(\frac{z}{2}\right)\right] \; ,
\end{equation}
and
\begin{equation}
\label{dilog}
\mbox{Li}_2(x) = \int\limits_0^x \drv y \,\frac{\ln(1-y)}{-y} \; .
\end{equation}

The RG-improved kernel (see point~(\emph{vi}) of Section\tref{ssec:jethad_settings}) can be obtained by making the following substitution
\begin{equation}
 \chi^{\rm NLL}(n,\nu) \,\to\, \chi^{\rm NLL}(n,\nu) + \chi^{\rm RG}(n,\nu) \,,
\label{chiRGsub}
\end{equation}
where
\begin{eqnarray}
\chi^{\rm RG}(n,\nu) &=& \sum_{\kappa=0}^{\infty} \left[\sqrt{\left(\frac{1+n}{2} + i \nu + \kappa - \sigma_n \bar \alpha_s\right)^2 + 2 \bar \alpha_s(1 + \rho_n \bar \alpha_s)} -\kappa - i\nu \right. \nonumber\\
&-& \left. \frac{1+n}{2} + \sigma_n \bar \alpha_s 
- \frac{\alpha_s(1 + \rho_n \bar \alpha_s)}{\kappa + \frac{1+n}{2} + i\nu}
- \frac{\bar \alpha_s^2 \sigma_n}{\left( \kappa + \frac{1+n}{2} + i\nu \right)^2}
+ \frac{\bar \alpha_s^2}{\left( \kappa + \frac{1+n}{2} + i\nu \right)^3} \right] \nonumber \\
&+& \, \{\nu \,\to\, -\nu \}
\end{eqnarray}
is the eigenvalue of collinear terms in the transverse-momentum representation, given in the form of a Bessel function\tcite{SabioVera:2005tiv,Vera:2006un,Vera:2007kn}. 
The $\rho_n$ and $\sigma_n$ coefficients read
\begin{eqnarray}
\rho_n &=& \frac{4 - \pi^2 + 5\beta_0/N_c}{12} - \frac{\pi^2}{24} + \frac{\beta_0}{4N_c} \left[ \psi(n+1) - \psi(1) \right] + \frac{1}{2} \psi^\prime (n+1) \\
&+& \frac{1}{8} \left[ \psi^\prime \left(\frac{n+1}{2}\right) - \psi^\prime \left(\frac{n+2}{2}\right) \right] - \frac{1}{36} \delta_n^0 \left(67 + 13 \frac{n_f}{N_c^3} \right) - \frac{47}{1800} \delta_n^2 \left(1 + \frac{n_f}{N_c^3} \right) \;,
\end{eqnarray}
and
\begin{eqnarray}
\sigma_n = - \frac{\beta_0}{8N_c} - \frac{1}{2} \left[ \psi(n+1) - \psi(1) \right] - \frac{1}{12} \delta_n^0 \left(11 + 2\frac{n_f}{N_c^3} \right) - \frac{1}{60} \delta_n^2 \left(1 + \frac{n_f}{N_c^3} \right) \;.
\end{eqnarray}


\setcounter{appcnt}{0}
\hypertarget{app:NLO_IF}{
\section*{Appendix~B: NLO impact factor}}
\label{app:NLO_IF}

We present the analytic expression for the NLO correction of the impact factor depicting the emission of a forward hadron $h$ at large transverse momentum

\begin{equation}
  \label{hadron_IF_NLO}
  \hat c_h(n,\nu,|\vec \kappa_{T_{\cal H}}|,x_{\cal H})=
  \frac{1}{\pi}\sqrt{\frac{C_F}{C_A}}
  \left(|\vec \kappa_{T_{\cal H}}|^2\right)^{i\nu-\frac{1}{2}}
  \int_{x_{\cal H}}^1\frac{\drv x}{x}
  \int_{\frac{x_{\cal H}}{x}}^1\frac{\drv \xi}{\xi}
  \left(\frac{x\xi}{x_{\cal H}}\right)^{2i\nu-1}
\end{equation}
  \[ \times \,
  \left[
  \frac{C_A}{C_F}f_g(x)D_g^h\left(\frac{x_{\cal H}}{x\xi}\right)C_{gg}
  \left(x,\xi\right)+\sum_{i=q\bar q}f_i(x)D_i^h
  \left(\frac{x_{\cal H}}{x\xi}
  \right)C_{qq}\left(x,\xi\right)
  \right.
  \]
  \[ \times \,
  \left.D_g^h\left(\frac{x_{\cal H}}{x\xi}\right)
  \sum_{i=q\bar q}f_i(x)C_{qg}
  \left(x,\xi\right)+\frac{C_A}{C_F}f_g(x)\sum_{i=q\bar q}D_i^h
  \left(\frac{x_{\cal H}}{x\xi}\right)C_{gq}\left(x,\xi\right)
  \right]\, ,
  \]

\begin{equation}
\stepcounter{appcnt}
\label{Cgg_hadron}
 C_{gg}\left(x,\xi\right) =  P_{gg}(\xi)\left(1+\xi^{-2\gamma}\right)
 \ln \left( \frac {|\vec \kappa_{T_{\cal H}}|^2 x^2 \xi^2 }{\mu_F^2 x^2}\right)
 -\frac{\beta_0}{2}\ln \left( \frac {|\vec \kappa_{T_{\cal H}}|^2 x^2 \xi^2 }
 {\mu^2_R x^2}\right)
\end{equation}
\[
 + \, \delta(1-\xi)\left[C_A \ln\left(\frac{s_0 \, x^2}{|\vec \kappa_{T_{\cal H}}|^2 \,
 x^2 }\right) \chi(n,\gamma)
 - C_A\left(\frac{67}{18}-\frac{\pi^2}{2}\right)+\frac{5}{9}n_f
 \right.
\]
\[
 \left.
 +\frac{C_A}{2}\left(\psi^\prime\left(1+\gamma+\frac{n}{2}\right)
 -\psi^\prime\left(\frac{n}{2}-\gamma\right)
 -\chi^2(n,\gamma)\right) \right]
 + \, C_A \left(\frac{1}{\xi}+\frac{1}{(1-\xi)_+}-2+\xi\bar\xi\right)
\]
\[
 \times \, \left(\chi(n,\gamma)(1+\xi^{-2\gamma})-2(1+2\xi^{-2\gamma})\ln\xi
 +\frac{\bar \xi^2}{\xi^2}{\cal I}_2\right)
\]
\[
 + \, 2 \, C_A (1+\xi^{-2\gamma})
 \left(\left(\frac{1}{\xi}-2+\xi\bar\xi\right) \ln\bar\xi
 +\left(\frac{\ln(1-\xi)}{1-\xi}\right)_+\right) \ ,
\]

\begin{equation}
\stepcounter{appcnt}
\label{Cgq_hadron}
 C_{gq}\left(x,\xi\right)=P_{qg}(\xi)\left(\frac{C_F}{C_A}+\xi^{-2\gamma}\right)\ln \left( \frac {|\vec \kappa_{T_{\cal H}}|^2 x^2 \xi^2 }{\mu_F^2 x^2}\right)
\end{equation}
\[
 + \, 2 \, \xi \bar\xi \, T_R \, \left(\frac{C_F}{C_A}+\xi^{-2\gamma}\right)+\, P_{qg}(\xi)\, \left(\frac{C_F}{C_A}\, \chi(n,\gamma)+2 \xi^{-2\gamma}\,\ln\frac{\bar\xi}{\xi} + \frac{\bar \xi}{\xi}{\cal I}_3\right) \ ,
\]

\begin{equation}
\stepcounter{appcnt}
\label{qg}
 C_{qg}\left(x,\xi\right) =  P_{gq}(\xi)\left(\frac{C_A}{C_F}+\xi^{-2\gamma}\right)\ln \left( \frac {|\vec \kappa_{T_{\cal H}}|^2 x^2 \xi^2 }{\mu_F^2x^2}\right)
\end{equation}
\[
 + \xi\left(C_F\xi^{-2\gamma}+C_A\right) + \, \frac{1+\bar \xi^2}{\xi}\left[C_F\xi^{-2\gamma}(\chi(n,\gamma)-2\ln\xi)+2C_A\ln\frac{\bar \xi}{\xi} + \frac{\bar \xi}{\xi}{\cal I}_1\right] \ ,
\]

\begin{equation}
\stepcounter{appcnt}
\label{Cqq_hadron}
 C_{qq}\left(x,\xi\right)=P_{qq}(\xi)\left(1+\xi^{-2\gamma}\right)\ln \left( \frac {|\vec \kappa_{T_{\cal H}}|^2 x^2 \xi^2 }{\mu_F^2 x_{\cal H}^2}\right)-\frac{\beta_0}{2}\ln \left( \frac {|\vec \kappa_{T_{\cal H}}|^2 x^2 \xi^2 }{\mu^2_R x_{\cal H}^2}\right)
\end{equation}
\[
 + \, \delta(1-\xi)\left[C_A \ln\left(\frac{s_0 \, x_{\cal H}^2}{|\vec \kappa_{T_{\cal H}}|^2 \, x^2 }\right) \chi(n,\gamma)+ C_A\left(\frac{85}{18}+\frac{\pi^2}{2}\right)-\frac{5}{9}n_f - 8\, C_F \right.
\]
\[
 \left. +\frac{C_A}{2}\left(\psi^\prime\left(1+\gamma+\frac{n}{2}\right)-\psi^\prime\left(\frac{n}{2}-\gamma\right)-\chi^2(n,\gamma)\right) \right] + \, C_F \,\bar \xi\,(1+\xi^{-2\gamma})
\]
\[
 +\left(1+\xi^2\right)\left[C_A (1+\xi^{-2\gamma})\frac{\chi(n,\gamma)}{2(1-\xi )_+}+\left(C_A-2\, C_F(1+\xi^{-2\gamma})\right)\frac{\ln \xi}{1-\xi}\right]
\]
\[
 +\, \left(C_F-\frac{C_A}{2}\right)\left(1+\xi^2\right)\left[2(1+\xi^{-2\gamma})\left(\frac{\ln (1-\xi)}{1-\xi}\right)_+ + \frac{\bar \xi}{\xi^2}{\cal I}_2\right] \; ,
\]
with $s_0$ an artificial normalization scale that genuinely emerge within the BFKL approach. We set $s_0 = m_{\perp_{\cal M}} m_{\perp_{\cal H}}$.
Furthermore, we define $\bar \xi \equiv 1 - \xi$ and $\gamma = - \frac{1}{2} + i \nu$. Expressions for the $P_{i j}(\xi)$ LO DGLAP splitting kernels are
\begin{eqnarray}
\stepcounter{appcnt}
\label{DGLAP_kernels}
 P_{gq}(z)&=&C_F\frac{1+(1-z)^2}{z} \; , \\ \nonumber
 P_{qg}(z)&=&T_R\left[z^2+(1-z)^2\right]\; , \\ \nonumber
 P_{qq}(z)&=&C_F\left( \frac{1+z^2}{1-z} \right)_+= C_F\left[ \frac{1+z^2}{(1-z)_+} +{3\over 2}\delta(1-z)\right]\; , \\ \nonumber
 P_{gg}(z)&=&2C_A\left[\frac{1}{(1-z)_+} +\frac{1}{z} -2+z(1-z)\right]+\left({11\over 6}C_A-\frac{n_f}{3}\right)\delta(1-z) \; .
\end{eqnarray}
The ${\cal I}_{1,2,3}$ functions read
\begin{equation}
\stepcounter{appcnt}
\label{I2}
{\cal I}_2=
\frac{\xi^2}{\bar \xi^2}\left[
\xi\left(\frac{{}_2F_1(1,1+\gamma-\frac{n}{2},2+\gamma-\frac{n}{2},\xi)}
{\frac{n}{2}-\gamma-1}-
\frac{{}_2F_1(1,1+\gamma+\frac{n}{2},2+\gamma+\frac{n}{2},\xi)}{\frac{n}{2}+
\gamma+1}\right)\right.
\end{equation}
\[
 \stepcounter{appcnt}
 \left.
 +\xi^{-2\gamma}\left(\frac{{}_2F_1(1,-\gamma-\frac{n}{2},1-\gamma-\frac{n}{2},\xi)}{\frac{n}{2}+\gamma}-\frac{{}_2F_1(1,-\gamma+\frac{n}{2},1-\gamma+\frac{n}{2},\xi)}{\frac{n}{2} -\gamma}\right)
\right.
\]
\[
 \left.
 +\left(1+\xi^{-2\gamma}\right)\left(\chi(n,\gamma)-2\ln \bar \xi \right)+2\ln{\xi}\right] \; ,
\]
\begin{equation}
\stepcounter{appcnt}
\label{I1}
 {\cal I}_1=\frac{\bar \xi}{2\xi}{\cal I}_2+\frac{\xi}{\bar \xi}\left[\ln \xi+\frac{1-\xi^{-2\gamma}}{2}\left(\chi(n,\gamma)-2\ln \bar \xi\right)\right] \; ,
\end{equation}
\begin{equation}
\stepcounter{appcnt}
\label{I3}
 {\cal I}_3=\frac{\bar \xi}{2\xi}{\cal I}_2-\frac{\xi}{\bar \xi}\left[\ln \xi+\frac{1-\xi^{-2\gamma}}{2}\left(\chi(n,\gamma)-2\ln \bar \xi\right)\right] \; ,
\end{equation}
and ${}_2F_1$ is the ordinary hypergeometric special function.

We introduced in Eqs.~(\ref{Cgg_hadron}) and~(\ref{Cqq_hadron}) the \emph{plus~prescription}
\begin{equation}
\label{plus-prescription}
\stepcounter{appcnt}
\int\limits^1_a \drv \xi \frac{F(\xi)}{(1-\xi)_+}
=\int\limits^1_a \drv \xi \frac{F(\xi)-F(1)}{(1-\xi)}
-\int\limits^a_0 \drv \xi \frac{F(1)}{(1-\xi)}\; ,
\end{equation}
where $F(\xi)$ is a generic function regular behaved when $\xi=1$.

\vspace{1.00cm}

\end{widetext}

\bibliography{references}

\end{document}